\documentclass[gmd, manuscript]{copernicus}

\usepackage{silence}
\usepackage{amsthm}
\usepackage{float}
\usepackage{subcaption}
\usepackage[]{graphicx}
\usepackage{pgfplots}
\pgfplotsset{compat=newest}
\usepackage{adjustbox}
\usepackage{xurl}
\usepackage{xspace}

\usepackage[utf8]{inputenc} 
\nolinenumbers

\newcommand{\plotbox}[1]{#1}
\newcommand{\ignore}[1]{}

\newcommand{\huhydro}{\ensuremath{\overline{hu}}\xspace}
\newcommand{\hhydro}{\ensuremath{\overline{h}}\xspace}

\newcommand{\huint}{\ensuremath{(hu)_{int}}\xspace}
\newcommand{\hint}{\ensuremath{h_{int}}\xspace}
\newcommand{\uint}{\ensuremath{u_{int}}\xspace}

\newcommand{\Sec}[1]{Section \ref{sec:#1}}

\newcommand{\Fig}[1]{Figure \ref{fig:#1}}

\begin{document}

\title{GeoFlood (v1.0.0): Computational model for overland flooding}

% \Author[affil]{given_name}{surname}

\Author[1]{Brian}{Kyanjo}
\Author[1]{Donna}{Calhoun}
\Author[2]{David L.}{George}

\affil[1]{Department of Mathematics, Boise State University, Boise, ID, 83706, USA}
\affil[2]{U.S. Geological Survey, Cascades Volcano Observatory, Vancouver, WA, USA}

%% The [] brackets identify the author with the corresponding affiliation. 1, 2, 3, etc. should be inserted.

%% If an author is deceased, please mark the respective author name(s) with a dagger, e.g. "\Author[2,$\dag$]{Anton}{Smith}", and add a further "\affil[$\dag$]{deceased, 1 July 2019}".

%% If authors contributed equally, please mark the respective author names with an asterisk, e.g. "\Author[2,*]{Anton}{Smith}" and "\Author[3,*]{Bradley}{Miller}" and add a further affiliation: "\affil[*]{These authors contributed equally to this work.}".

\correspondence{Brian Kyanjo (briankyanjo@u.boisestate.edu)}

\runningtitle{GeoFlood: Computational model for overland flooding}

\runningauthor{Brian Kyanjo, Donna Calhoun, David L. George}

\received{}
\pubdiscuss{} %% only important for two-stage journals
\revised{}
\accepted{}
\published{}

%% These dates will be inserted by Copernicus Publications during the typesetting process.

\firstpage{1}

\maketitle

\begin{abstract}
	This paper presents GeoFlood, a new open-source software package for solving the shallow-water equations (SWE) on a quadtree hierarchy of mapped, logically Cartesian grids managed by the parallel, adaptive library ForestClaw \citep{ca-bu:2017}. The GeoFlood model is validated using standard benchmark tests from \cite{neelz2013benchmarking} as well as the historical Malpasset dam failure. The benchmark test results are compared against those we obtained from GeoClaw \citep{clawpack} and the software package HEC-RAS (Hydraulic Engineering  Center - River Analysis System, Army Corp of Engineers) \citep{brunner2018}. The Malpasset outburst flood results are compared with those presented in \cite{george2011adaptive} (obtained from the GeoClaw software), model results from \citet{hervouet1999malpasset}, and empirical data. The comparisons validate GeoFlood's capabilities for idealized benchmarks compared to other commonly used models as well as its ability to efficiently simulate highly dynamic floods in complex terrain, consistent with historical field data. Because it is massively parallel and scalable, GeoFlood may be a valuable tool for efficiently computing large-scale flooding problems at very high resolutions.
\end{abstract}

% \copyrightstatement{TEXT} %% This section is optional and can be used for copyright transfers.

\introduction  %% \introduction[modified heading if necessary]
\label{sec:intro}

Overland flooding simulation is critical to society for hazard mitigation because, among natural hazards, flooding is a leading cause of casualties and property damage.  Examples include the 1959 Malpasset flood, caused by the Malpasset dam failure, which killed at least 423 people, injured 83 and caused nearly 103.5 million francs in damages \citep{boudou2017analysis}; the 1993 Mississippi River floods resulting from extreme weather and hydrologic conditions killed at least 47 people and caused nearly 20 billion dollars in damages \citep{missipi}; the 2013 Colorado floods triggered by heavy rains resulting from a slow cold front colliding with warm humid monsoonal air killed at least 9 people and cost an estimated 4 billion dollars in damages \citep{coloflood}. The potential for future devastating floods caused by the failure of still-operational dams remains. For instance, a failure of the Mosul dam in Iraq, considered the most dangerous dam in the world, could cause a catastrophic flood affecting millions of people and costing billions of dollars in damages \citep{mosul2017}. Simulating the inundation extent and timing of these potential events aids the engineering of mitigation strategies in emergency planning and infrastructure development.

Numerical simulation of advancing water over topography is a powerful tool for understanding and predicting the behavior of overland flooding in complex environments. However, floods occur on large spatial domains over many hours and must be represented with suitable yet tractable mathematical models. Historically, researchers have utilized one-dimensional (1D) channel-flow models and, more recently, the two-dimensional (2D) shallow-water equations (SWE)---a system of hyperbolic partial differential equations (PDEs) for depth-averaged conservation of mass and momentum.  While solving a full three dimensional model of overland flooding might capture more flow detail, computationally efficient and robust models for the SWE are capable of handling overland flows in complex terrain. Nevertheless, developing numerical models based on SWE remains a challenging problem with an active research community. In the last few decades, shock-capturing finite-volume methods have become the dominant class of numerical schemes for this problem, due to their accurate and robust nature for hyperbolic problems. In addition to the codes described in this paper, researchers have developed a variety of such schemes solved on a variety of mesh structures; for example, \cite{valiani2002case} designed a classical Godunov scheme to solve the SWE on a static fitted mesh, \cite{kirstetter} presented an open-source Saint-Venant model (a 2D finite volume solver) for solving SWE on adaptively refined meshes. In addition, \cite{YU2021126262} created a 2D shallow-water equation model using finite volume techniques to model overland flows in rural and urban areas.  More recently,  \cite{w12082120}, \cite{ali:2023}, and \cite{mrousseau:2015} devloped similar SWE-based models for modeling overland flow.

In this paper, we present GeoFlood, a new computational model that employs the wave-propagation algorithms (WPA) utilized in Clawpack \citep{clawpack}, the augmented Riemann solvers available in the software GeoClaw \citep{george2008}, and the parallel, adaptive library ForestClaw \citep{ca-bu:2017}, to solve the shallow-water equations on mapped, logically Cartesian adaptive meshes (\Sec{software}).  

 GeoFlood demonstrates the ability and benefits of simulating overland flows using a parallel tree-based, adaptive mesh refinement (AMR) structure, verified through comparisons with GeoClaw and HEC-RAS (Hydraulic Engineering  Center - River Analysis System) results for some standard benchmark problems \citep{neelz2013benchmarking}.  The model is also validated against \cite{george2011adaptive} GeoClaw results for the Malpasset dam break, which provides an ideal test case for the model because it involves nearly instantaneous dam-break initial conditions (the dam collapsed suddenly and catastrophically) followed by downstream flow through complex irregular terrain (\Sec{malpasset}). This instantaneous, catastrophic event is a well-known benchmark problem in the field of overland flooding due in part to the existence of extensive downstream field data, including timing information. \ignore{ see, for instance, \cite{hervouet1996numerical,hervouet1999malpasset,johnstone2005, valiani2002case, george2011adaptive}. }

\section{Software Overview}
\label{sec:software}
In this section we summarize the software packages and libraries that provide components from which GeoFlood is built. We also briefly describe other overland flood modeling packages (GeoClaw and HEC-RAS) that we used for comparison and validation. 

\subsection{GeoFlood's fundamental building libraries} 
GeoFlood incorporates components from other open-source software projects along with new code that integrates these diverse libraries. This section provides an overview of these previously developed codes. \Sec{GF} provides a broad overview of the standalone package, GeoFlood, including novel features that facilitate simulating and analyzing results for overland flooding applications.
 
\subsubsection{Clawpack and GeoClaw}
\label{sec:GeoClaw}
The GeoClaw software is a submodule of Clawpack \citep{clawpack}, an open-source software package for solving general hyperbolic systems of PDEs. GeoClaw was initially developed as an extension to Clawpack for tsunami modeling \citep{george2006finite}, but has since been extended to overland flooding problems \citep{george2011adaptive} and hurricane-generated storm surges \citep{mandli2013numerical}. GeoClaw combines finite-volume wave-propagation algorithms in Clawpack \citep{leveque2002}, Riemann solvers for shallow-water wave equations \citep{george2008}, patch-based adaptive mesh refinement (AMR) schemes \citep{berger1984adaptive,berger1989local} with interpolation schemes developed for free-surface flows over topography, and methods for integrating general sets (overlapping, misaligned, etc.) of digital elevation models (DEMs) in a manner that preserves volume conservation in the adaptive setting. See \citet{george2006finite,leveque2011} for details. Clawpack and GeoClaw have undergone extensive development over the past several decades \citep{mandli2016clawpack} and are actively maintained by the Clawpack development team. 

\subsubsection{p4est}
\label{sec:p4est}
The p4est code is a robust library for the parallel computation of adaptive, hierarchical tree meshes, including efficient parallel algorithms for creating, refining, and distributing those meshes. The p4est mesh-management library distributes a quadtree or octree mesh across multiple processors using a Message-Passing Interface (MPI), providing a scalable and fault-tolerant framework for large-scale simulations. It is designed to be compatible with various parallel-computing architectures and can scale to millions of processor cores \citep{Burstedde}.

\subsubsection{ForestClaw}
\label{sec:fclaw}
The ForestClaw library is built as a PDE layer on top of the p4est library for parallel tree-mesh management.   While it can be used with any Cartesian-based patch solver, ForestClaw makes extensive use of the wave-propagation algorithms in Clawpack for solving a variety of hyperbolic problems.  The resulting ForestClaw library is an adaptive, parallel, multi-block structured finite volume code that parallelizes the solution of hyperbolic PDEs on mapped, logically Cartesian meshes \citep{ca-bu:2017}. The GeoClaw extension of ForestClaw incorporates the SWE solvers and AMR interpolation schemes from GeoClaw and is expected to produce numerically consistent results, but on tree-based meshes. This GeoClaw extension serves as the basis for the new standalone code called GeoFlood, the software package that is the focus of this paper.

\subsection{GeoFlood overview}
\label{sec:GF}
GeoFlood is a standalone package that uses the ForestClaw library, the Riemann solvers in GeoClaw, and other features tailored toward modeling problems in overland flooding. GeoFlood offers several advantages over GeoClaw and introduces enhancements to ForestClaw targeted toward improved overland flood modeling. Like GeoClaw, GeoFlood can restrict and optimize grid refinement to user-specified spatial and temporal regions of interest.  A key advantage of GeoFlood over GeoClaw, however,  is that it can be run efficiently on large distributed parallel platforms.   The tree-based communication patterns inherited from ForestClaw and p4est allow for simplified load balancing and a decentralized, distributed regridding algorithm.   The multi-resolution grid hierarchy in a typical GeoFlood mesh is composed of composite structures of non-overlapping fixed-sized grids, each stored as a block in a quadtree multi-block forest.  This allows for easy data storage on each processor and for fast neighbor searches. The Cartesian grid layout of each patch in a quadrant simplifies communication among patches. 

GeoFlood is written in C, C++ and Fortran and makes use of Python scripts available in Clawpack and GeoClaw for providing input parameters.  The build system is managed using CMake, which facilitates the set-up of general overland flooding problems. This design offers the flexibility to integrate alternative Riemann solvers, other numerical methods, or custom scripts for a specific application. Python scripts are provided that can download and handle topography files, specific problem parameters, retrieve initial conditions, and generate a GeoFlood configuration file from user-defined settings and inputs. The configuration file is then read by GeoFlood via the command line while running in serial or parallel mode. Documentation on the installation and operation of GeoFlood is provided on the GeoFlood Wiki \citep{gfloodwiki}.

The GeoFlood model can generate a series of frames of the simulation domain in Cartesian or latitude and longitude coordinates.  A Python script reads the output frames and generates Keyhole Markup Language \citep{KML-OGC} files that can be read in the  Google Earth browser. In cases where the computational domain coordinates are not latitude and longitude, an automated Python routine has been designed to read user-specified ground control points within the domain and georeference them to a latitude and longitude coordinate system, even when the coordinate projection is unknown (\Sec{malpasset}). This allows the GeoFlood simulation frames to be visualized on Google Earth.

\subsection{Software for comparison with GeoFlood: GeoClaw and HEC-RAS}

We used GeoClaw and HEC-RAS for validation of GeoFlood on several idealized benchmark problems and a field-scale flood.  

Although most commonly used for tsunami simulations, GeoClaw has been used for outburst-flood simulations that utilize AMR to dynamically resolve the evolving flood domain. For example, \cite{george2011adaptive} simulated the outburst flood resulting from the $1959$ Malpasset dam failure. The results for high-water marks and flood arrival times at nine gauge locations were validated against field data and experimental results from a scaled laboratory model \citep{frazaodam,morris2000concerted}. \cite{spero2022simulating} further tested GeoClaw's overland flooding capabilities by extending it to the simulation of the $1976$ Teton dam rupture. The results were in agreement with historical observations as well as HEC-RAS \ignore{(\Sec{hec-ras})} simulations of the same event.

% \subsubsection{HEC-RAS}
\label{sec:hec-ras}
HEC-RAS consists of 1D and 2D hydraulic modeling software developed by the U.S. Army Corps of Engineers \citep{brunner2002hec}. It utilizes a variety of numerical schemes for different applications. The model we test against uses an implicit finite-volume method to solve the shallow-water wave equations on uniformly structured grids and is capable of modeling steady or unsteady flows, and sediment transport in complex terrain. HEC-RAS is widely used in government and industry for levee breach analysis and is considered the industry standard for floodplain modeling \citep{brunner2018}. It has been used extensively to simulate dam-break floods, including the $1976$ Teton dam failure \citep{spero2022simulating}, the $2006$ Ukai dam failure \citep{patel2017assessment}, and the Temenggor dam failure \citep{shahrim2020dam}. In this paper, we compare GeoFlood results with those computed using the Eulerian-Lagrangian SWE solver (SWE-ELM) available in HEC-RAS 6.3.1. 

% \label{sec:geoflood}

\section{Governing models and numerical algorithms}
\label{sec:numerical_methods}
\subsection{The shallow-water equations}
The dynamics of floods in rugged terrain varies in three dimensions (3D); however, the depth-averaged (2D) shallow water equations are widely considered to be a suitable and tractable approximation for determining the extent and timing of a flood scenario for the assessment of hazards \citep[e.g.,][]{george2011adaptive,bai2016depth,altaie2018,qin2018}. These equations are derived by integrating the 3D Euler (inviscid Navier-Stokes) equations over the vertical $z$-direction from the solid bed to the free surface of the flow and applying kinematic boundary conditions at these surfaces.
When combined with the simplifying assumption of vertically hydrostatic pressure (by neglecting higher-order terms in the asymptotic expansion of the vertical momentum equations under the long-wave or shallowness assumption \citep[e.g.,][]{vreugdenhil1994numerical}), this leads to the nonlinear shallow water equations (SWE) for conservation of mass and momentum. 

The governing equations used in GeoFlood are a SWE system of hyperbolic PDEs given by
\begin{subequations} \label{shallow}
  \begin{eqnarray}
    \frac{\partial h}{\partial t} + \frac{\partial (hu)}{\partial x} + \frac{\partial (hv)}{\partial y} &=& 0, \label{shallow1} \\
    \frac{\partial (hu)}{\partial t} + \frac{\partial (hu^2 + \frac{1}{2}gh^2)}{\partial x} + \frac{\partial (huv)}{\partial y} &=& -gh\frac{\partial b}{\partial x} - S_{fx}, \label{shallow2} \\
    \frac{\partial (hv)}{\partial t} + \frac{\partial (huv)}{\partial x} + \frac{\partial (hv^2 + \frac{1}{2}gh^2)}{\partial y} &=& -gh\frac{\partial b}{\partial y} - S_{fy},  \label{shallow3}
    \end{eqnarray}
    \label{shallow123}
\end{subequations}
where $h(x,y,t)$ is the water depth, $u(x,y,t)$ and $v(x,y,t)$ are the depth-averaged horizontal velocities in the $x$ and $y$ directions respectively, $g$ is the gravitational acceleration, $S_{fx}$ and $S_{fy}$ are the friction slopes in the $x$ and $y$ directions, respectively, and  $b(x,y)$ is the bed elevation. The friction slopes are commonly obtained from the empirical resistance relationships in the Manning equations \citep[e.g.,][]{molls1998} which are given by
\begin{equation}\label{eq:manning}
   \begin{aligned}
S_{fx} &= n^2guh^{-4/3} \sqrt{u^2 + v^2},\\
S_{fy} &= n^2gvh^{-4/3} \sqrt{u^2 + v^2},
\end{aligned} 
\end{equation}
where $n$ is Manning's roughness coefficient depicting the roughness of the bed surface.

\subsection{Finite-volume discretizations}
Finite-volume discretizations are widely used for hyperbolic systems generally \citep[e.g.,][]{leveque2002} and overland flood modeling in particular because they provide a framework that is robust in the presence of drying regions, can capture discontinuities (weak solutions to equations \eqref{shallow123}) such as hydraulic bores or non-smooth topography, can be made well-balanced with respect to near-steady flows, and can resolve the inundating shoreline and run-up features; see for instance   \cite{george2008,george2011adaptive},  \cite{zhao2022novel}, \cite{song2011,song2012}, \cite{caleffi2003}, and \cite{yoshioka2014}.

In one space dimension, the SWE can be expressed in compact conservative form as
\begin{equation}
    q_t + f(q)_x = \Psi(x,q),
    \label{conservative_}
\end{equation}
where the vector $q$ represents conserved quantities $q = [h,hu]^{T}$, $f(q)$ represents the corresponding fluxes, 
\begin{equation}
f(q) = 
\begin{bmatrix}
h u \\ hu^2 + \frac{1}{2}g h^2
\end{bmatrix},
\end{equation}
and $\Psi(x,q)$ includes bathymetry and friction source terms.
The system \eqref{conservative_} can be expressed in quasi-linear form as,
\begin{equation}
    q_t + A(q) q_x = \Psi(x,q),
    \label{conservative}
\end{equation}
where $A(q)$ is the flux-Jacobian given by
\begin{equation}
A(q):=f'(q) = 
\begin{bmatrix}
0 & 1\\ -u^2 + gh & 2u
\end{bmatrix}.
\end{equation} 
Hyperbolicity of the 1D shallow-water equations is conferred by the eigenstructure of $A(q)$, as its eigenvalues (\emph{i.e.}, wave speeds, $u\pm\sqrt{gh}$), are real and distinct for $h>0$.
\par
Consider $C_{i} = [x_{i-\frac{1}{2}},x_{i+\frac{1}{2}}]$ to be the $i^{th}$ grid cell, the average value over the $i^{th}$ cell at time $t_{n}$ is given by Equation \eqref{wpa0}.
	\begin{equation}
		Q_{i}^{n} \approx \dfrac{1}{\Delta x} \int_{C_{i}}q(x,t_{n})dx,
		\label{wpa0}
	\end{equation}
where $\Delta x$ is the cell size. The vector $q$ represents an exact spatially-varying cell solution at time $t_{n}$. The wave-propagation algorithm \citep{leveque1997wave,leveque2002} updates the numerical solution from $Q_{i}^{n}$ to $Q_{i}^{n+1}$ by solving Riemann problems at the boundaries of cell $C_i$ and directly re-averaging the resulting waves onto adjacent grid cells.  

The solution update, neglecting the source terms, can be accomplished using the first-order method of the form 
\begin{equation}
    Q_{i}^{n+1} = Q_{i}^{n} -\frac{\Delta t }{\Delta x } \left( \mathcal{A^{-}}\Delta Q_{i+\frac{1}{2}}^{n} + \mathcal{A^{+}}\Delta Q_{i-\frac{1}{2}}^{n} \right),
    \label{wpa_update}
\end{equation}
where $\Delta Q_{i+\frac{1}{2}}^{n} = Q_{i}^{n}-Q_{i-1}^{n}$, and the fluctuations (flux-differences) $\mathcal{A^{-}}\Delta Q_{i+\frac{1}{2}}^{n}$  and $\mathcal{A^{+}}\Delta Q_{i-\frac{1}{2}}^{n}$ represent propagating waves (jump-discontinuities) carrying units of flux. These waves contribute the net effect of all left- and right-going waves propagating into the cell $C_i$ from its right and left boundaries respectively. Additional higher-order correction terms can be added to the wave-propagation method \eqref{wpa_update} to achieve second-order accuracy in smooth regions:
\begin{equation}
    Q_{i}^{n+1} = Q_{i}^{n} -\frac{\Delta t }{\Delta x } \left( \mathcal{A^{-}}\Delta Q_{i+\frac{1}{2}}^{n} + \mathcal{A^{+}}\Delta Q_{i-\frac{1}{2}}^{n} \right) -\frac{\Delta t }{\Delta x } \left( \Tilde{F}_{i+\frac{1}{2}}^{n} - \Tilde{F}_{i-\frac{1}{2}}^{n} \right). 
    \label{wpa_update_corrections}
\end{equation}
The second-order correction fluxes $\Tilde{F}_{i\pm\frac{1}{2}}^{n}$ are obtained from the Riemann solutions and are adjusted using limiters that ensure a total-variation diminishing (TVD) update, preserving the first-order method's convergence to discontinuous weak solutions satisfying the integral form of the system \eqref{conservative_} (\emph{i.e.,} shock-capturing). The effect of the source term, $\Psi(x,q)$, is traditionally integrated separately by using a splitting or fractional-step method, in conjunction with the homogeneous update \eqref{wpa_update_corrections}, or is incorporated directly into Riemann solutions in various ways to achieve well-balancing \cite[e.g.,][]{leveque2002,bale2003,george2008}. For details on the extension of \eqref{wpa_update_corrections} to two dimensions, see \cite{leveque2002}.

\subsection{Augmented Riemann Solver}
Modeling flooding extent in highly variable and irregular topography is challenging due to the need to numerically balance large flux gradients (from variable depth) and source terms resulting from variable topography. The problem is further complicated by the time-varying solution domain caused by moving wet-dry boundaries. The GeoFlood code employs an approximate Riemann solver developed by \citet{george2008} that solves an augmented SWE system that includes the momentum flux and topographic bed ($b$) as state variables in order to determine stationary and propagating waves in the approximate Riemann solution. It provides a numerical update similar to the f-wave formulation of the wave-propagation algorithm \citep{leveque2002,bale2003}, where the effect of the topographic source term (the right side of the shallow-water equations \eqref{shallow123}) is incorporated into the Riemann solution and, consequently, into \eqref{wpa_update_corrections}. This approach eliminates the need for a poorly balanced fractional-step treatment of the source term. The Riemann solver provides well-balanced resolution of pools or steady-flow---a numerically challenging but common flow condition for overland flooding over topography. The augmented solver employs a wave-speed estimate formulation that reduces to the HLLE (Harten, Lax, van Leer, and Einfeldt) wave-speed estimates \citep{ein-god:1988, ein-god:1991} in interior (wet) regions, ensuring entropy-satisfying depth positivity and accurate shock capturing based on Roe averages \citep{ro:1981}. The wave-speeds reduce to those of dam-break front speeds for moving wet-dry boundaries. Additionally, the solver handles flows against structures or highly irregular topography by solving artificial Riemann test problems (\emph{i.e.}, problems in which the data in one cell are manipulated to determine the physically relevant solution in a neighboring cell) (see \citet{leveque2002,george2008,george2011adaptive}). 

Coupling these capabilities with block-structured AMR capabilities provided by the ForestClaw library gives GeoFlood the ability to robustly handle situations with complex topography and abruptly moving wet-dry fronts in simulating overland flows.

\section{Adaptive mesh refinement using quadtree meshing}
\label{sec:amr}
The multi-resolution grid hierarchy in ForestClaw is a composite structure of non-overlapping fixed-sized grids (e.g. $32~\times~32$), each stored as a leaf in a quad or octree forest. The multi-block structure available in the forest-of-trees paradigm inherited from p4est makes it possible to solve problems on, for example, the cubed-sphere. 

The refinement strategy uses a constant refinement scale factor of $2$, ensuring that progression through levels is sequential (e.g., Level 0 to Level 1 to Level 2). Skipping levels is not permitted. For a grid block of fixed size $m_x \times m_y$ (dimensions in the $x$ and $y$ directions) within a $m_i \times m_j$ arrangement of blocks (e.g. trees) the total grid size at the coarsest level (Level 0) is 
$(m_i \cdot m_x) \times (m_j \cdot m_y)$. The effective resolution (e.g. total number of cells in a uniformly refined grid) at  refinement level $l$ is $(m_i \cdot m_x \cdot 2^l) \times (m_j\cdot m_y  \cdot 2^l )$. At the finest level of refinement ($l_{\text{max}}$), the effective resolution ($E$) is $(m_i \cdot m_x \cdot 2^{l_{\text{max}}}) \times (m_j \cdot m_y \cdot 2^{l_{\text{max}}})$. \ignore{For a domain with physical dimensions $L_x \times L_y$ (dimensions in $x$ and $y$ directions), the resolution per grid cell is given by $L_x/(m_i \cdot m_x \cdot 2^{l_{\text{max}}})$.}  Ideally, the block arrangement is chosen so that for non-square domains of dimensions $L_x \times L_y$, mesh cells are approximately square.

\begin{figure}[ht]
	\begin{subfigure}[b]{0.45\textwidth}
	  \centering
	  \plotbox{\includegraphics[width=1.0\textwidth]{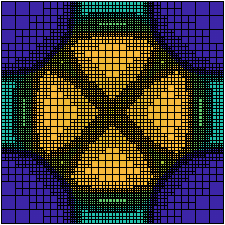}}
	  \caption{Shallow-water wave equation simulation in a box.  Wave structure shows reflections off of solid wall boundaries. }
	  \label{fig:SWE_soln}
	\end{subfigure}
	\hfil
 % \medskip
	\begin{subfigure}[b]{0.45\textwidth}
	  \centering
	  \plotbox{\includegraphics[width=\textwidth,clip=true,trim=1cm -5cm 2cm 1cm]{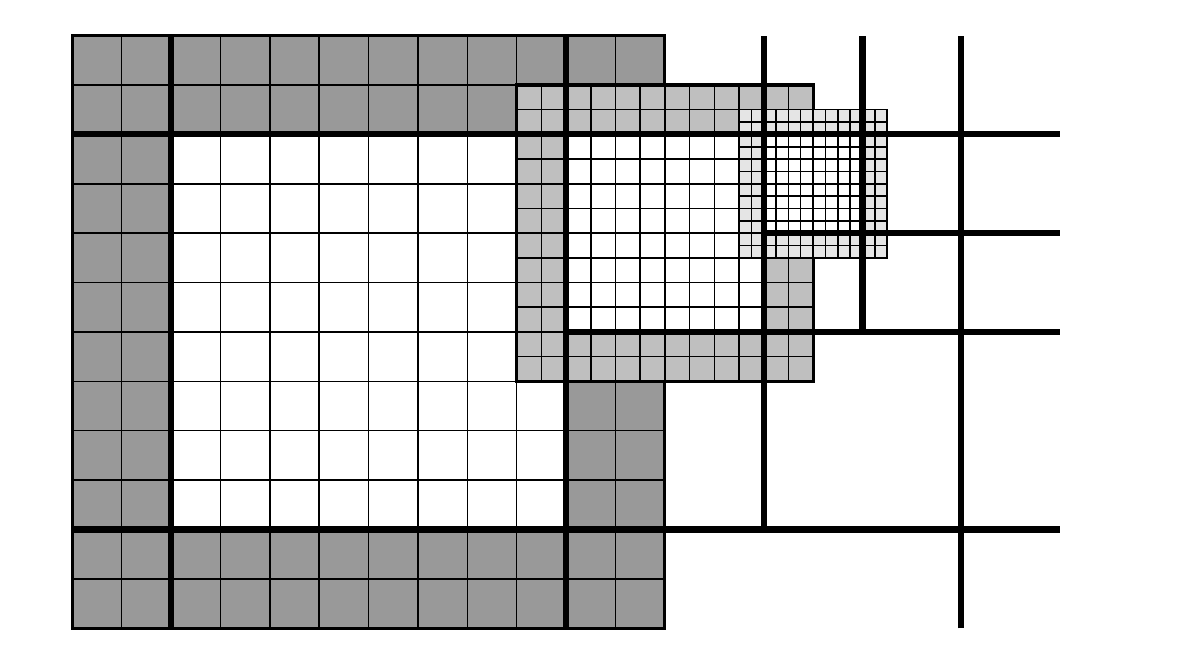}}
	  \caption{Quadtree of patches \citep{ca-bu:2017}}
	  \label{fig:quad}
	\end{subfigure}
	\caption{The left figure depicts an adaptively refined ForestClaw solution to the shallow-water equations on a Cartesian grid in the quadtree layout on a single block. The right figure depicts three adjacent adaptive levels, each with an $8~\times~8$ simulation grid (with thick borders) and a layer of ghost cells (shaded gray). Quadrant boundaries are indicated by thick lines.}
	\label{fig:Cartesian_mesh}
  \end{figure}

\subsection{GeoFlood refinement criteria}
A new AMR strategy based on flags imposed by different refinement criteria has been developed in the GeoFlood code. These include: 1) Water depth criteria, where refinement is permitted only in wet cell regions by imposing a flag on cells with water depths greater than a certain threshold value and the refinement level is determined by the water depth. 2) Velocity-depth product flag, used to force refinement in shallow regions where the flow changes rapidly such as near river banks or shorelines. 3) Velocity criteria, which assume that the magnitude of the water velocity in both $x-$ and $y-$ directions is greater than a certain threshold value. 4) Flood source flags, used to force refinement in regions containing the flood source, i.e., the dam in the case of a dam break. This allows the code to refine the flood source at high resolution to capture the flood details along the floodplain and allows the specification of regions to be refined to a given desired resolution by user-specified coordinates and minimum and maximum refinement levels. 5) Flow-grades flag, where refinement is enforced to given levels for depths or velocities greater than user-defined thresholds. This enables the code to refine regions containing lakes, seas, or rivers in the floodplain at varying intermediate levels compared to the flowing material.

\section{Benchmark Test Cases}
We selected a series of benchmark scenarios, designed by the United Kingdom Environment Agency \citep{neelz2013benchmarking}, to evaluate the capabilities of GeoFlood within the context of flood risk management. This evaluation focused on the model's precision in replicating flood progression and the extent of inundation across diverse physical landscapes and configurations. The performance of GeoFlood in these benchmark tests was evaluated by comparing results with those obtained from both HEC-RAS and GeoClaw under identical test conditions. These test cases have also been employed previously in the benchmarking of various other models, including HEC-RAS. For a more comprehensive overview of these benchmarking studies, see, for example, \citet{brunner2018}, \citet{cea2020benchmarking}, and \citet{neelz2013benchmarking}.

\subsection{Test Case 1: Speed of Flood Propagation over an Extended Floodplain }
\label{sec:flood_speed}
The first benchmark was designed to test a model's ability to resolve an advancing flood front over a floodplain in two-dimensions. The test itself consists of a relatively large, flat (uniform elevation) bed with inflow along a relatively short inlet on one boundary. Although the test involves highly idealized geometry and inflow, it is designed to accentuate model differences or error because the flow is very thin (< 1 m in depth) and it advances for several hours over a relatively large domain (1000 m by 2000 m). Errors in the propagation of a thin wet-dry front are enhanced over large domains and long computation times. Furthermore, the leading edge of an advancing front is prone to spurious numerical oscillations in depth and speed \citep[e.g.,][]{brunner2018, george2008}. Lastly, the problem is nearly radially symmetric which facilitates assessments of 2D algorithms at arbitrary angles on non-radially symmetric meshes.

\subsubsection{Problem Setup}
\label{sec:flood_speed_setup}
\begin{figure}[H]
  \centering
  \includegraphics[width=0.4\textwidth,clip=true,trim=0cm 1.5cm 0cm 2cm]{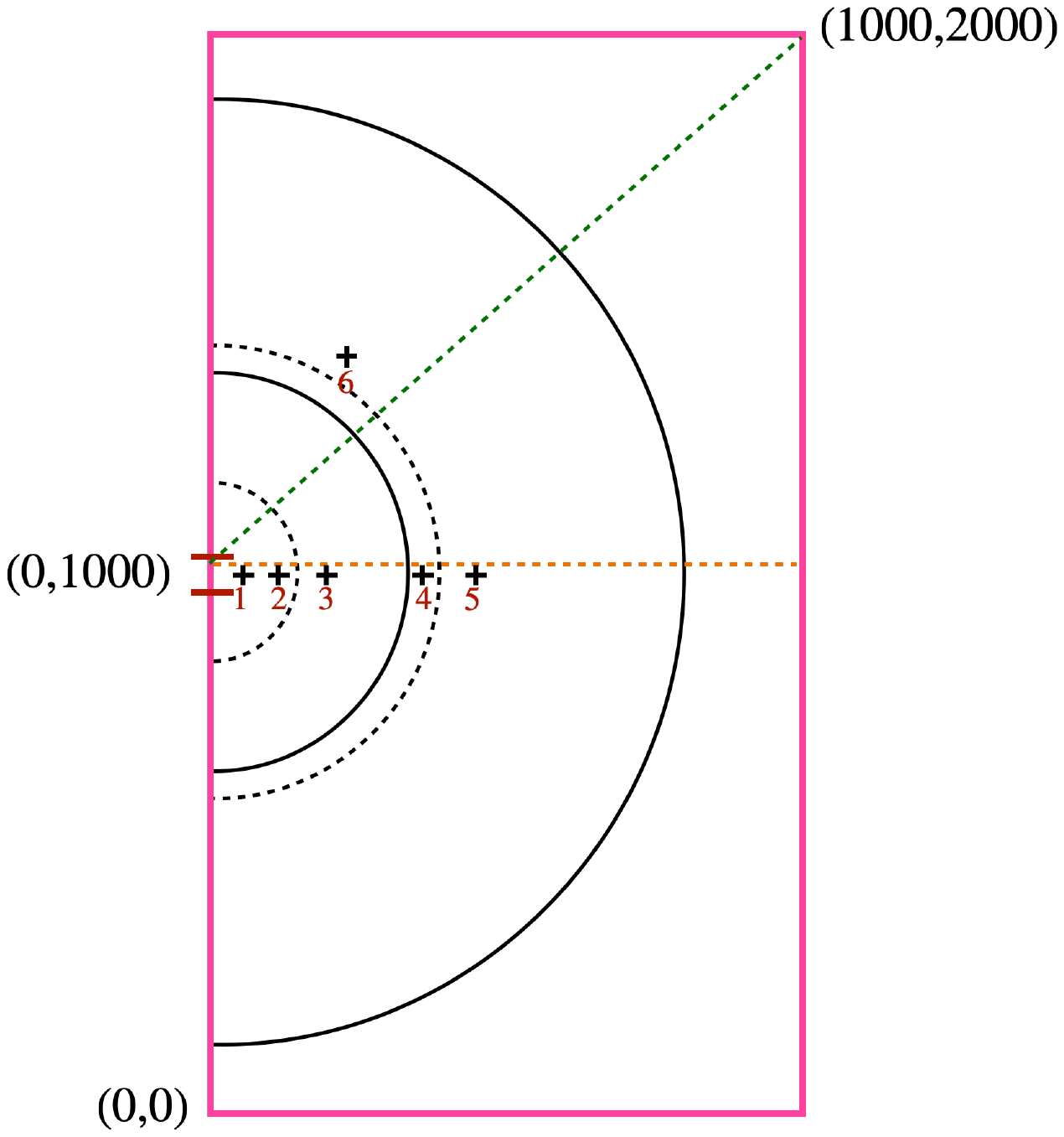}
  \caption{Spatial domain displaying the inflow location (boundary condition) along $20$ m of the left boundary ($x=0$ and $990\le y \le 1010$ m). Level-set contours are shown for the depths $h=10$ and $h=20$ cm at $t=1$~\unit{hour} (black dashed curves) and $t=3$~\unit{hours} (black solid curves), indicating the nearly radially symmetric solution. The diagonal (green dashed) and horizontal (orange dashed) lines depict the two transects considered in \Fig{transect}. The six control points ($+$ symbols) indicate the location for time series depicted in \Fig{depth_} and \Fig{depth_45}.}
  \label{fig:test4bed}
\end{figure}

\begin{figure}[ht]
  \centering
  \includegraphics[width=0.45\textwidth]{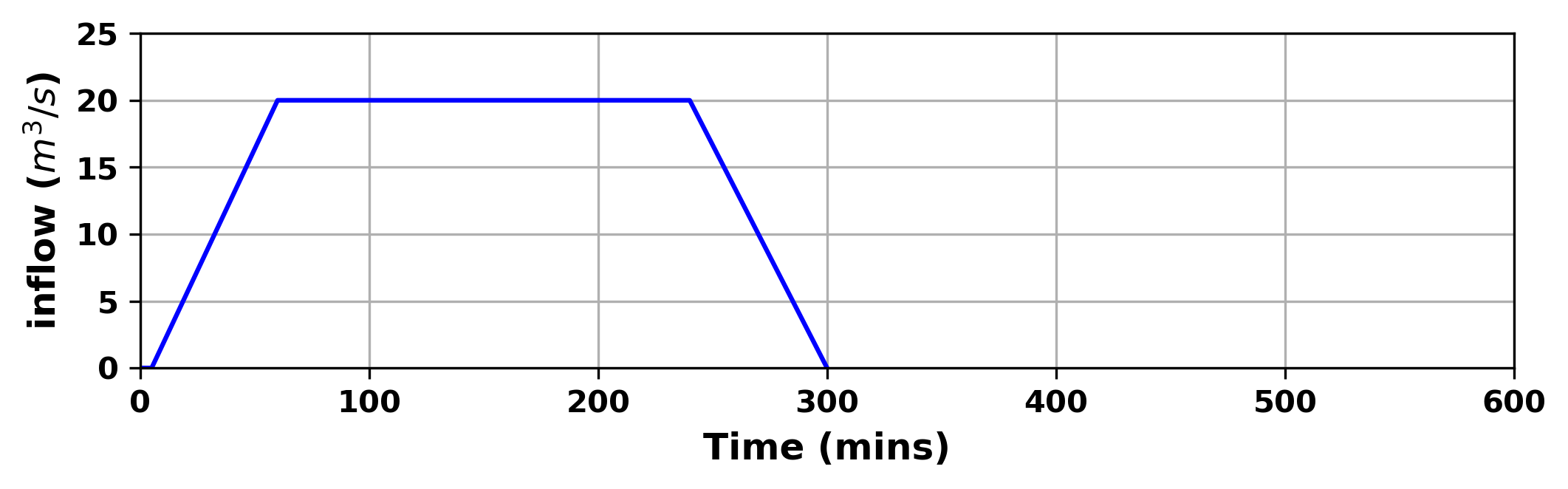}
  \caption{Inflow hydrograph imposed at the inlet boundary}
  \label{fig:test4bc}
\end{figure}

The computational domain is a $1000 \times 2000$ m rectangular region. An inflow hydrograph specifying discharge is imposed along a $20$-m length of the left boundary at its midline (\Fig{test4bed}). The total inflow discharge (volume flux) rises linearly from 0 to a peak value of $20$~m$^3/$s during the first $60$ minutes, remains constant for the next $180$ minutes, then drops linearly to zero for the final $60$ minutes (\Fig{test4bc}).  Dividing the total inflow discharge by the channel length provides the inflow momentum density, \huhydro$(t)$. Before each numerical timestep we set the momentum in ghost cells (exterior cells along the the channel left of the physical boundary) with $hu^n =$\huhydro$(t^n)$.
 
  % \vspace{-0.5cm}

To determine the depth \hhydro$^n$ for the ghost cells at the inflow location, we solve

\begin{equation}
\frac{\huhydro^n}{\hhydro^n} = 2\sqrt{g\hhydro^n} + \uint^n - 2\sqrt{g\hint^n},
\label{eq:riemann}
\end{equation}
where $\uint^n = \huint^n/\hint^n$, and $\hint^n$ and $\huint^n$ refer to the numerical solution determined in the interior values adjacent to the ghost cells.  To avoid numerical difficulties, if $\hint^n$ is non-zero but less than a minimum tolerance $\tau$, $0 < \tau \ll 1$, we set

\begin{equation}
\hint^n = \max \left( \left(\frac{\huhydro^n}{\sqrt{g}} \right)^{2/3}, \tau\right),
\end{equation}
before solving for \hhydro$^n$. If the solution to \eqref{eq:riemann} led to the condition, $\hhydro^n > \hint^n$, we recomputed \hhydro$^n$ by solving 
\begin{equation}
        \frac{\huhydro^n}{\hhydro^n} = \uint^n + (\hhydro^n - \hint^n)\sqrt{\frac{g}{2}\left(\frac{1}{\hhydro^n} + \frac{1}{\hint^n} \right)},
        \label{eq:2-shock}
    \end{equation}
utilizing Hugoniot-loci for a single incoming shock rather than the Riemann invariants implied by \eqref{eq:riemann} \citep[e.g.,][]{leveque2002}. Equations \eqref{eq:riemann} and \eqref{eq:2-shock} were solved using the Newton Raphson method to determine the depth \hhydro$^n$ in ghost cells. The Newton solver was initialized using an initial estimate $(\hhydro^n)^{0} = \left[\frac{\huhydro^n}{\sqrt{g}}F\right]^{2/3}$, where the constant $F$ is an initial rough estimate of the Froude number, which we set to 0.5 for the Riemann invariant problem and to 1 for the single shock problem. 

Following the cessation of inflow after 300 minutes ($\huhydro(t^n) = 0$), we filled ghost cell values at the channel edge by applying a wall boundary condition using interior values $\hint^n$ and $\huint^n$. All other boundaries are closed, and the initial condition considered is a dry bed.

The spatial domain was discretized into $200~\times~400$ cells with a uniform grid spacing of $5$ m in both the $x-$ and $ y-$directions. The following numerical configurations were used: a wet dry threshold of $\tau = 0.0001$ m, an adaptive time step with a maximum CFL (Courant-Friedrichs-Lewy)  of $0.9$ to ensure numerical stability, a final time of $6$ hours, and a Manning coefficient of $0.03$, following \citep{neelz2013benchmarking}.

\subsubsection{Test Case 1: Simulation Results}
\begin{figure}[H]
  \centering
   \begin{subfigure}[b]{8.3cm}
        \centering
	  \includegraphics[width=\textwidth,clip=true,trim=0cm 0cm -1cm 0cm]{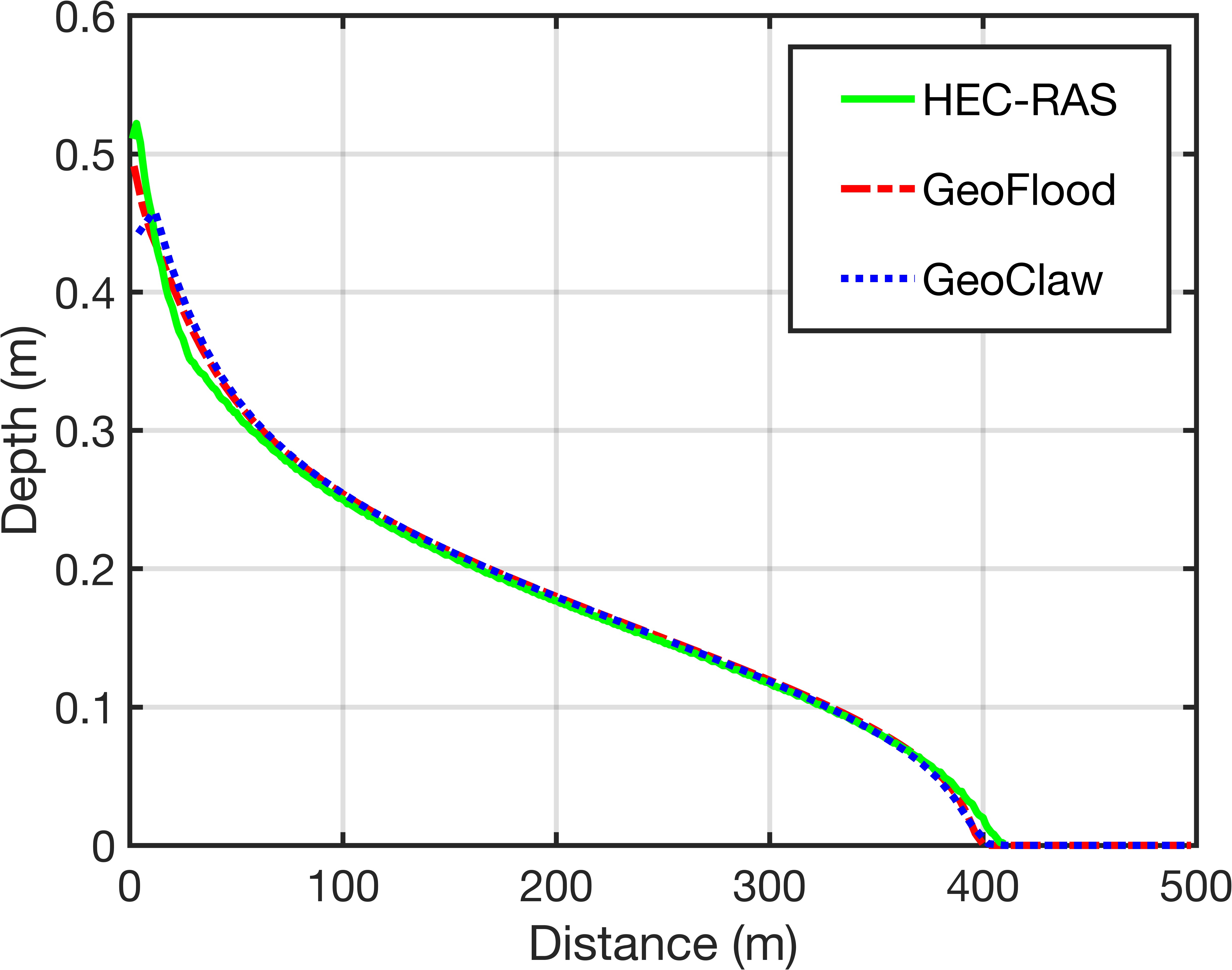}
	  \caption{Horizontal transect at $7$~\unit{m} above the domain central line}
	  \label{fig:depth_}
	\end{subfigure}
 \hfil
	\begin{subfigure}[b]{8.3cm} %<-- -0.3cm since the image seems bigger than the other
	  \centering
	  \includegraphics[width=\textwidth,clip=true,trim=0cm 0cm -1cm 0cm]{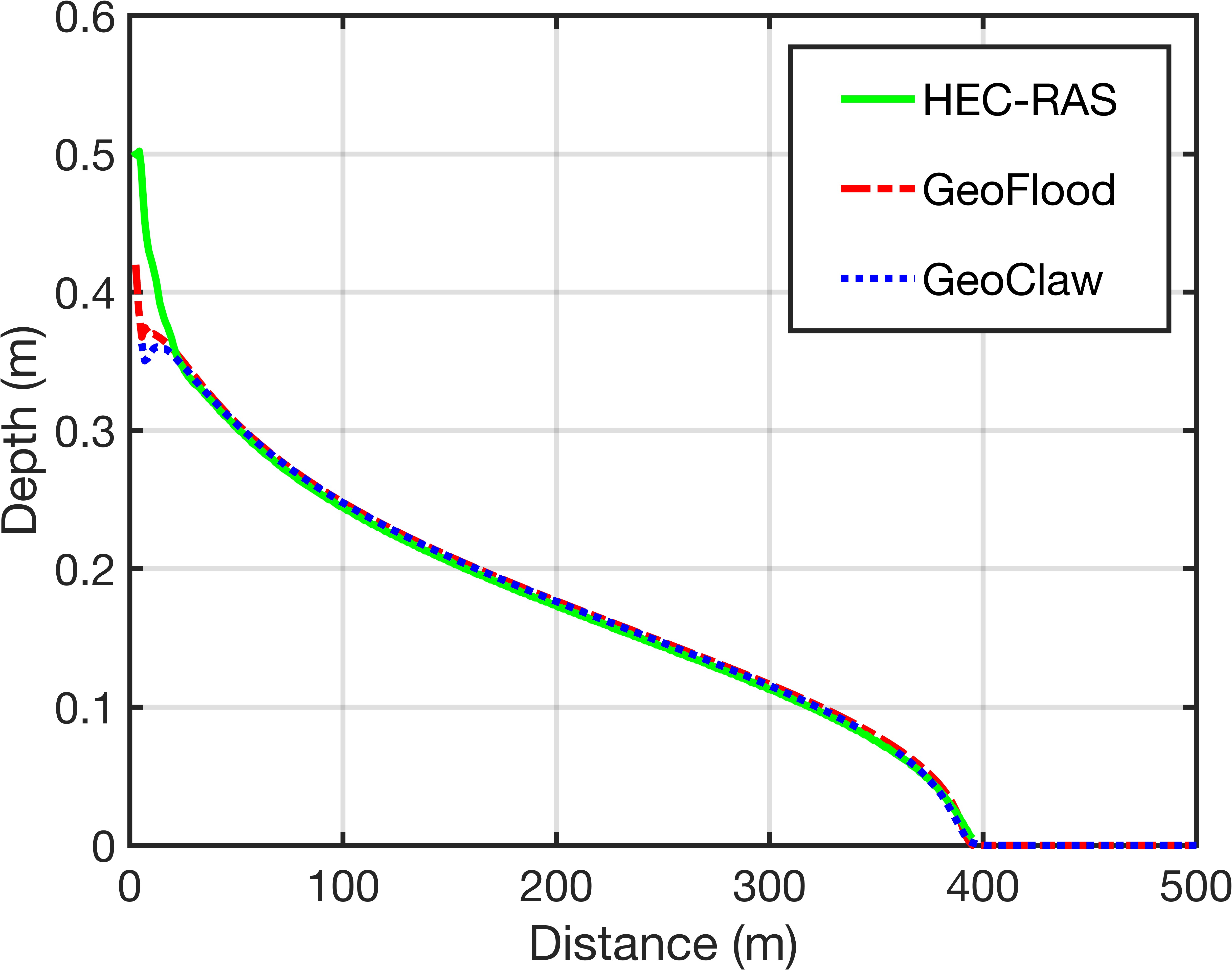}
	  \caption{Transect tilted at $45^{\circ}$ to the horizontal}
	  \label{fig:depth_45}
	\end{subfigure}
	\caption{Cross-section of depths along (a) a horizontal line $7$~\unit{m} above the horizontal central line through the domain and (b) tilted at $45^{\circ}$ to the horizontal at time $t = 1$~\unit{hour} for both HEC-RAS, GeoFlood, and GeoClaw. Refer to \Fig{test4bed} for a depiction of the horizontal and tilted lines.}
	\label{fig:transect}
\end{figure}

GeoFlood produced results that are numerically consistent with the previously benchmarked models, GeoClaw and HEC-RAS.  In \Fig{depth_} and \Fig{depth_45}, the three modeled results for the depth $h$ along two transects (see \Fig{test4bed}) are shown for comparison. The grid-aligned (horizontal) transect shown in \Fig{depth_} and the grid-diagonal transect shown in \Fig{depth_45} reveal that none of the models exhibit significant spurious grid-alignment effects or oscillations at the flow-front. The only significant discrepancy between the models occurs in the depth near the inflow boundary. There are two possible explanations for the discrepancy that do not relate to the interior solution. First, we plotted the GeoFlood and GeoClaw solutions on rays (horizontal (\Fig{depth_}) and diagonal (\Fig{depth_45}) originating at (0, 1007), due to the spatial arrangement of the block-structured AMR which we chose to resolve the inflow boundary segment. Second, and probably more significantly, an inflow discharge hydrograph does not uniquely determine boundary conditions necessary to solve the shallow-water equations, even given equivalent numerical discharge values, $hu$. Physical or mathematical constraints are required to determine the inflow stage and velocity \citep[e.g.,][]{brunner2018,george2008}. The boundary conditions employed in GeoClaw and GeoFlood (\Sec{flood_speed_setup}) are distinct from those employed in HEC-RAS, which may be based on approximate steady channel flow assumptions commonly used in flood modeling, but are not described in numerical detail in HEC-RAS documentation. The boundary discrepancy in stage is a result of how the value for discharge is enforced in different models, not a difference in model accuracy because the exact solution is ambiguous. Nevertheless, equivalent inflow discharges in the three models lead to consistent results downstream of the boundary. 

\begin{figure}[H]
	\centering
	\begin{subfigure}[b]{0.215\textwidth}
		\centering
		\plotbox{\includegraphics[width=\textwidth]{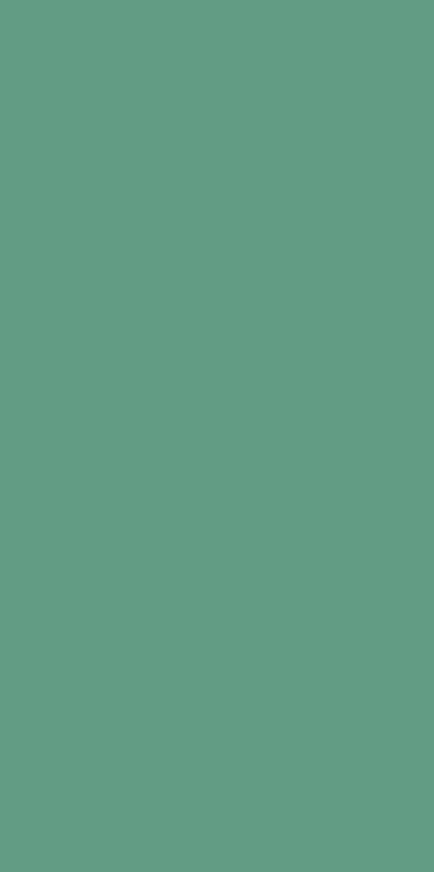}}
		\caption{At $t=0~\unit{s}$}
		\label{fig:case_t0}
	\end{subfigure}
	\hfil
	\begin{subfigure}[b]{0.215\textwidth}
		\centering
		\plotbox{\includegraphics[width=\textwidth]{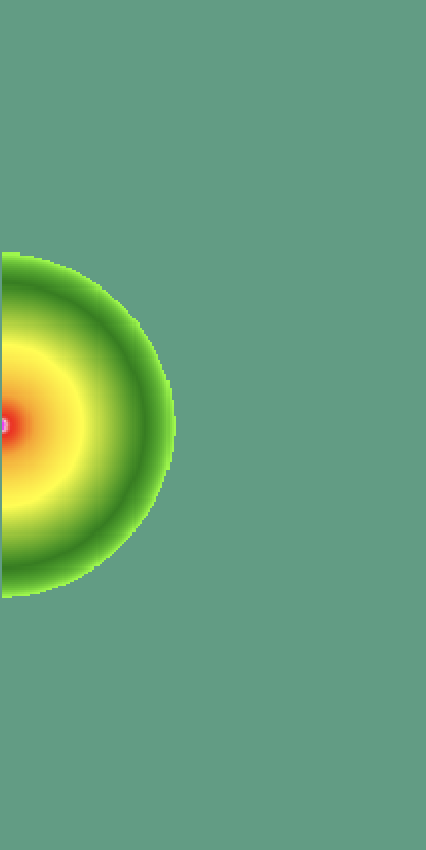}}
		\caption{At $t=1~\unit{hour}$}
		\label{fig:case_t1}
	\end{subfigure}
	\hfil
	\begin{subfigure}[b]{0.215\textwidth}
		\centering
		\plotbox{\includegraphics[width=\textwidth]{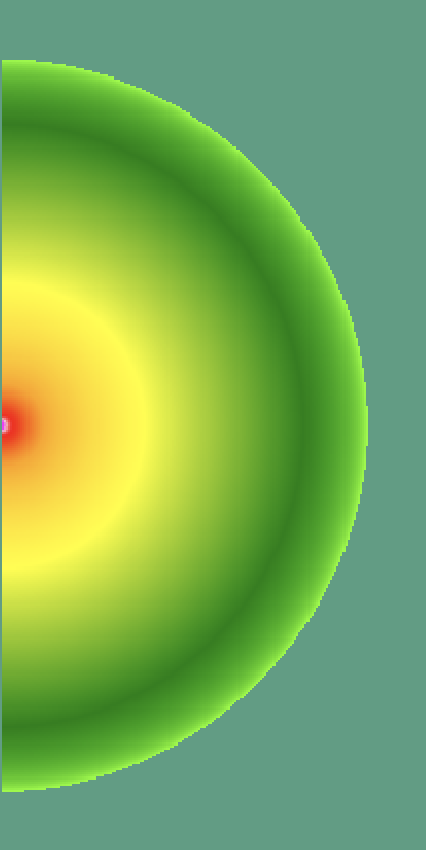}}
		\caption{At $t=2.5~\unit{hours}$}
		\label{fig:case_t25}
	\end{subfigure}
    \medskip
    %\begin{adjustbox}{raise=0.46cm, height=3.2cm,width=1.3cm}\begin{subfigure}[b]{0.05\textwidth}
    \begin{adjustbox}{raise=1.1cm, height=7cm,width=1.6cm}\begin{subfigure}[b]{0.05\textwidth}
		% \centering
		%\plotbox{\includegraphics[width=\textwidth]{images/flood_plain_color.png}}
        \plotbox{\includegraphics[width=\textwidth]{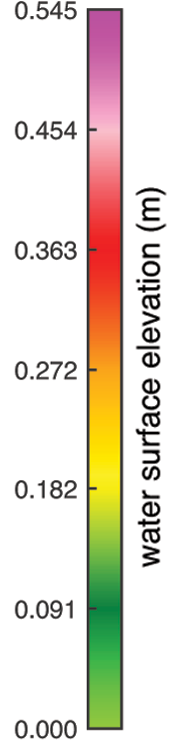}}
	\end{subfigure}\end{adjustbox}\\
 \medskip 
		\begin{subfigure}[b]{0.22\textwidth}
		% \centering
		\plotbox{\includegraphics[width=\textwidth]{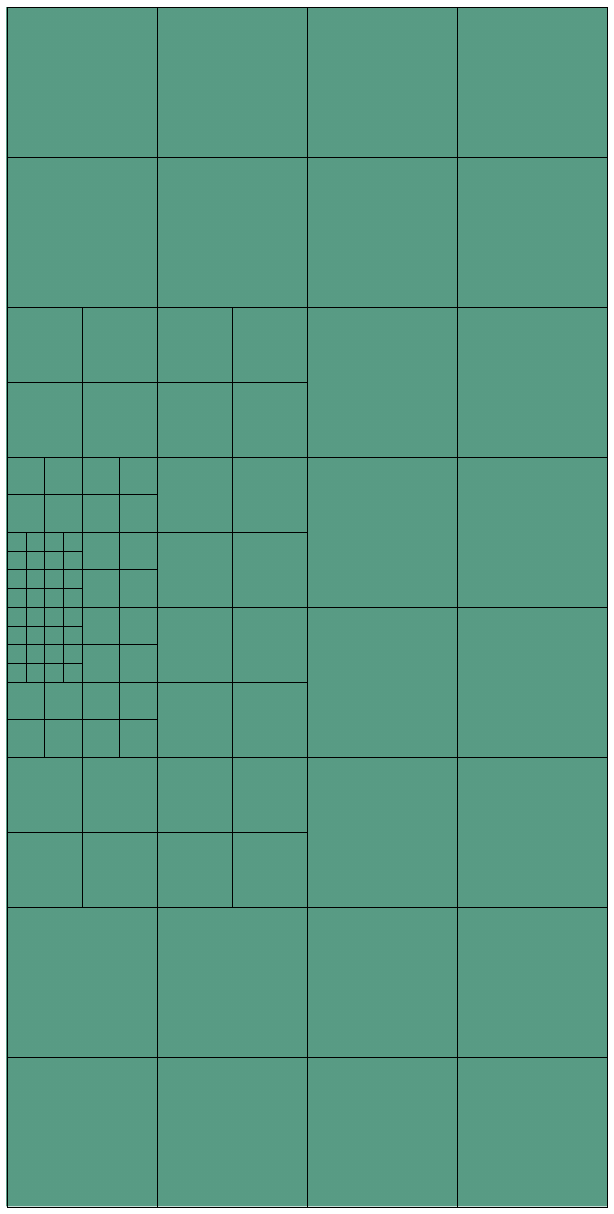}}
		\caption{At $t=0~\unit{s}$}
		\label{fig:flood_t0}
	\end{subfigure}
	\hfil
	\begin{subfigure}[b]{0.22\textwidth}
		\centering
		\plotbox{\includegraphics[width=\textwidth]{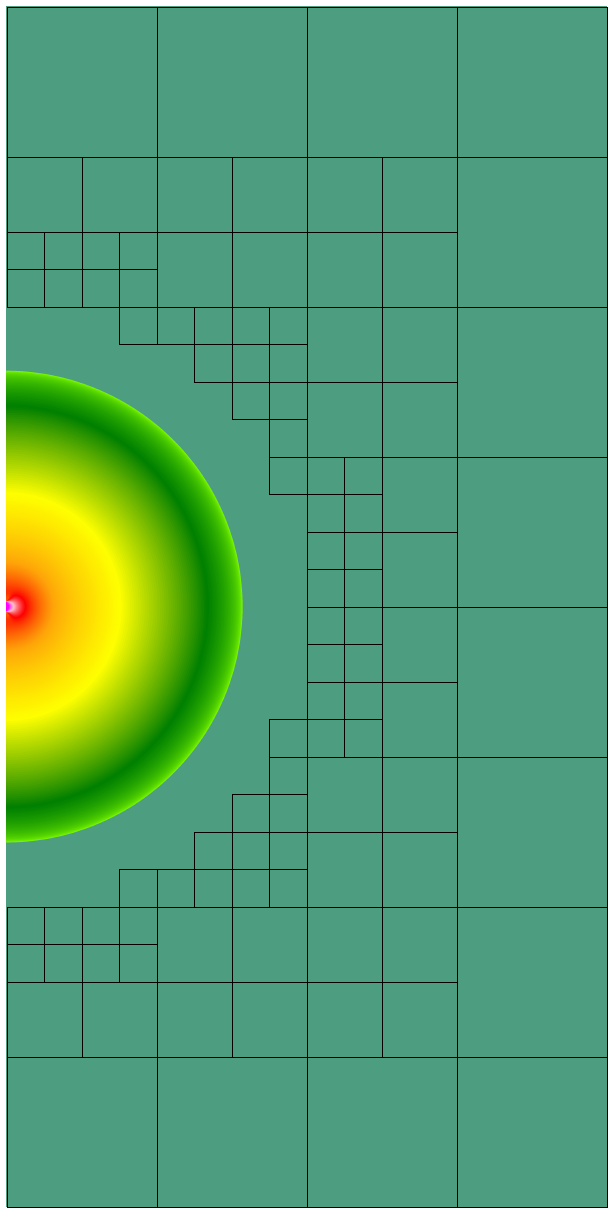}}
		\caption{At $t=1~\unit{hour}$}
		\label{fig:flood_t1}
	\end{subfigure}
	\hfil
	\begin{subfigure}[b]{0.22\textwidth}
		\centering
		\plotbox{\includegraphics[width=\textwidth]{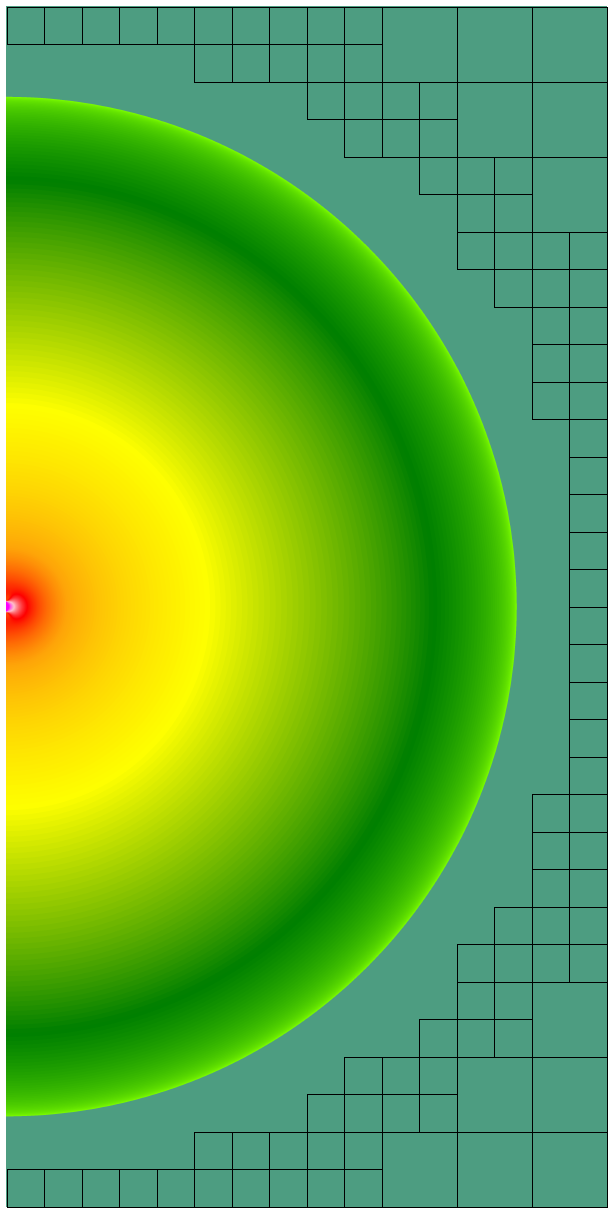}}
		\caption{At $t=2.5~\unit{hours}$}
		\label{fig:flood_t25}
	\end{subfigure}
        \medskip
        %\begin{adjustbox}{raise=1.1cm, height=5.25cm,width=1.2cm}\begin{subfigure}[b]{0.05\textwidth}
        \begin{adjustbox}{raise=1.1cm, height=7cm,width=1.6cm}\begin{subfigure}[b]{0.05\textwidth}
		% \centering
		%\plotbox{\includegraphics[width=\textwidth]{images/flood_colorbar.pdf}}
        \plotbox{\includegraphics[width=\textwidth]{images/geoflood_flood_speed_colorbar.png}}
	\end{subfigure}\end{adjustbox}
	   \caption{Overhead map perspectives of the water surface elevation at various times for both the HEC-RAS (a-c) and GeoFlood (d-f) simulations. The HEC-RAS simulation was performed on a $5$ \unit{m} uniformly structured grid with $200 \times 400$ grid cells while GeoFlood was simulated on an adaptively refined grid with max-level = $4$, min-level = $1$, starting on the coarsest mesh of $50 \times 50$ level 0 grid blocks in a $2 \times 4$ block arrangement to finest mesh of $1600 \times 3200$ grid cells at $0.6$ \unit{m} grid cell resolution. At $t = 1$ and $t = 2.5~\unit{hours}$, grid lines for HEC-RAS and the maximum refinement level in GeoFlood are omitted from the plots.}
    \label{fig:case_2_results}
  \end{figure}

 In another test, higher refinement (minimum cell size of 0.6 m) of the propagating flood was employed in the GeoFlood simulation (\Fig{case_2_results}). Refinement through AMR allows a more optimal spatio-temporal allocation of computational resources by neglecting the resolution of dry regions in the domain. An equivalent grid resolution in HEC-RAS with the same time-step stability criterion would require approximately 580 times ($(5/0.6)^3$) more computational expense than the 5 m simulation shown in \Fig{case_2_results}.

\begin{figure}[t]
		\centering % <-- added
		\begin{subfigure}{0.33\textwidth}
			\includegraphics[width=\linewidth]{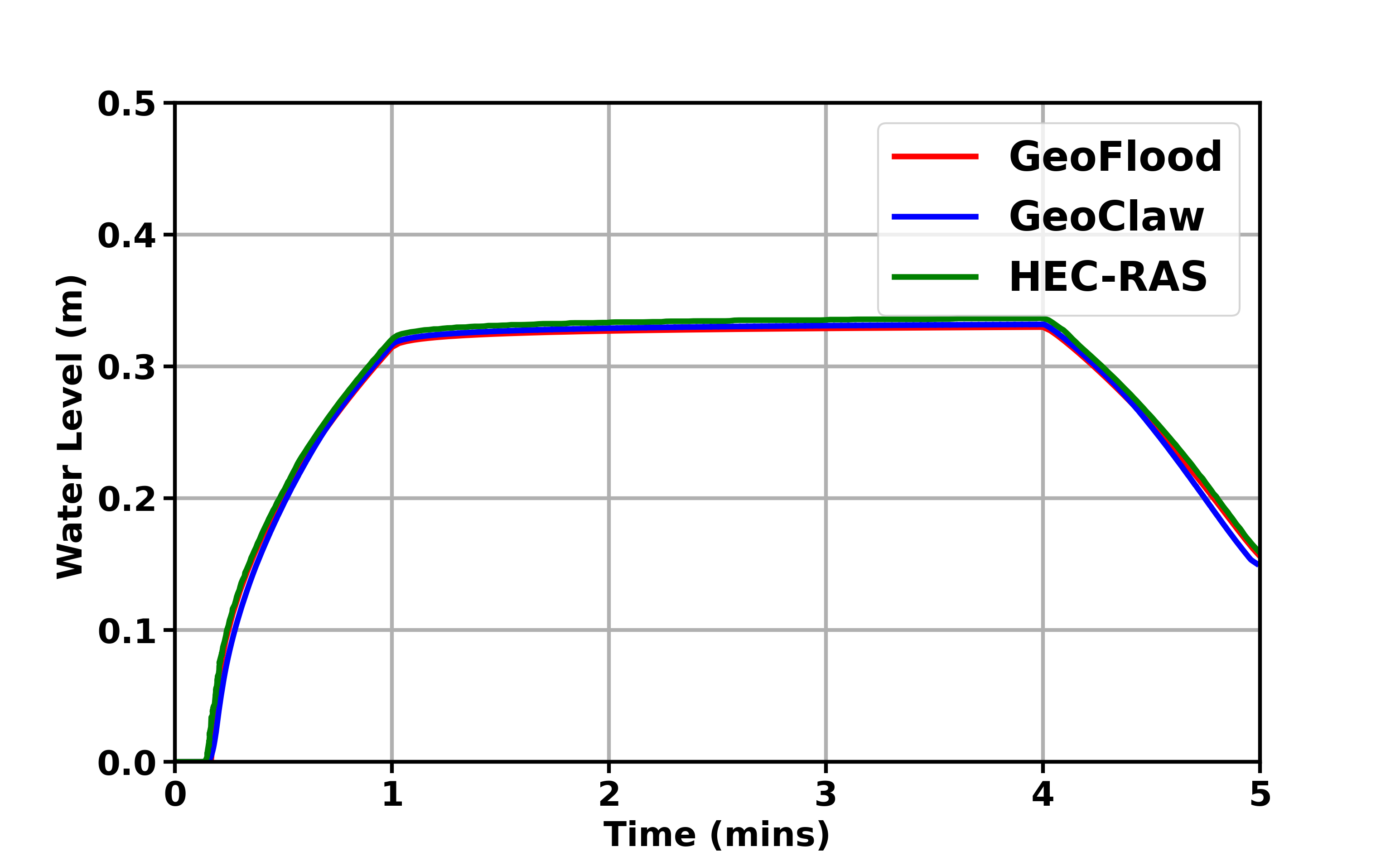}
			\caption{Water level at point 1}
			\label{fig:eta_p1_flood}
		\end{subfigure}\hfil % <-- added
		\begin{subfigure}{0.33\textwidth}
			\includegraphics[width=\linewidth]{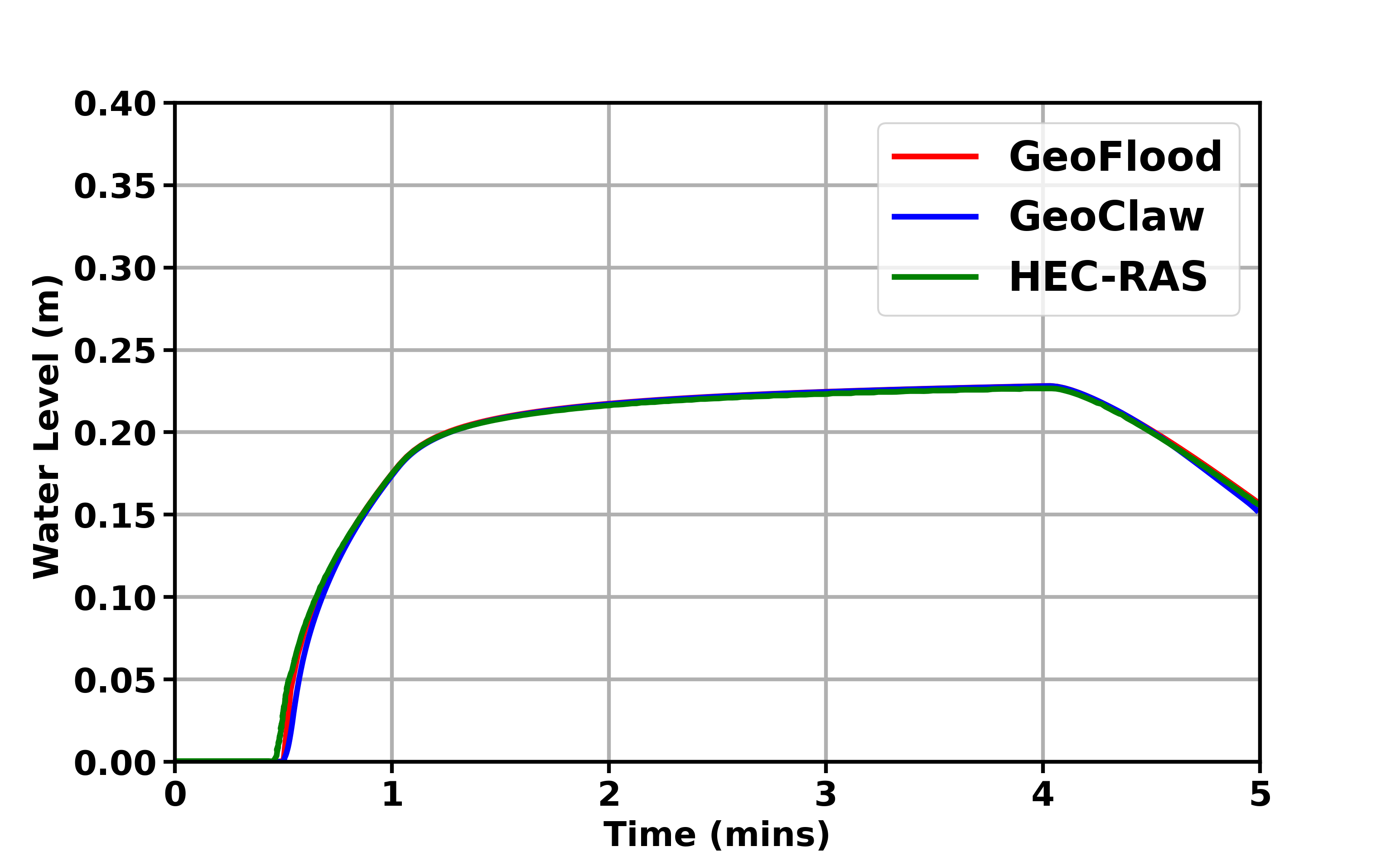}
			\caption{Water level at point 3}
			\label{fig:eta_p3_flood}
		\end{subfigure}\hfil % <-- added
		\begin{subfigure}{0.33\textwidth}
			\includegraphics[width=\linewidth]{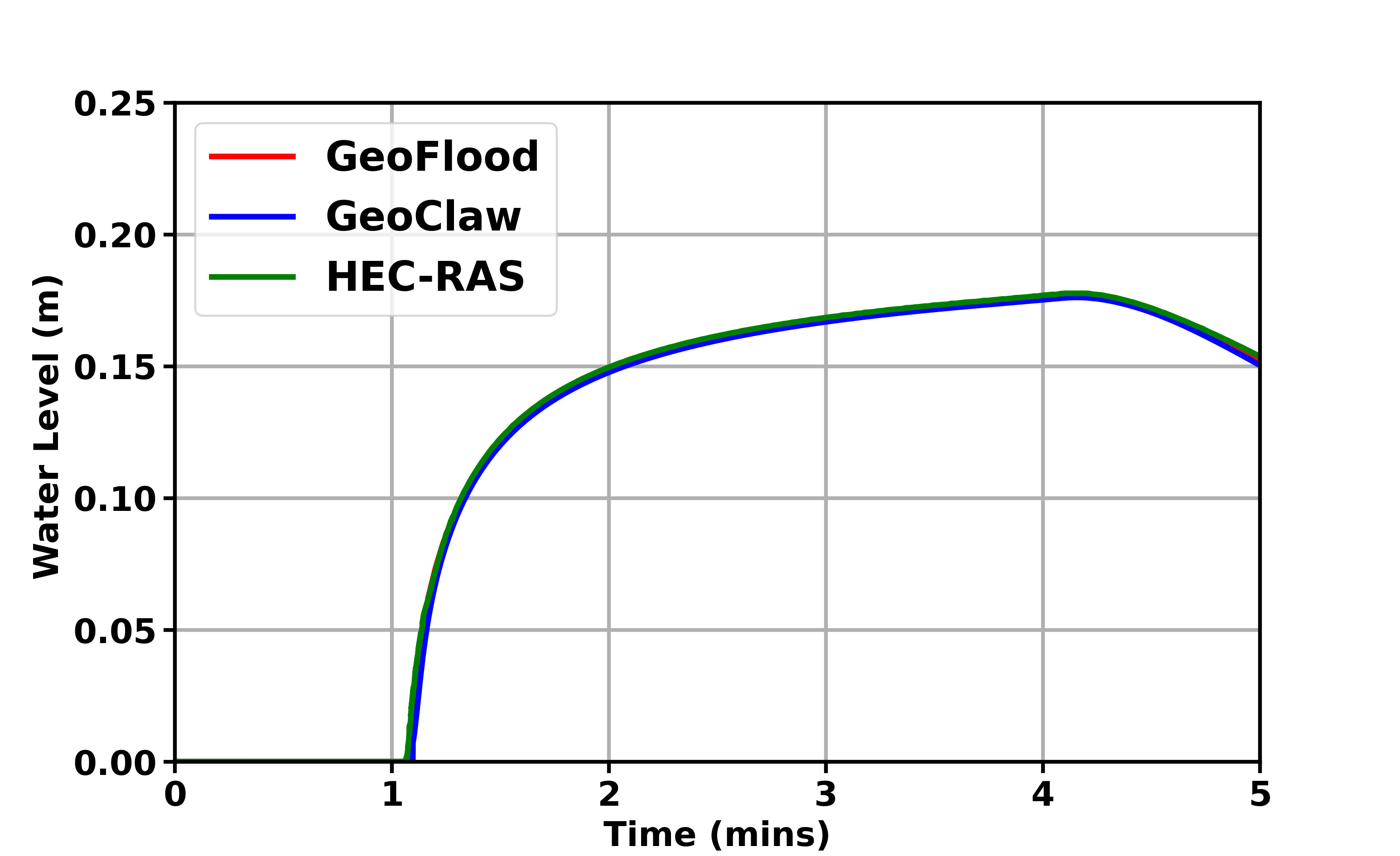}
			\caption{Water level at point 6}
			\label{fig:eta_p6_flood}
		\end{subfigure}
		
		\medskip
		\begin{subfigure}{0.33\textwidth}
			\includegraphics[width=\linewidth]{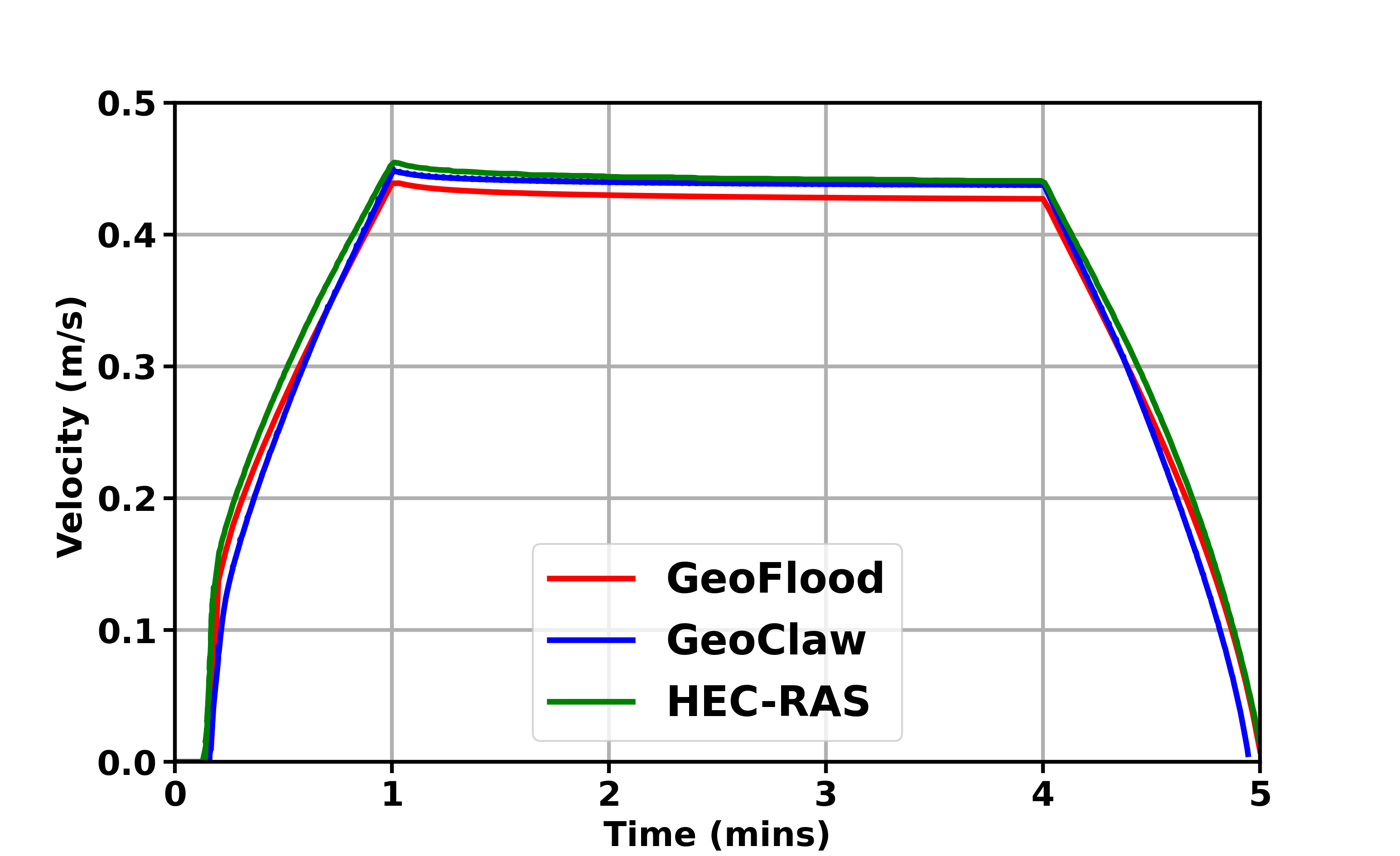}
			\caption{Velocity at point 1}
			\label{fig:u_p1_flood}
		\end{subfigure}\hfil % <-- added
		\begin{subfigure}{0.33\textwidth}
			\includegraphics[width=\linewidth]{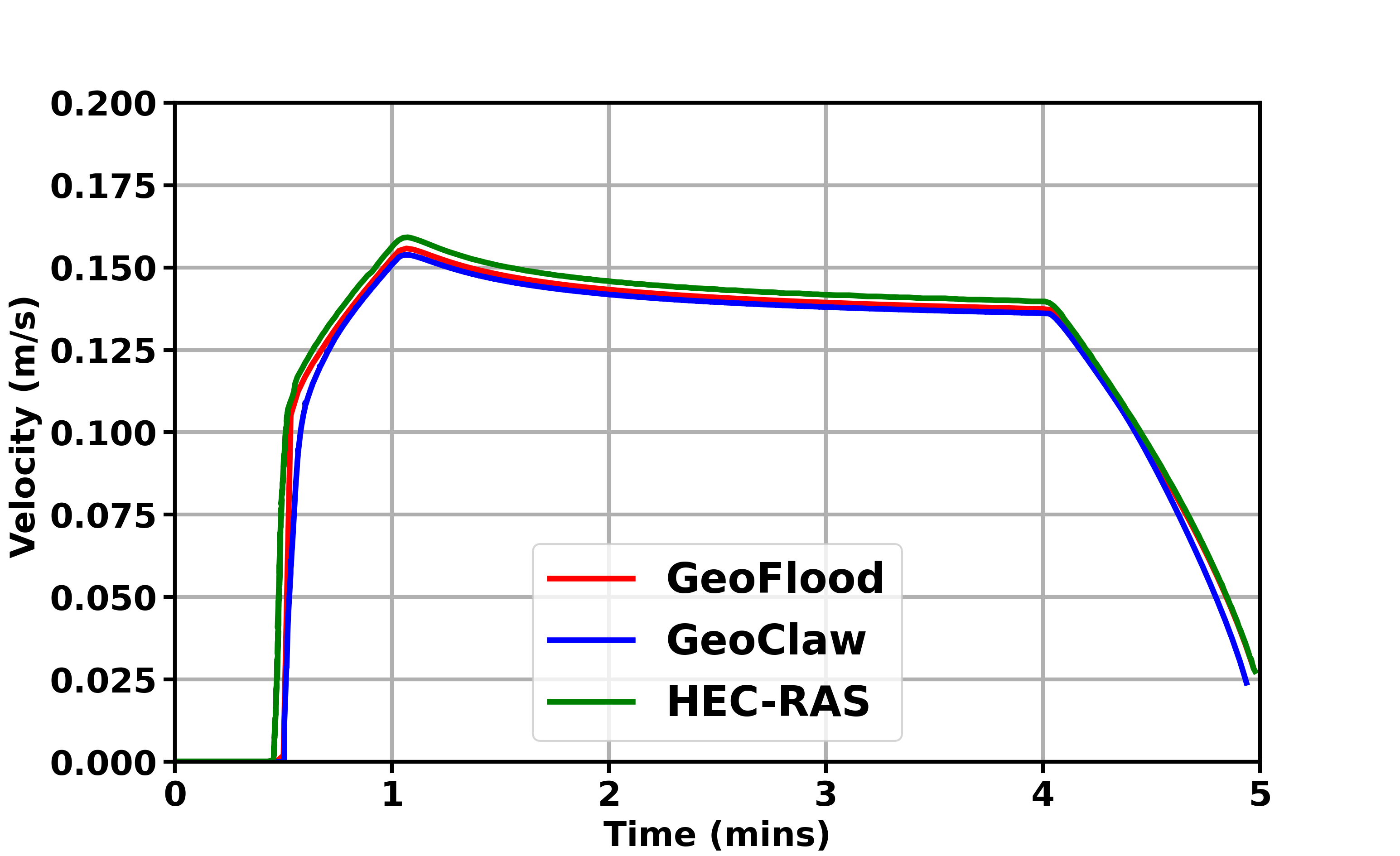}
			\caption{Velocity at point 3}
			\label{fig:u_p3_flood}
		\end{subfigure}\hfil % <-- added
		\begin{subfigure}{0.33\textwidth}
			\includegraphics[width=\linewidth]{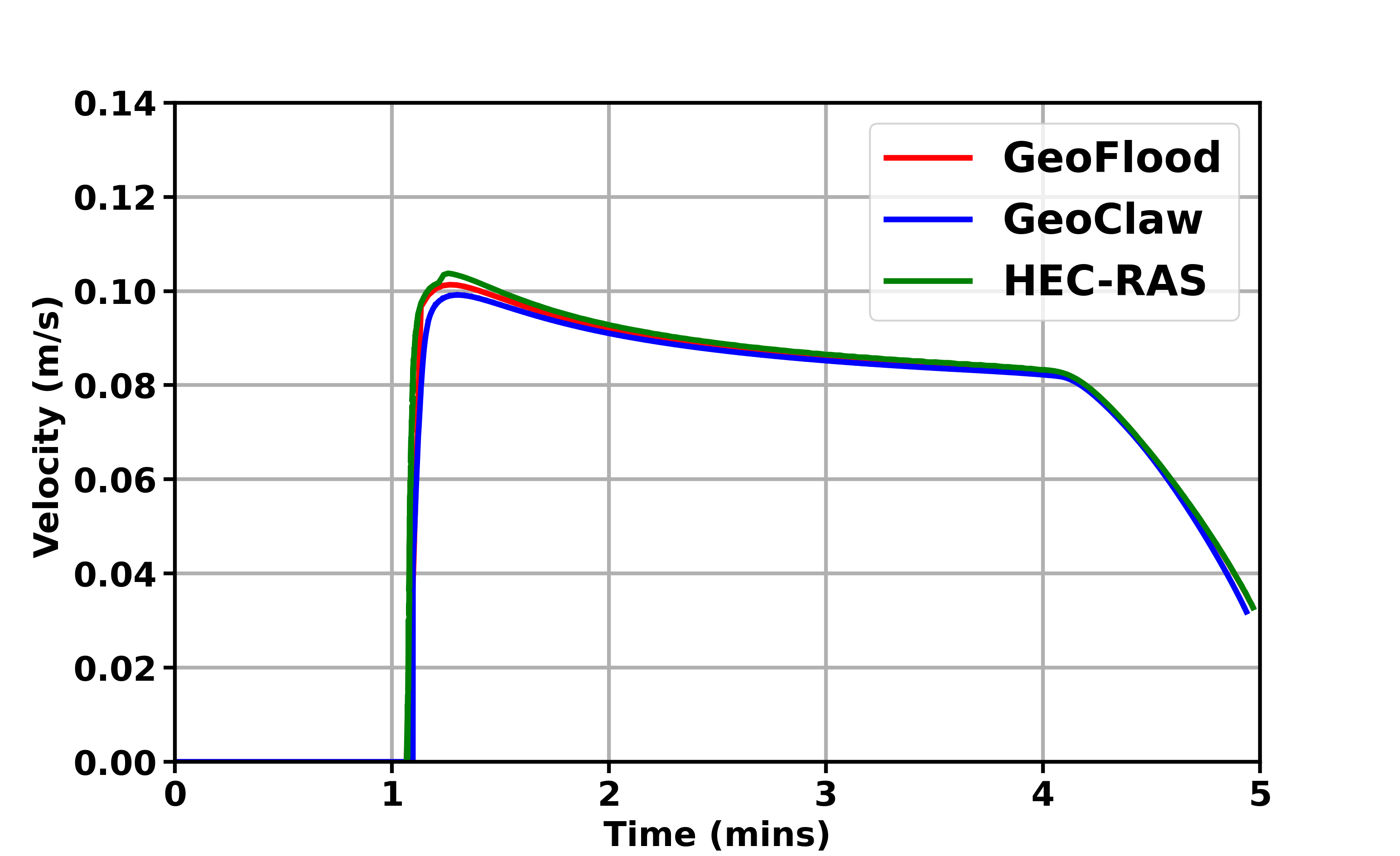}
			\caption{Velocity at point 6}
			\label{fig:u_p6_flood}
		\end{subfigure}
		\caption{Temporal evolution of the water surface elevation (a-c) and velocity (d-f) at various control points (\Fig{test4bed}) for GeoFlood compared to GeoClaw and HEC-RAS.}
		\label{fig:flood_speed}
	\end{figure}

 The temporal evolution of flow depth and velocity are compared at several control points (\Fig{flood_speed} with locations shown in \Fig{test4bed}) demonstrating that GeoFlood produces results consistent with the previously validated codes, GeoClaw and HEC-RAS.
 
\subsection{Test Case 2: Filling of Floodplain Depressions }
\label{filligdep}
The second benchmark problem we consider is similar to Test Case 1, in that inflow discharge is specified leading to inundating flow into a highly idealized floodplain (topographic surface). However, unlike Test Case 1, the floodplain contains isolated depressions, some of which fill slowly over several hours. This test is designed to evaluate a model's predictive accuracy in determining the inundation extent and final depth of flooding over a long period of time under conditions of low-momentum flow across intricate topographical landscapes. The primary focus of this assessment is on the final depths in individual depressions rather than their peak levels. Similar to Test Case 1, the problem accentuates model differences and sensitivities. 

\subsubsection{Problem Setup}
\begin{figure}[H]
% Figure 7
\centering
\begin{subfigure}{0.35\textwidth}
    \includegraphics[width=\textwidth,clip=true,trim=-1cm -3cm -1cm -1cm]{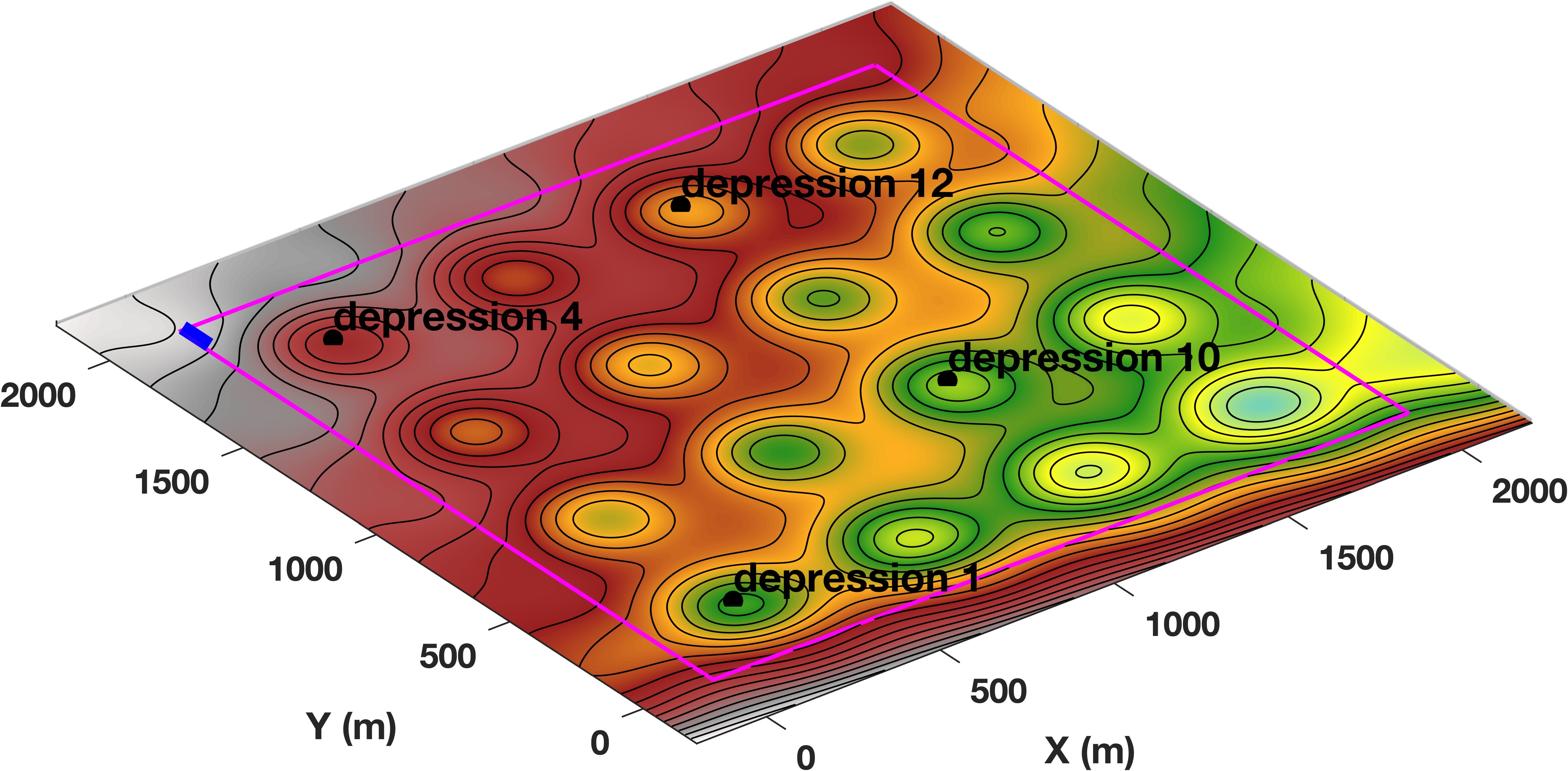}
    \caption{Topography with equal aspect ratio in $x$, $y$ and $z$ directions depicting the depressions}
    \label{fig:topo4}
\end{subfigure}
\hfil
\begin{subfigure}{0.35\textwidth}
    \includegraphics[width=\textwidth]{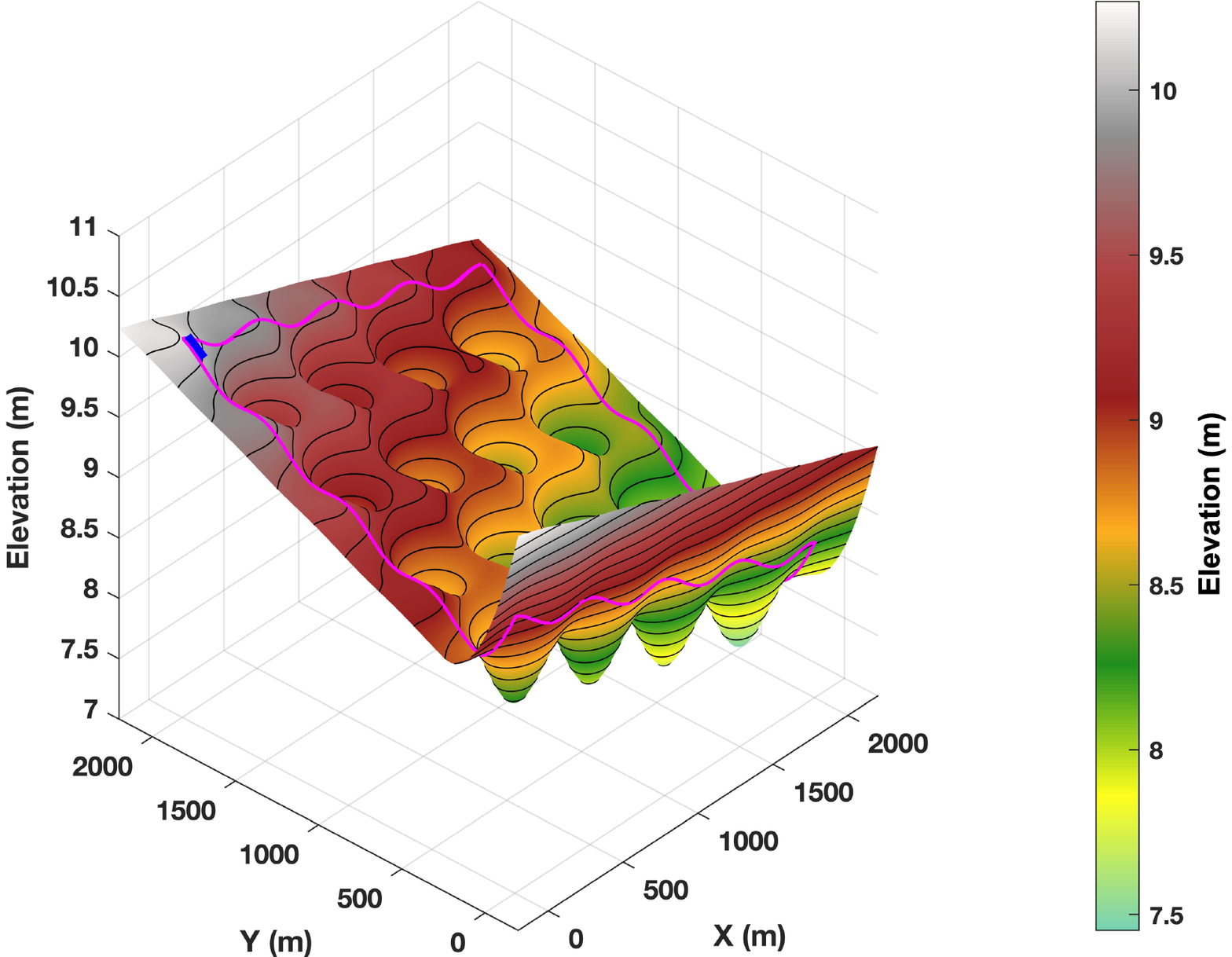}
    \caption{Topography with exaggerated aspect ratio in $z$ to show the $0.5$\unit{m} depressions}
    \label{fig:topo4_}
\end{subfigure}
\hfil
\begin{subfigure}{0.45\textwidth}
    \includegraphics[width=\textwidth]{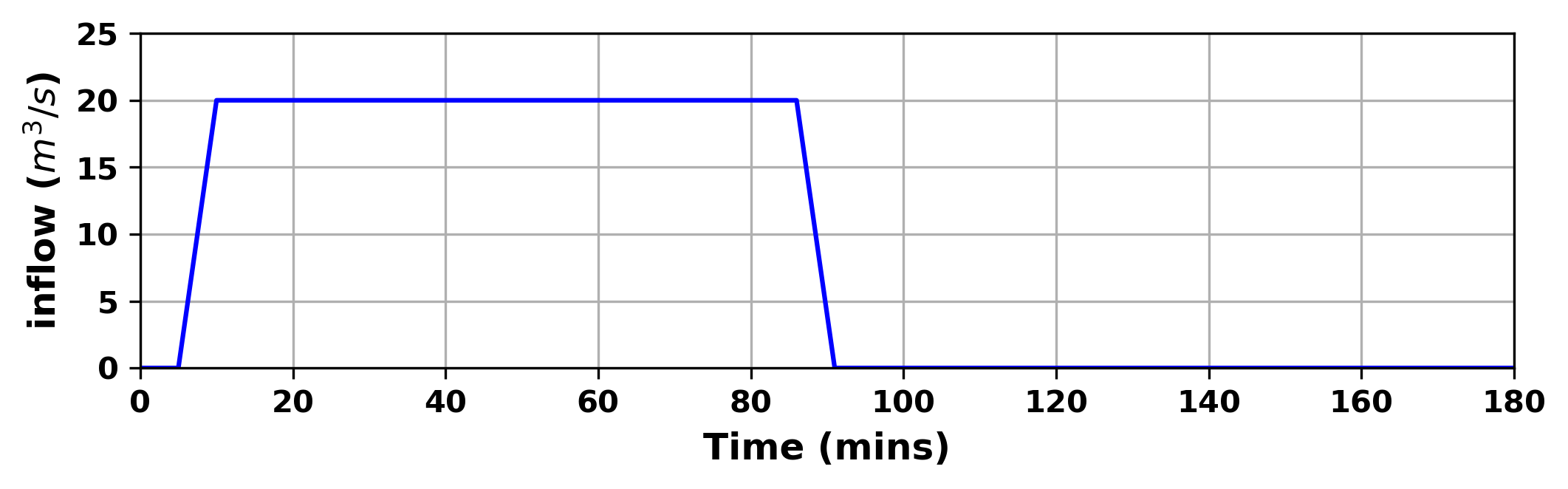}
    \caption{In flow hydrograph imposed at the inlet boundary.}
    \label{fig:bc2}
\end{subfigure}
\caption{Topography (a-b) showing the inflow location (blue thick line at the northwest corner) and ground elevation with contour lines at $0.5$~m intervals. The depressions are numbered sequentially $1-16$, increasing in the positive $Y$-direction followed by the positive $X$-direction. Labeled depressions $1,4,10,$ and $12$, indicate where numerical time series are compared in \Fig{case_1_results}. (c) inflow hydrograph for total discharge through the inflow boundary segment.}
\label{fig:DEM_file_1}
\end{figure}

The computational domain of this test is a square with a side length of $2000$~m. \Fig{topo4} depicts a $4~\times~4$ matrix of $0.5$~m deep depressions with smooth topographic transitions obtained by multiplying sinusoids in the north-to-south and west-to-east directions. The underlying average slope is $1:1500$ in the north-south direction and $1:3000$ in the west-east direction, with an elevation drop of about $2$~m along the northwest-to-southeast diagonal.  At the upstream boundary (northwest corner of the domain), an inlet hydrograph with a peak discharge of $20$ m$^3$/s and a time base of approximately $85$~minutes is imposed along a $100$-m-long line (blue thick line) that runs north to south \Fig{topo4}. All other boundaries were treated as solid-wall boundaries, and the initial condition assumes a dry bed. The inflow hydrograph boundary condition was implemented as described in Section \ref{sec:flood_speed}.

The model simulates two days ($48$ hours) to achieve a final hydrostatic steady state. The following numerical configurations were used: a wet-dry threshold of $\tau=0.0001\unit{m}$, an adaptive time step with a maximum CFL of $0.9$, and a Manning coefficient of $0.03$ for comparison with \citet{brunner2018}. 

\subsubsection{Test Case 2: Simulation Results}
\begin{figure}[H]
	\centering
	\begin{subfigure}[b]{0.25\textwidth}
		\centering
		\plotbox{\includegraphics[width=\textwidth]{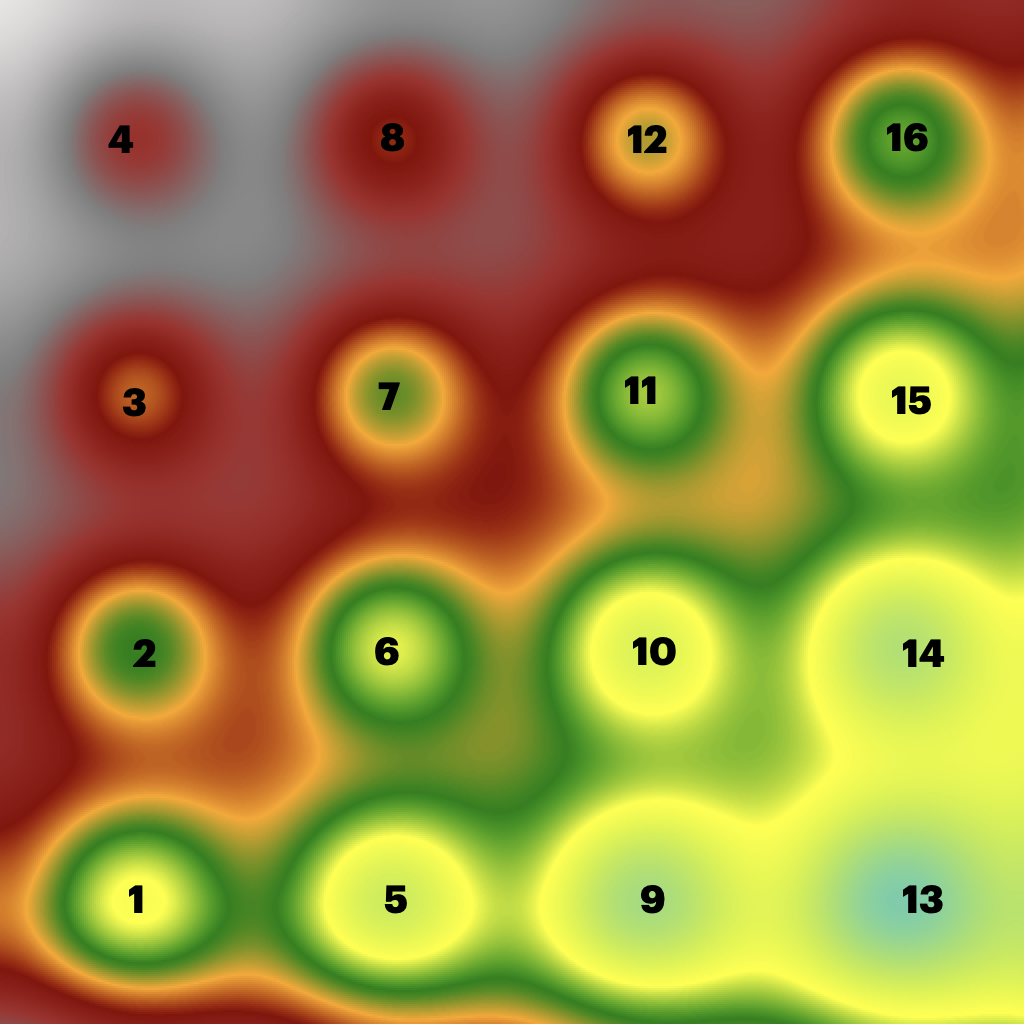}}
		\caption{At $t=0~\unit{s}$}
		\label{fig:filling_0_hec}
	\end{subfigure}
	\hfil
	\begin{subfigure}[b]{0.25\textwidth}
		\centering
		\plotbox{\includegraphics[width=\textwidth]{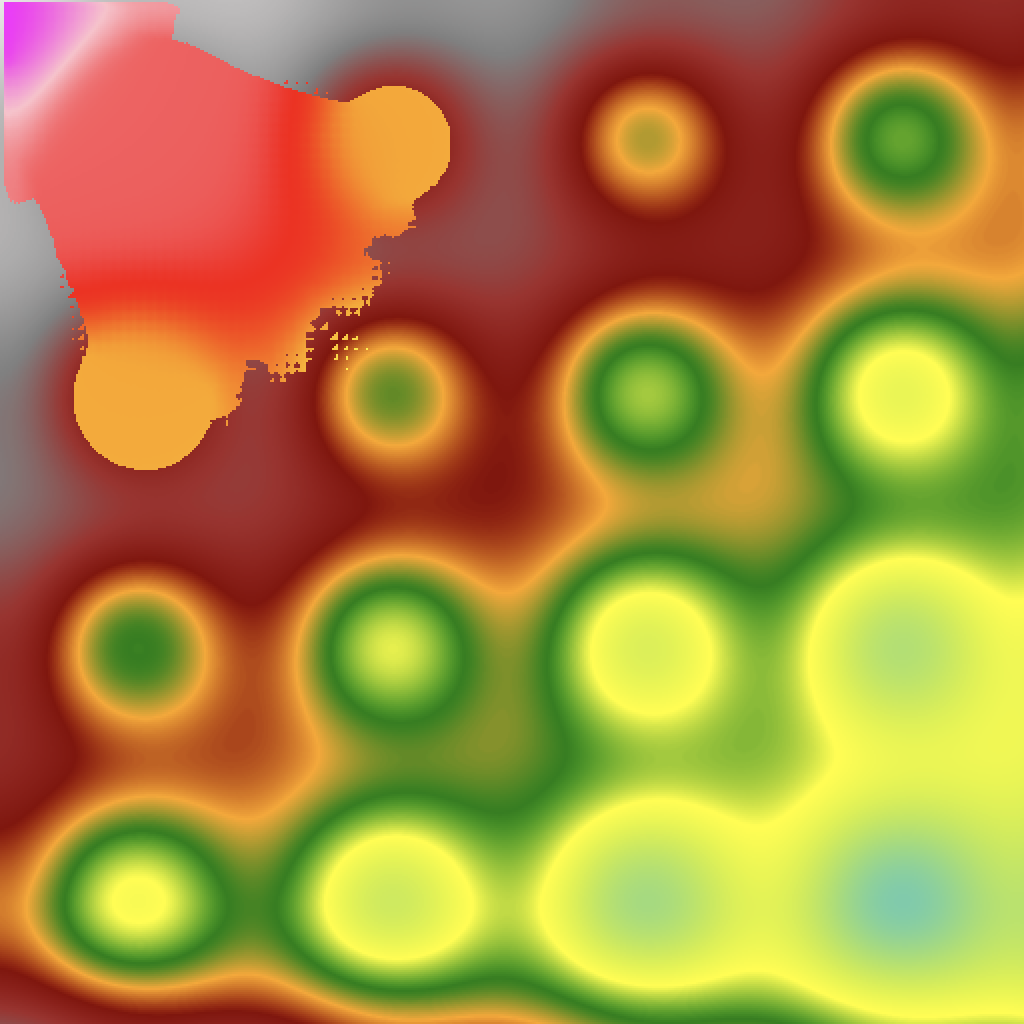}}
		\caption{At $t=1~\unit{hour}$}
		\label{fig:filling_1_hec}
	\end{subfigure}
	\hfil
	\begin{subfigure}[b]{0.25\textwidth}
		\centering
		\plotbox{\includegraphics[width=\textwidth]{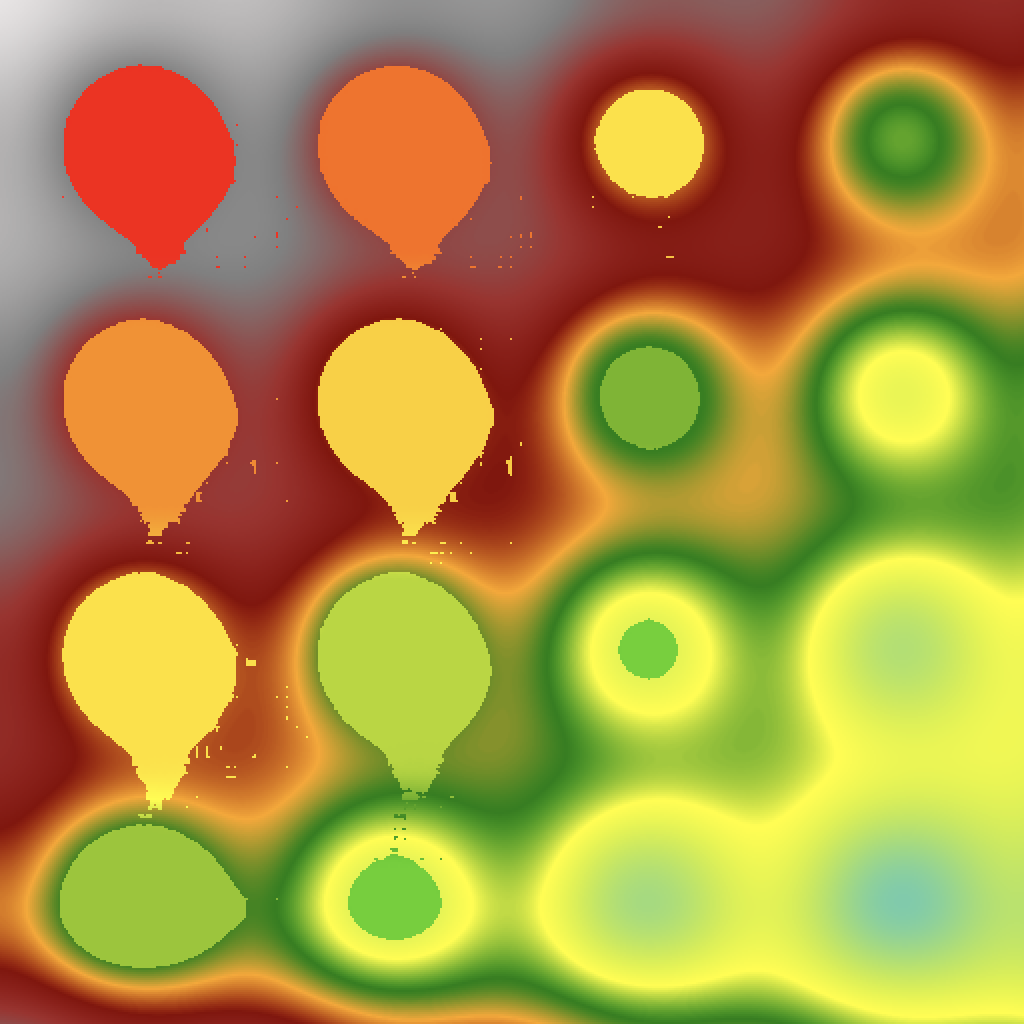}}
		\caption{At $t=20~\unit{hours}$}
		\label{fig:filling_20_hec}
	\end{subfigure}
  \medskip
  %\begin{adjustbox}{raise=0.5cm, height=3.0cm,width=1.3cm}\begin{subfigure}[b]{0.05\textwidth}
		% \centering
		%\plotbox{\includegraphics[width=\textwidth]{images/filling_color_bar.png}}
        %\plotbox{\includegraphics[width=\textwidth]{images/filling_geo_colobarr_0.png}}
	%\end{subfigure}\end{adjustbox}
    \begin{adjustbox}{raise=0.8cm, height=5.2cm,width=1.3cm}\begin{subfigure}[b]{0.05\textwidth}
		% \centering
		\plotbox{\includegraphics[width=\textwidth]{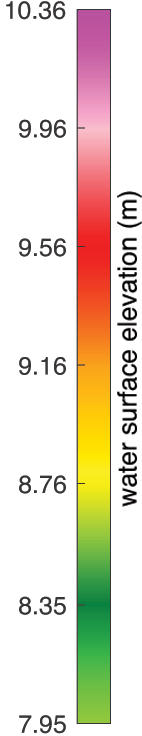}}
	\end{subfigure}\end{adjustbox} 
 % \medskip 
		\begin{subfigure}[b]{0.25\textwidth}
		\centering
		\plotbox{\includegraphics[width=\textwidth,clip=true,trim=0cm 0cm 0cm 0cm]{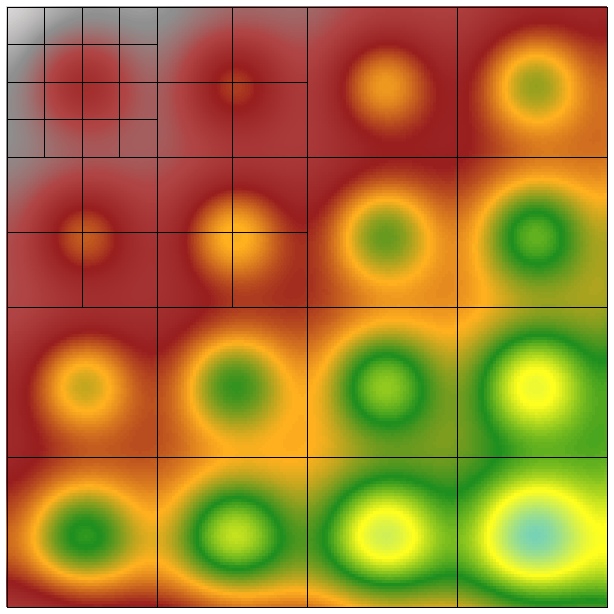}}
		\caption{At $t=0~\unit{s}$}
		\label{fig:filling_0}
	\end{subfigure}
	\hfil
	\begin{subfigure}[b]{0.25\textwidth}
		\centering
		\plotbox{\includegraphics[width=\textwidth,clip=true,trim=0cm 0cm 0cm 0cm]{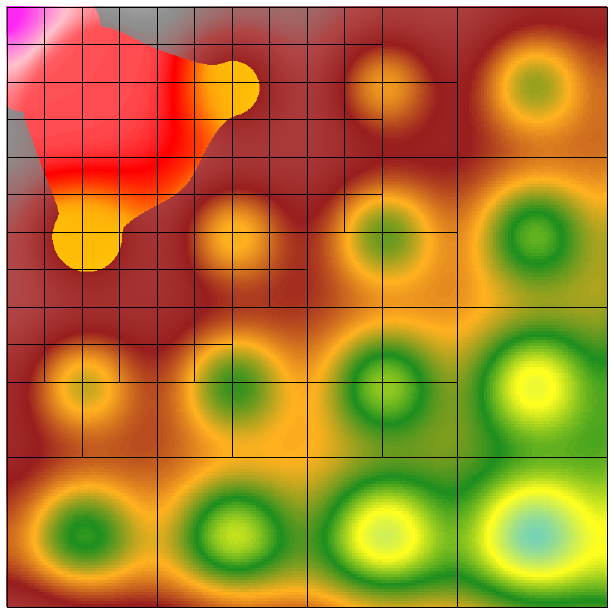}}
		\caption{At $t=1~\unit{hour}$}
		\label{fig:filling_2}
	\end{subfigure}
	\hfil
	\begin{subfigure}[b]{0.25\textwidth}
		\centering
		\plotbox{\includegraphics[width=\textwidth,clip=true,trim=0cm 0cm 0cm 0cm]{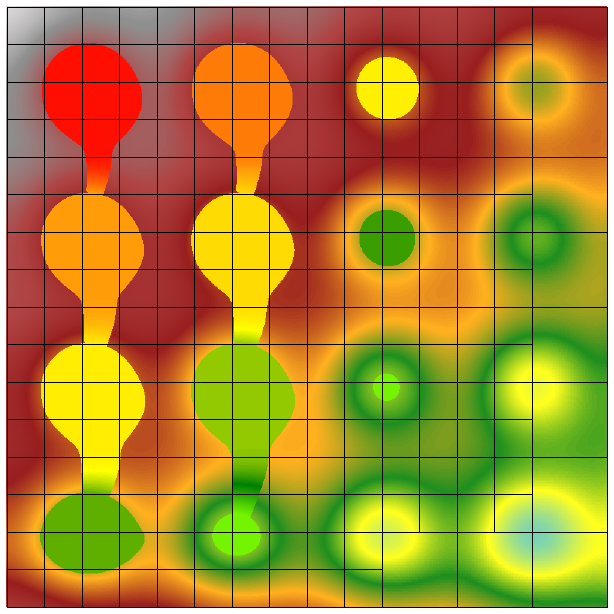}}
		\caption{At $t=20~\unit{hours}$}
		\label{fig:filling_20}
	\end{subfigure} 
    \begin{adjustbox}{raise=0.8cm, height=5.2cm,width=1.3cm}\begin{subfigure}[b]{0.05\textwidth}
		% \centering
		\plotbox{\includegraphics[width=\textwidth]{images/filling_geo_colobarr_0.png}}
	\end{subfigure}\end{adjustbox}
	   \caption{Overhead maps showing the surface elevation (water plus topography, $h+b$) at various times for both HEC-RAS (a-c) and GeoFlood (d-f). }
    \label{fig:case_1_results}
\end{figure}

\begin{figure}[H]
		\centering % <-- added
		\begin{subfigure}[b]{0.33\textwidth}
			\includegraphics[width=\linewidth]{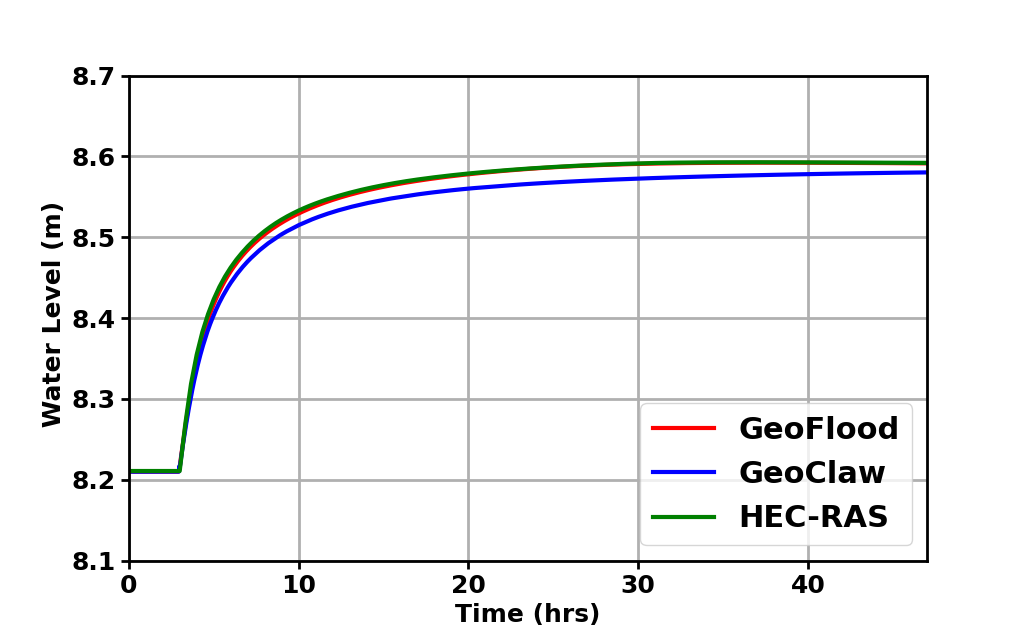}
			\caption{Depression 1}
			\label{fig:case_1_1}
		\end{subfigure}\hfil % <-- added
		\begin{subfigure}[b]{0.33\textwidth}
			\includegraphics[width=\linewidth]{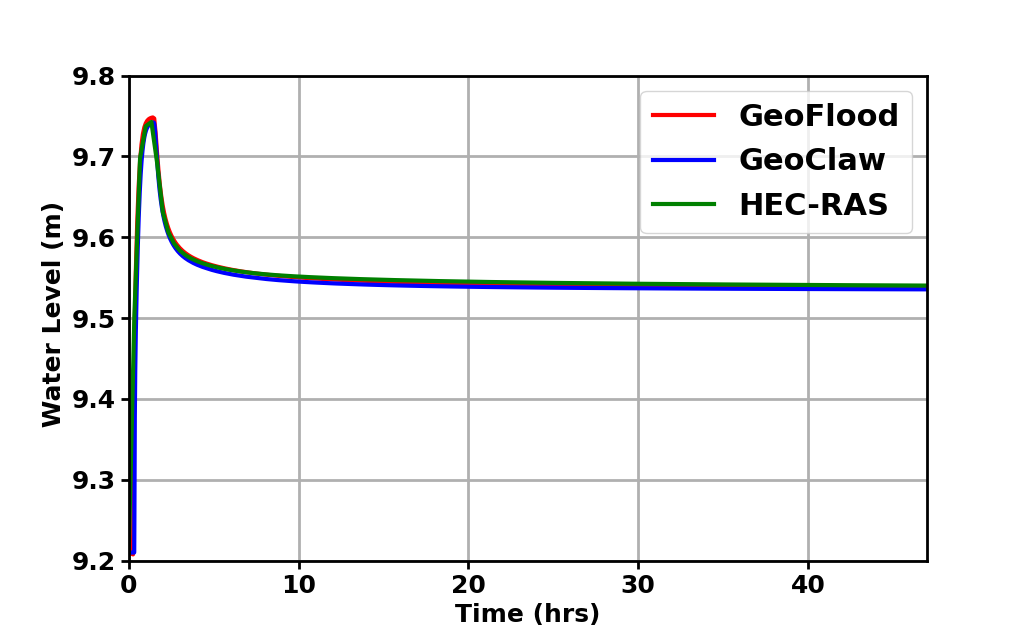}
			\caption{Depression 4}
			\label{fig:case_1_4}
		\end{subfigure}\hfil % <-- added
		\begin{subfigure}[b]{0.33\textwidth}
			\includegraphics[width=\linewidth]{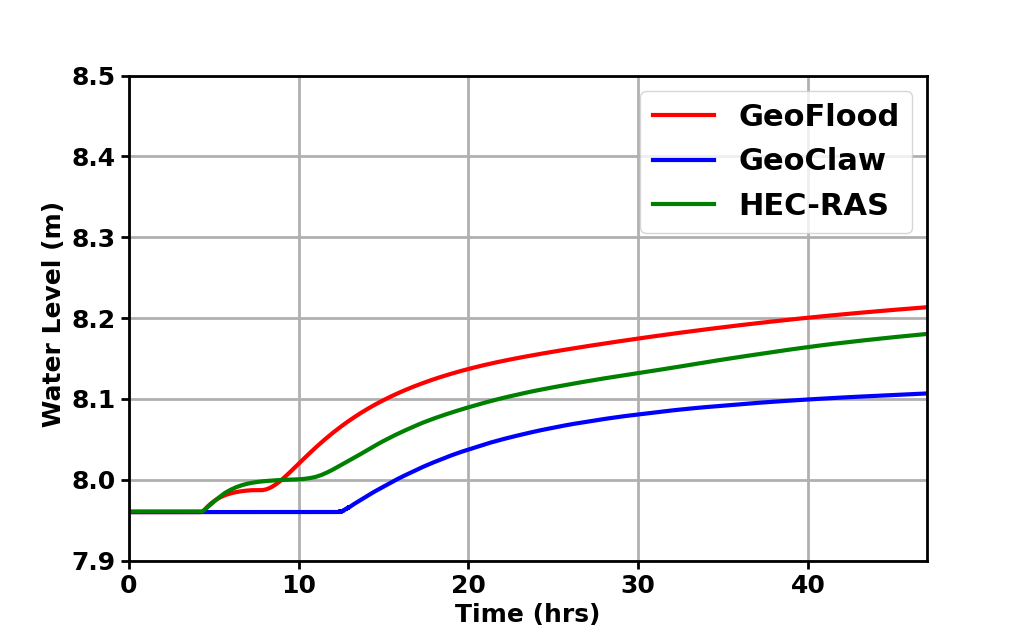}
			\caption{Depression 5}
			\label{fig:case_1_5}
		\end{subfigure}
		% \medskip
		\begin{subfigure}[b]{0.33\textwidth}
			\includegraphics[width=\linewidth]{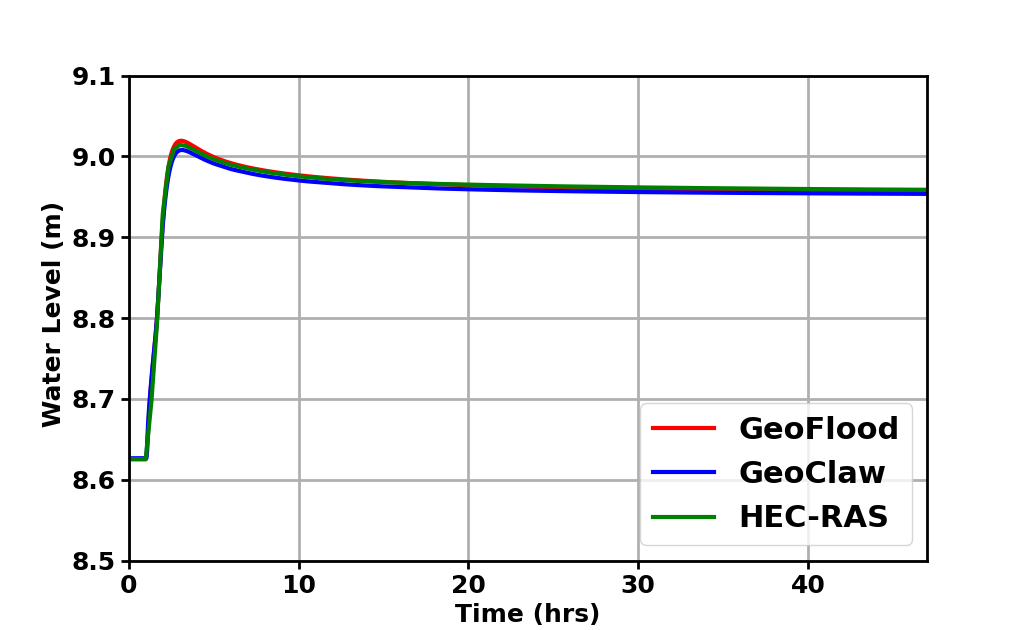}
			\caption{Depression 7}
			\label{fig:case_1_7}
		\end{subfigure}\hfil % <-- added
		\begin{subfigure}[b]{0.33\textwidth}
			\includegraphics[width=\linewidth]{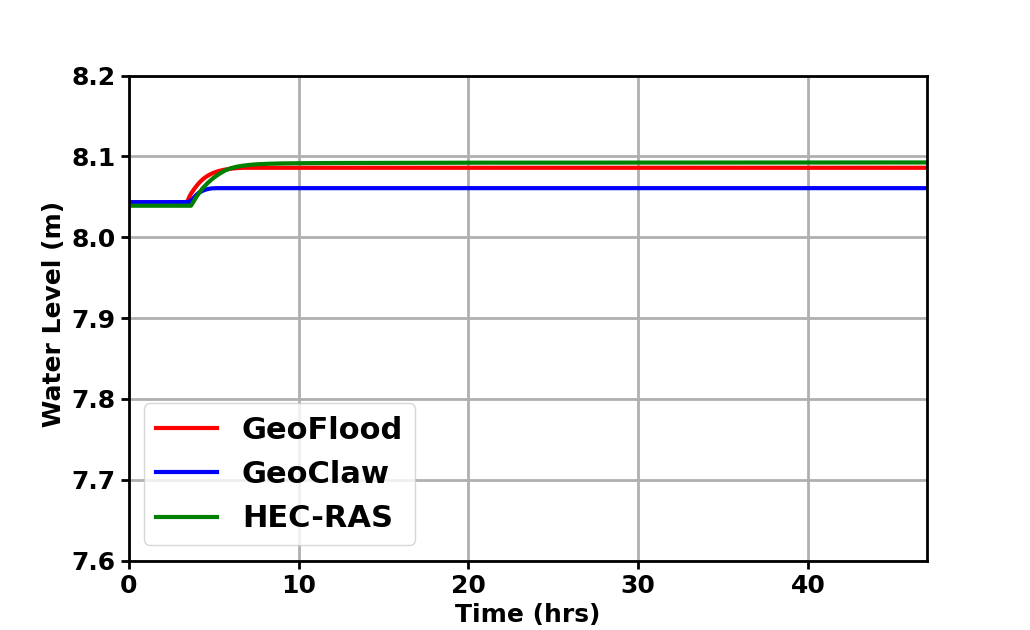}
			\caption{Depression 10}
			\label{fig:case_1_10}
		\end{subfigure}\hfil % <-- added
		\begin{subfigure}[b]{0.33\textwidth}
			\includegraphics[width=\linewidth]{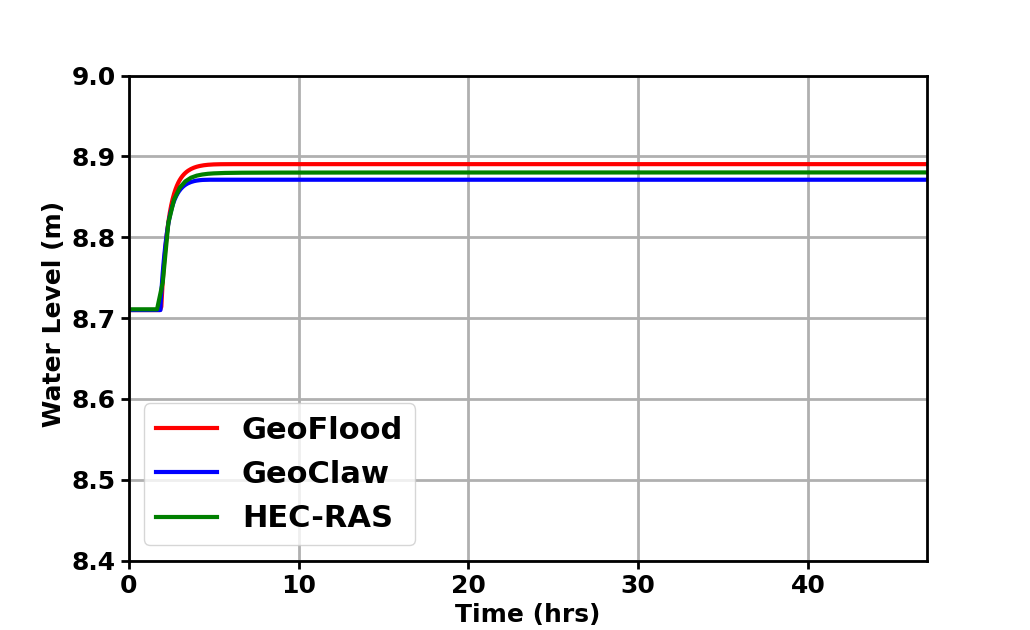}
			\caption{Depression 12}
			\label{fig:case_1_12}
		\end{subfigure}
		\caption{Temporal evolution of the water level at various depressions for GeoFlood compared to GeoClaw and the HEC-RAS model. See figure \Fig{case_1_results} for locations of depressions on the simulation grid.}
		\label{fig:filling_depressions}
	\end{figure}

GeoFlood generated results that were comparable to those of HEC-RAS and GeoClaw and predicted the flood extent at all depressions. However, different flood arrival times and initial depths were observed at depressions 5 and 10 for all models, indicating a sensitivity that we attribute to extremely shallow flow between the depressions over the thresholds and relatively slight differences in wave dynamics. We attribute the slight difference between the water elevations observed in the depressions 10 and 12 to the shallow depths and the bowl-shaped topography at these depressions. Figures \ref{fig:filling_0} - \ref{fig:filling_20} show the adaptively refined solution simulated by GeoFlood at times $t=0s$, $t =2$ hours, and $t = 2.5$ \unit{hours}, respectively, at mesh refinement levels, 0, 1, and 2, on a $50 \times 50$ grid of 4 level 0 blocks in each of the x- and y-directions,  yielding a 10 m grid resolution at level 0 and 2.5m at level 2. We compare the results with those from HEC-RAS on a uniform $200 \times 200$ grid of 10 m resolution. \Fig{filling_2} shows that during the influx, a transient water level peak was observed close to the inflow. Following the cessation of the inflow, each depression's water level steadily dropped until it eventually reached the level of the lowest "sill" separating it from adjacent depressions, as shown in \Fig{filling_20} after 20 \unit{hours}. \Fig{filling_depressions} depicts the water-surface elevation time series at six depressions.

% [b]{0.3\textwidth}
\subsection{Test Case 3:  Dam break}
The third benchmark test is designed to assess a model's ability to accurately simulate transcritical flows, hydraulic jumps, and wakes behind obstacles. This test case involves the rupture of a dam in a laboratory scale model and is presented in \citet{frazao2002dam}. Physical parameters have been scaled by $20$ times relative to the experimental set-up in order to replicate the scale of a real-world dam break scenario.
\subsubsection{Problem Setup}
\begin{figure}[ht]
\centering
  \includegraphics[width=\textwidth,clip=true,trim=0cm 2cm -3cm 2.5cm]{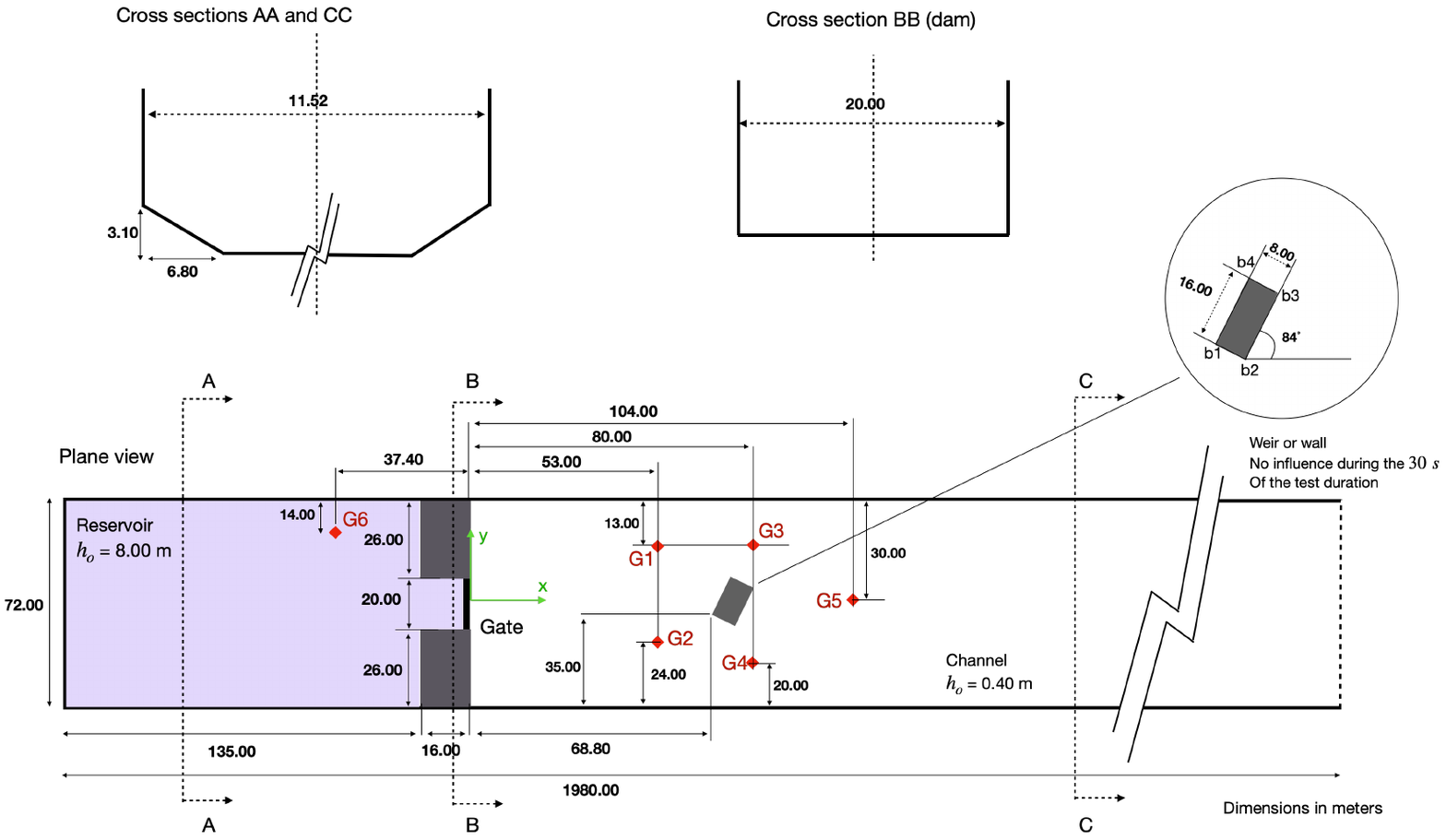}
  \caption{Geometry and dimensions for the experimental dam break test case, scaled-up 20 times. Adapted from \citet{neelz2013benchmarking}.} 
  \label{fig:dambreak}
\end{figure}

\begin{figure}[H]
    \centering
    \begin{subfigure}{0.85\textwidth}
        \centering
        \includegraphics[width=\textwidth,clip=true,trim=0cm 0cm 0cm 0cm]{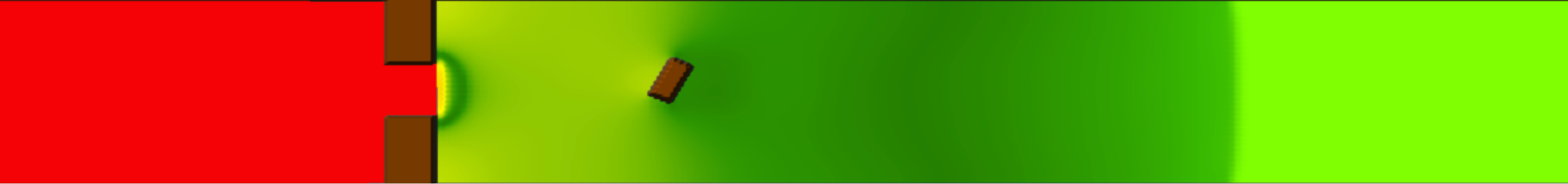}
        \caption{Dam-break HEC-RAS Simulation after $1$~\unit{minute}}
        \label{fig:damhec}
    \end{subfigure} \begin{adjustbox}{raise=0.45cm, height=2.5cm,width=1.3cm}\begin{subfigure}[b]{0.05\textwidth}
		% \centering
		\plotbox{\includegraphics[width=\textwidth]{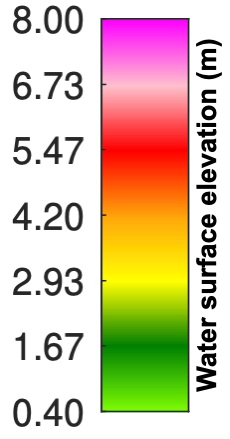}}
	\end{subfigure}\end{adjustbox}
    \begin{subfigure}{0.85\textwidth}
        \centering
        \includegraphics[width=\textwidth,clip=true,trim=0cm 0cm 0cm 0cm]{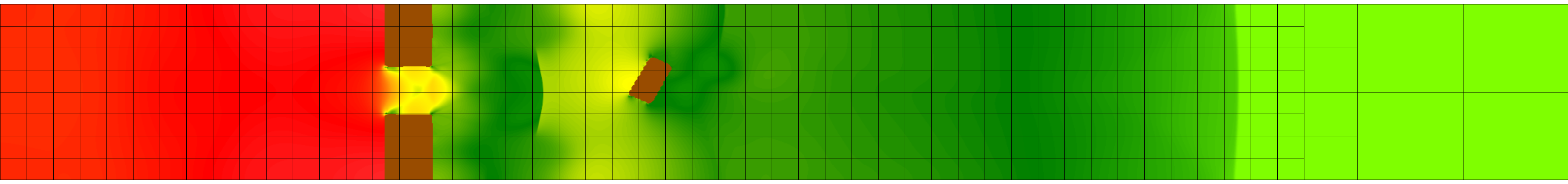}
        \caption{Dam-break GeoFlood Simulation after $1$~\unit{minute}}
        \label{fig:damgeo}
    \end{subfigure}
    \medskip
    \begin{adjustbox}{raise=0.45cm, height=2.5cm,width=1.3cm}\begin{subfigure}[b]{0.05\textwidth}
        % \centering
        \plotbox{\includegraphics[width=\textwidth]{images/dam_colorbar.png}}
    \end{subfigure}\end{adjustbox}
    %\begin{adjustbox}{raise=0.6cm, height=3.0cm,width=1.3cm}\begin{subfigure}[b]{0.05\textwidth}
        %\centering
        %\plotbox{\includegraphics[width=\textwidth]{images/colorbar_dam_geo.png}}
     %   \plotbox{\includegraphics[width=\textwidth]{images/colorbar_dam_geo.png}}
    %\end{subfigure}\end{adjustbox}
    \caption{Flow depth and wave structure $1$~\unit{minute} after the dam break (Test Case 3) for the (a) HEC-RAS and (b) GeoFlood simulations, on uniform and adaptively refined grids, respectively. Color scale indicates the water surface elevation.}
    \label{fig:dam_break}
\end{figure}

 In the experiment, a dam-break wave was generated by the nearly instantaneous opening of the gate at the end of laboratory-scale reservoir. The wave then collided with an oblique rectangular obstacle positioned downstream of the dam, producing a hydraulic jump just upstream of the obstacle and a wake zone downstream, as shown in Figures \ref{fig:dambreak} and \ref{fig:dam_break}.
% \vspace{-1.0cm}
The scaled-up computational domain was $1980$ \unit{m} long and $72$ \unit{m} wide (\Fig{dambreak}). The simulation was run for 30 minutes. The following numerical configurations were used: a wet-dry threshold of $\tau = 0.0001$, an adaptive time step with a maximum CFL of $0.9$, and a Manning coefficient of $0.05$.

\subsubsection{Test Case 3: Simulation Results}

\begin{figure}[H]
		\centering % <-- added
		\begin{subfigure}{0.33\textwidth}
			\includegraphics[width=\linewidth]{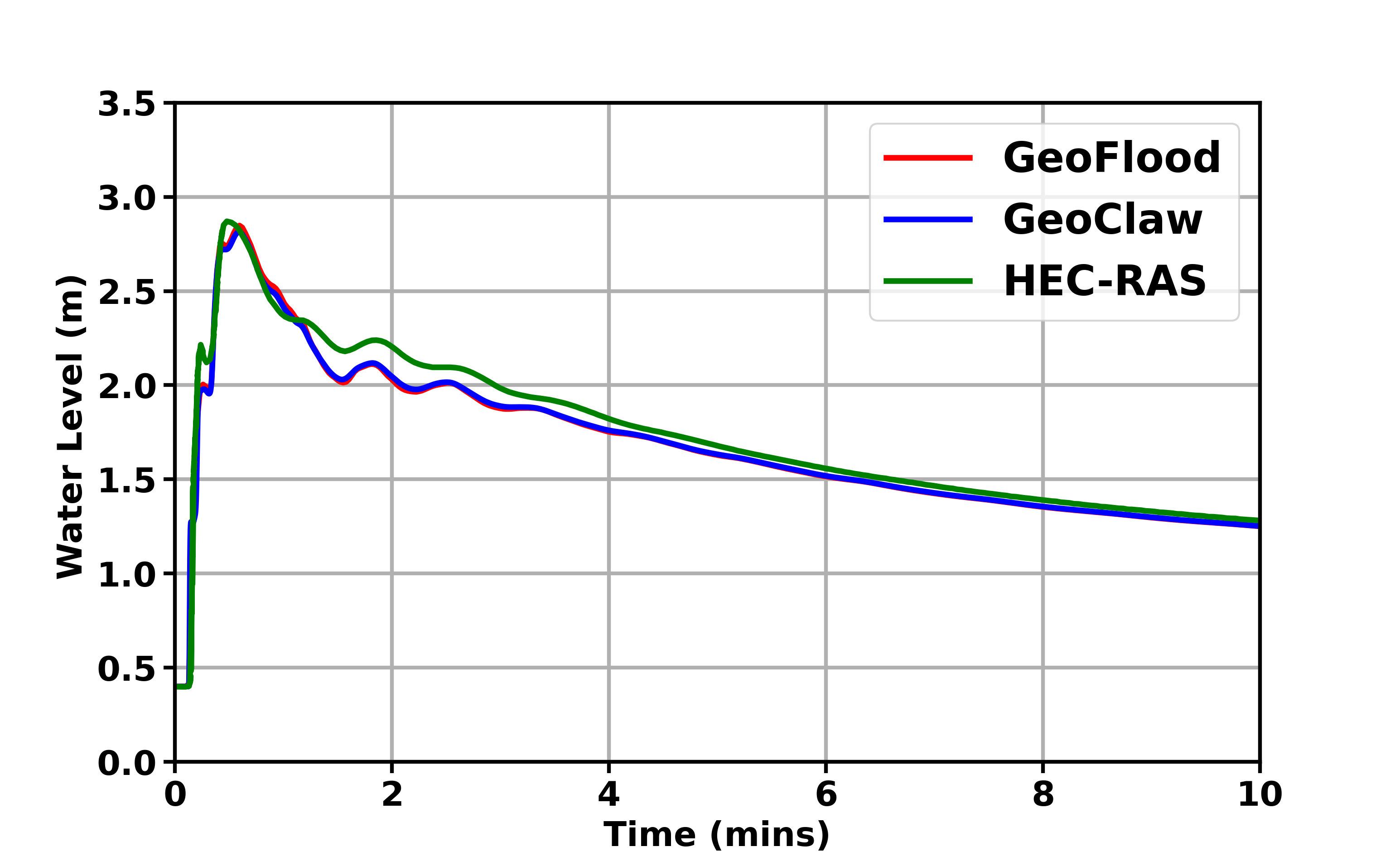}
			\caption{Water elevation at point 1}
			\label{fig:eta_p1}
		\end{subfigure}\hfil % <-- added
		\begin{subfigure}{0.33\textwidth}
			\includegraphics[width=\linewidth]{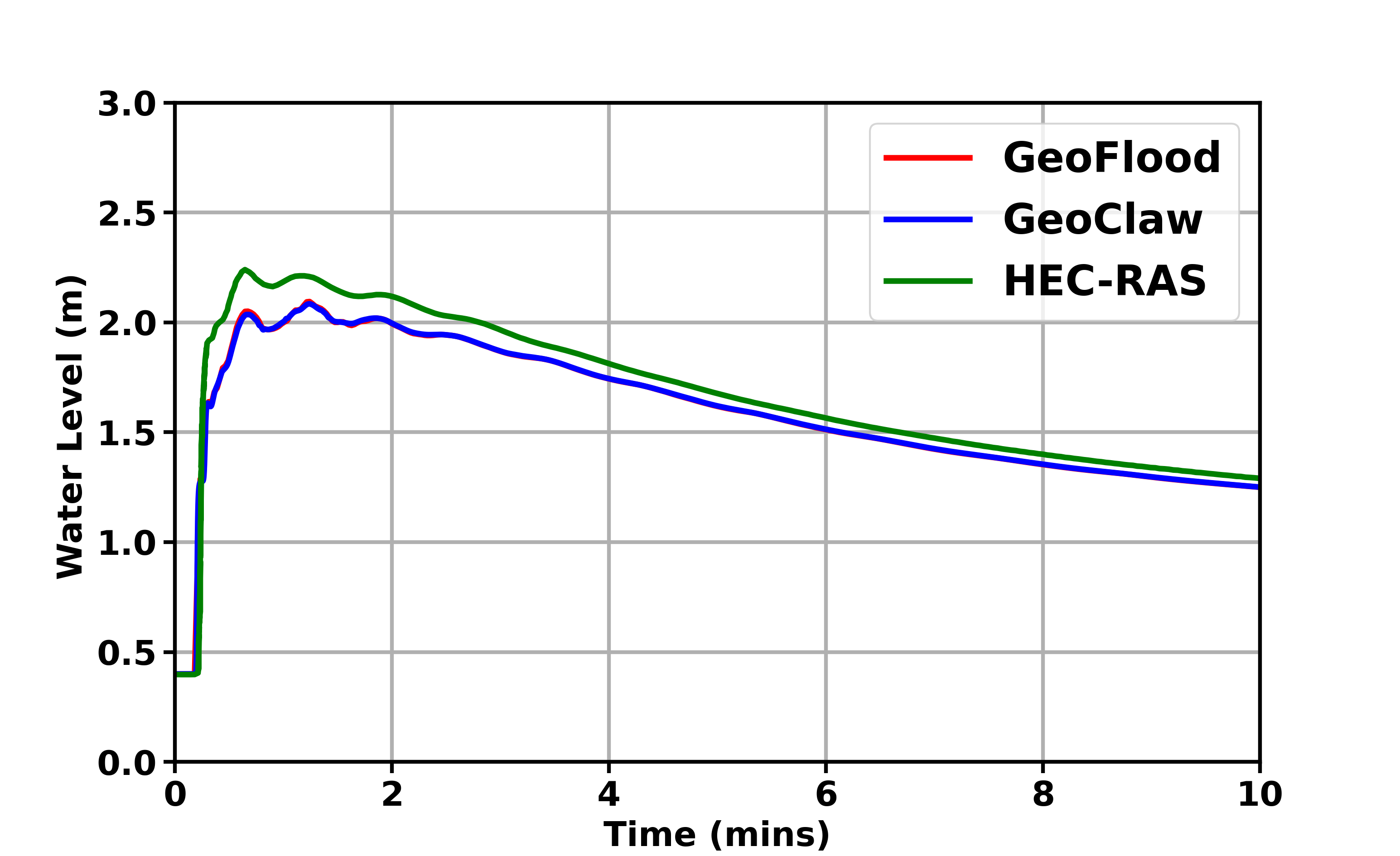}
			\caption{Water elevation at point 4}
			\label{fig:eta_p4}
		\end{subfigure}\hfil % <-- added
		\begin{subfigure}{0.33\textwidth}
			\includegraphics[width=\linewidth]{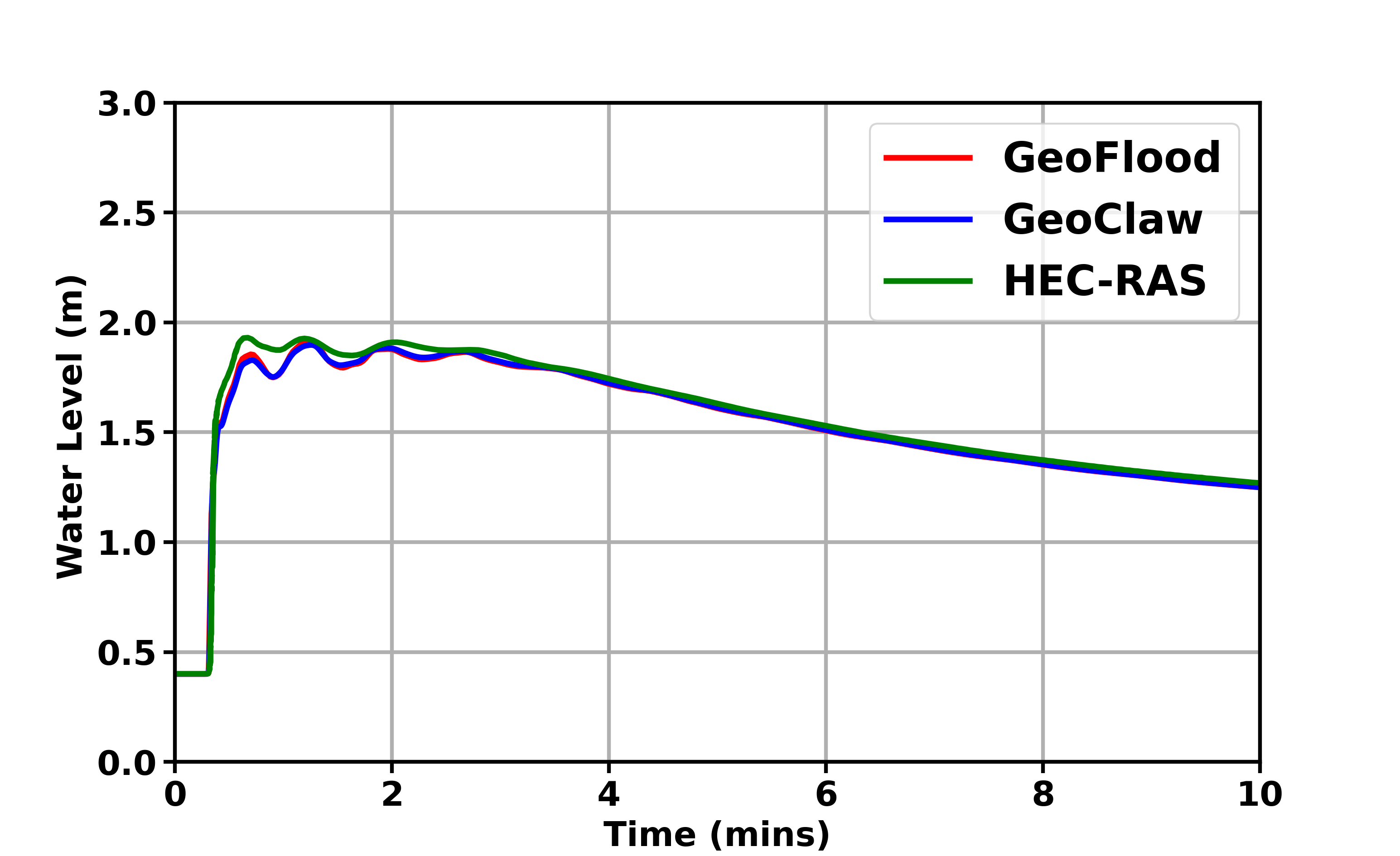}
			\caption{Water elevation at point 5}
			\label{fig:eta_p5}
		\end{subfigure}
		
		\medskip
		\begin{subfigure}{0.33\textwidth}
			\includegraphics[width=\linewidth]{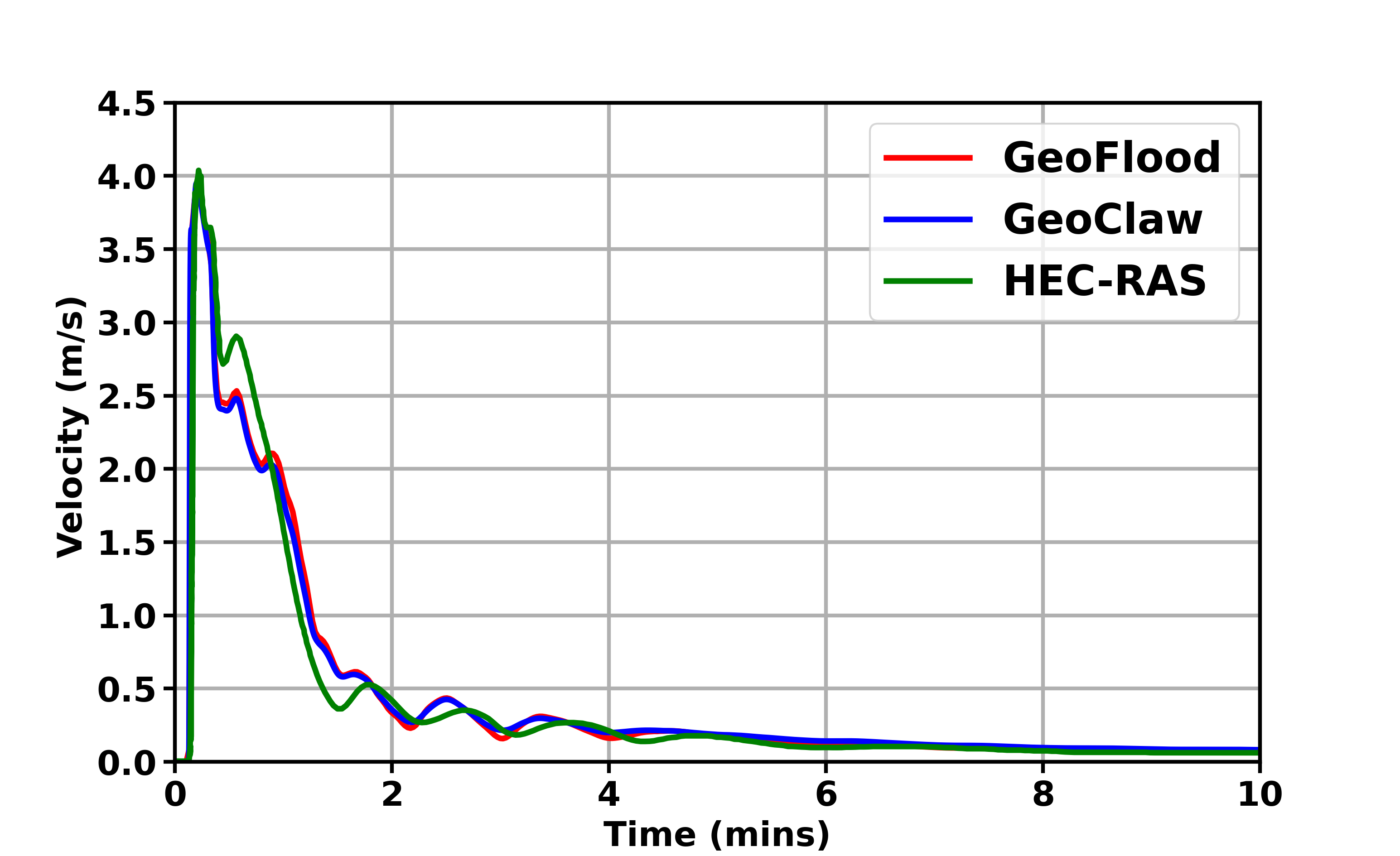}
			\caption{Velocity at point 1}
			\label{fig:u_p1}
		\end{subfigure}\hfil % <-- added
		\begin{subfigure}{0.33\textwidth}
			\includegraphics[width=\linewidth]{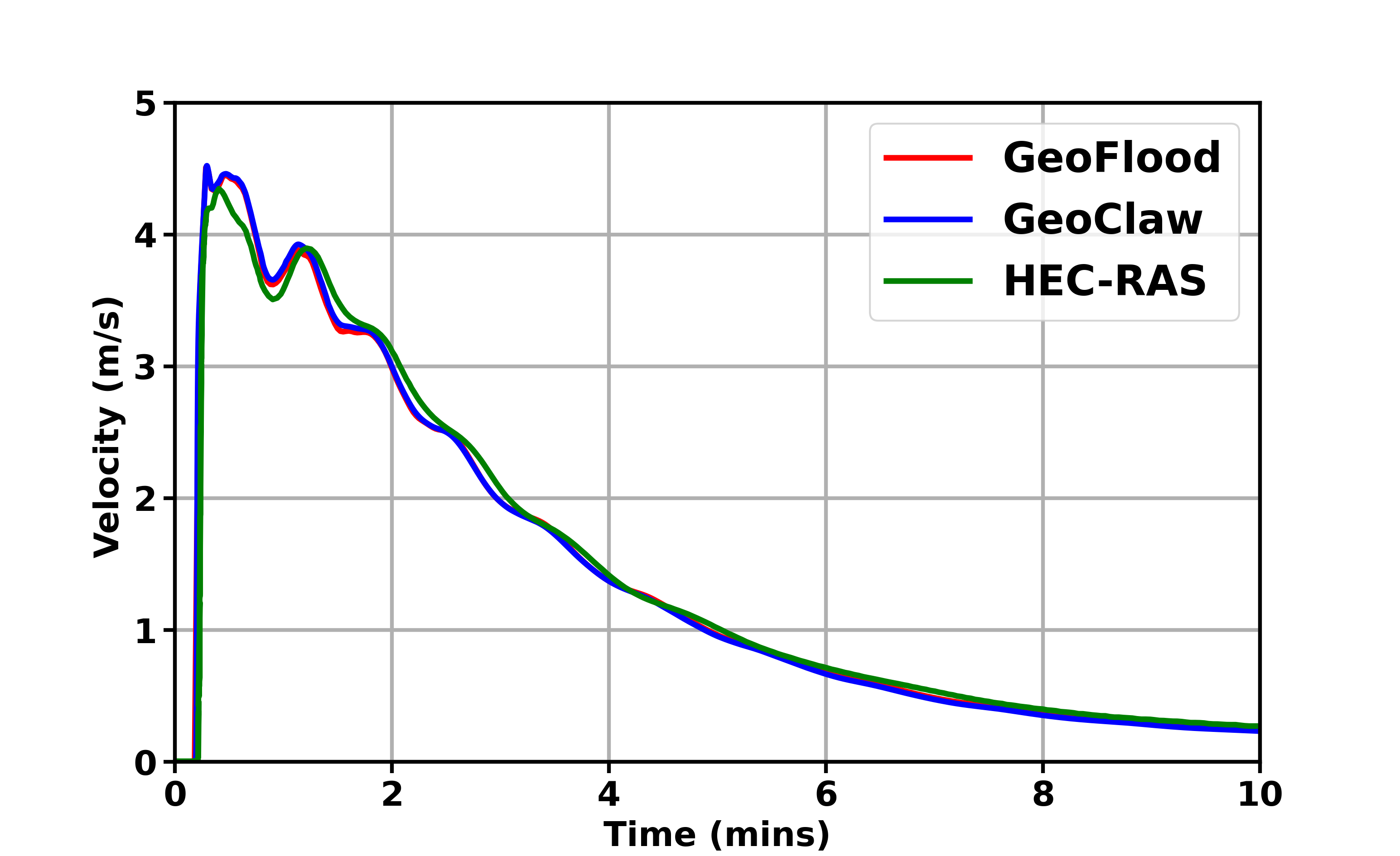}
			\caption{Velocity at point 4}
			\label{fig:u_p4}
		\end{subfigure}\hfil % <-- added
		\begin{subfigure}{0.33\textwidth}
			\includegraphics[width=\linewidth]{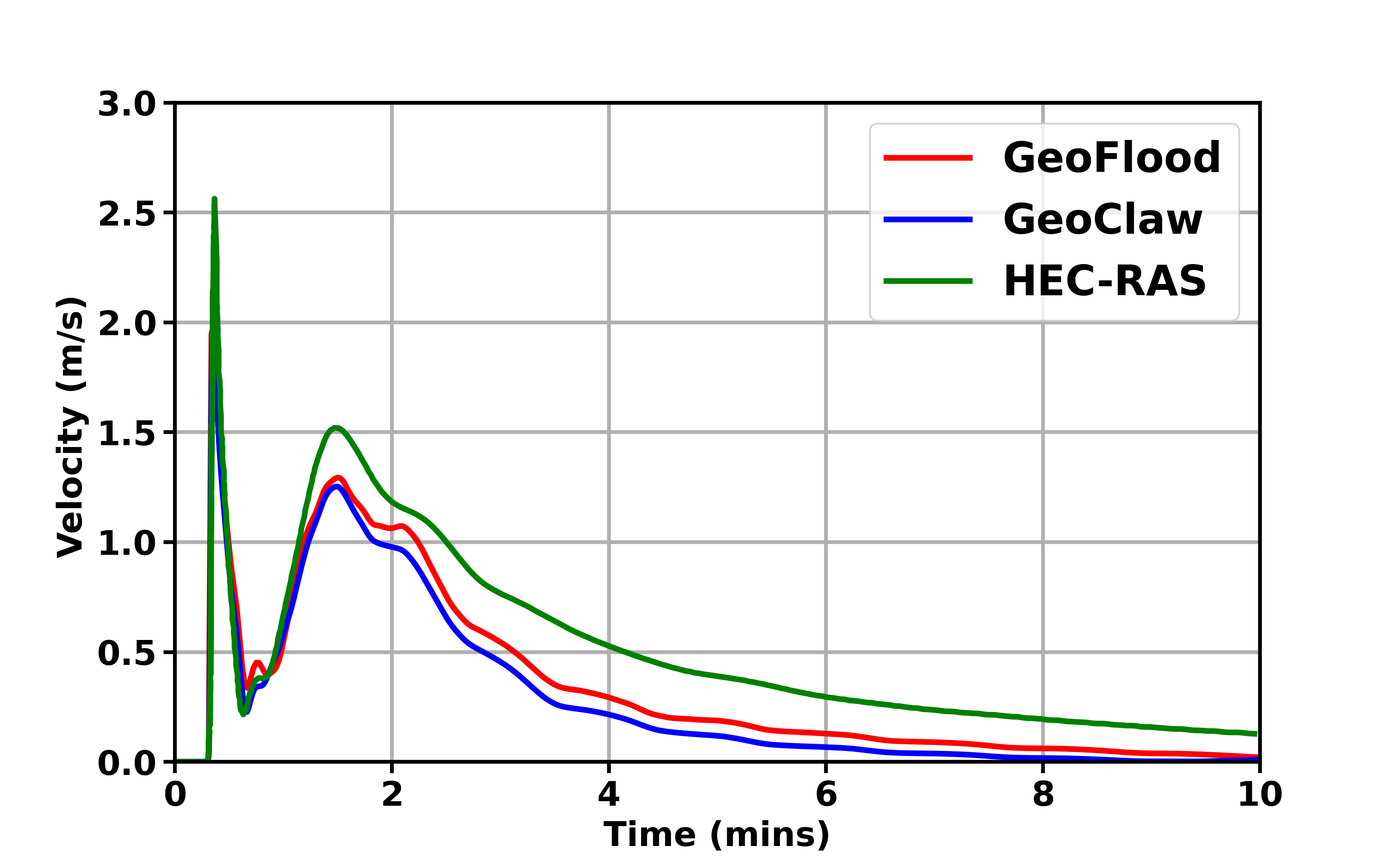}
			\caption{Velocity at point 5}
			\label{fig:u_p5}
		\end{subfigure}
		\caption{Temporal evolution of the water level (a-c) and velocity (d-f) at various control points for GeoFlood, GeoClaw, and HEC-RAS. The locations for the time series shown, point 1 (a and d), point 4 (b and e), and point 5 (c and f) are shown and labeled G1, G4 and G5, respectively in \Fig{dambreak}}, 
		\label{fig:dambreak_plots}
	\end{figure}

The advancing waves downstream of the obstacle are qualitatively similar in the GeoFlood and HEC-RAS simulations, but the GeoFlood simulation produces a more pronounced upstream-directed stationary bore above the obstacle (where flow transitions from super- to sub-critical). Nevertheless, comparisons of the temporal variation of the water-surface elevation and velocity for GeoFlood, GeoClaw, and HEC-RAS demonstrate generally consistent predictions at gauge points (\Fig{dambreak_plots}). The HEC-RAS simulation produces a sharp discontinuity remaining at the location of the dam and relatively smooth flow near the obstacle. The HEC-RAS solution was computed on a uniform $990 \times 36$ grid, while the GeoFlood solution used an adaptive grid starting with coarse $36 \times 36$ level $0$ grid blocks in a $27 \times 1$ block arrangement at $2$ m grid resolution refining to a 0.3 m per grid cell resolution.

%% ------ new section --------------------
\section{Malpasset Outburst-Flood Simulations}
\label{sec:malpasset}
For our final test case, we assess GeoFlood's ability to simulate a historical dam break and overland flood event.  The availability of field data, laboratory-scale model data, and previous numerical results provide a means for verification and validation of GeoFlood for a real-world problem that is numerically challenging due to highly-energetic flows in steep and irregular terrain. We also use this test case to benchmark GeoFlood's parallel computational efficiency, and to demonstrate the usage of GeoFlood's AMR schemes for a large-scale, real-world problem.
\subsection{Historical Background} 
\begin{figure}[H]
    \includegraphics[width=0.45\linewidth]{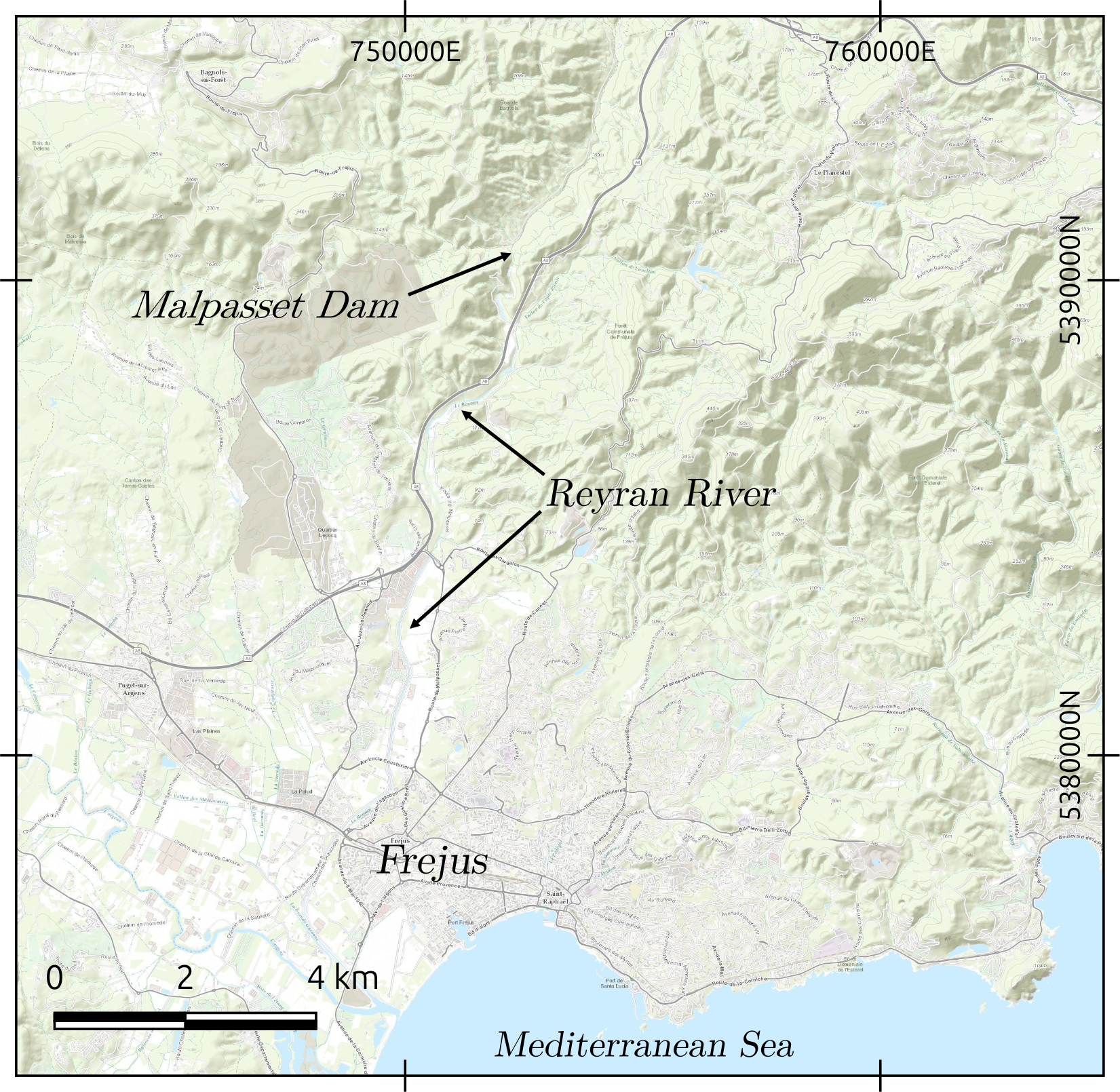}\hfil
    \includegraphics[width=0.45\linewidth]{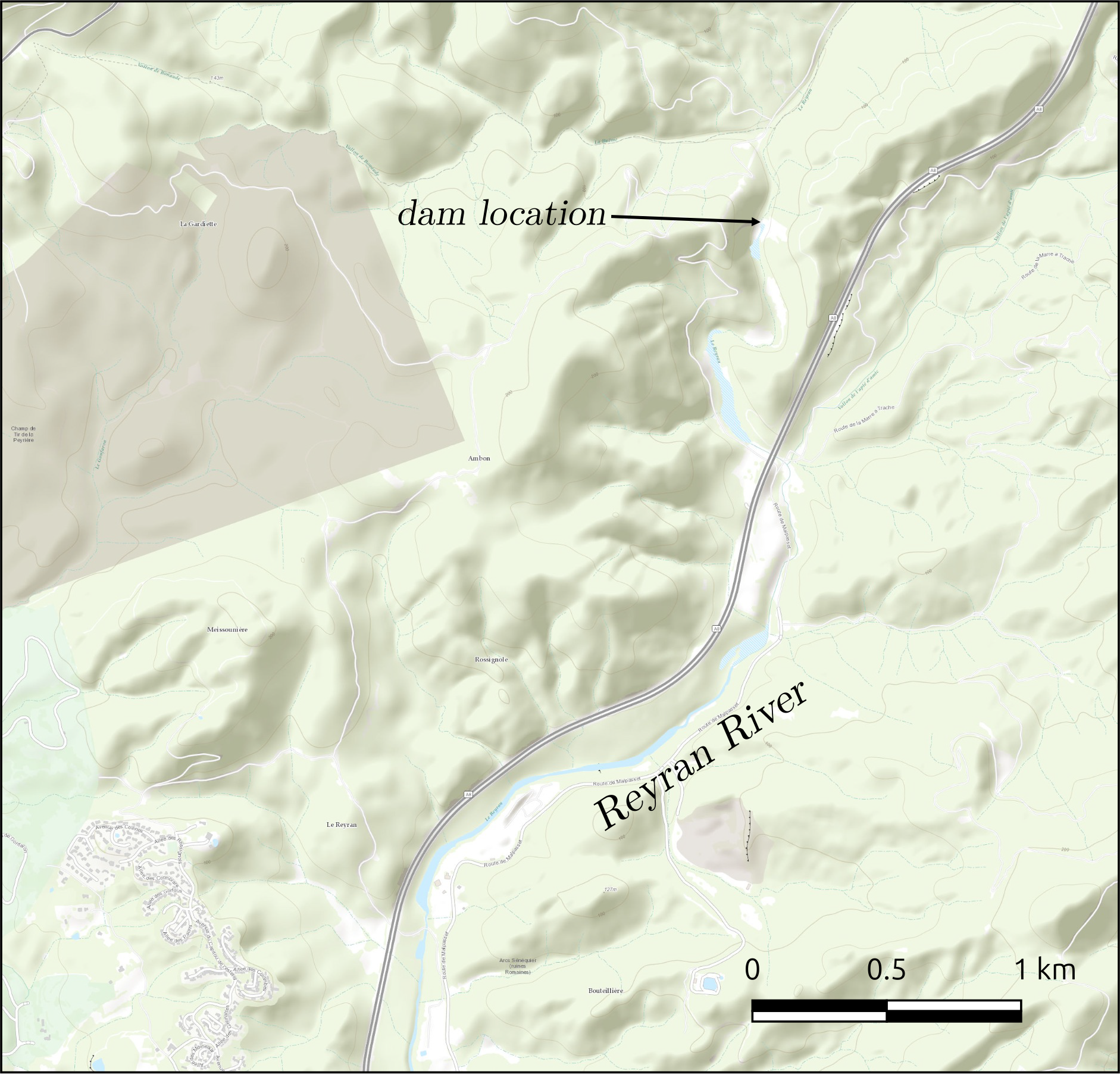}\hfil
    \caption{Overhead map of the area surrounding Fr\'{e}jus, France (a). The reservoir impounded by the Malpasset Dam lay in the river valley to the northeast of the dam. Close-up (b) shows the Reyran River gorge features below the dam (b). Base images: ESRI World Topo.}
    \label{fig:map_qgis}
    \end{figure}

The Malpasset Dam is situated approximately $12$~\unit{km} upstream from the town of Fr\'{e}jus, France (\Fig{map_qgis}). This thin-arch dam was constructed in a narrow gorge above the Reyran River valley in order to impound a reservoir with a storage capacity of $55106~\unit{m}^3$. Upon reaching reservoir capacity, the dam failed suddenly and catastrophically on December 2, $1959$ at $21$:$14$~\unit{hours} (generating an acoustic shock wave witnessed by residents in Fr\'{e}jus, suggesting a nearly instantaneous failure). The dam had a maximum height of $66.5$~\unit{m} and a crest span of $223$~\unit{m}. Only remnants of the dam's arch remained after the failure, with significant erosion of the adjacent rock bank. Subsequent investigations suggest the arch dislodged from its base leading to a rapid sequential collapse \citep{valiani2002case}.

The catastrophic breach led to a rapid flood wave that descended through a channelized ravine, eventually inundating the wide floodplain adjacent to the Mediterranean and surrounding Fr\'{e}jus. The event resulted in $433$ fatalities and significant infrastructure damages, including the obliteration of a $1.5$-\unit{km} section of highway and an adjoining bridge, and extensive flooding of Fr\'{e}jus. The downstream displacement of massive concrete blocks indicated the power of the flood. The flood waves rose to $\approx 20$~\unit{m} above the original riverbed. A complete description of this event can be found in \cite{boudou2017analysis}.

\subsection{Topographic features and model domain}
Downstream of the dam, the Reyran River passes through a pronounced depression, then winds through a series of steep, fast-flowing serpentine bends before descending through a $\approx 4$~\unit{km} gorge, from which it ultimately emerges onto the broad, low-lying alluvial fan in the Reyran River valley surrounding Fr\'{e}jus (\Fig{map_qgis}). Highly energetic flows through these features pose numerical challenges due to large flow gradients and topographic source terms, making them a valuable test for model robustness and stability \citep[e.g.,][]{george2011adaptive, hervouet1999malpasset}.

Our simulation dimensions are $17500 \times 9000$~\unit{m}. The digital elevation model (DEM) for the GeoFlood simulations originated from a benchmarking exercise in 1999, sponsored by CADAM (Concerted Action on Dam-break Modelling) project, a European research group \citep{frazaodam} which included a set of $13541$ irregularly spaced elevation points covering the study domain. The points were interpolated to overlapping sets of regularly-spaced Cartesian grids with 2, 5, and 20 m resolutions by \citet{george2011adaptive} and used in this study. 
\par
GeoFlood incorporates GeoClaw's topographic processing schemes, which allow users to list multiple overlapping or nested DEMs of different resolutions without requiring preprocessing. The AMR interpolation schemes are designed to preserve the integral of a unique topographic interpolant defined by the DEMs, which is necessary for volume conservation upon dynamic refinement of a submerged area \citep{leveque2011}. Nevertheless, it is generally recommended to maintain the same resolution for rapid flows in steep terrain \citep{george2011adaptive}.

\subsection{Initial and Boundary Conditions}
The sea level and the initial reservoir level are assumed to be constant and are set at $0$ and $100$~\unit{m} above sea level, respectively.  Although the outlet gate near the bottom of the dam was open during the event, we neglected pre-event streamflow in the channel---the bottom was considered dry. The actual pre-event streamflow discharge is unknown, but we assumed that it was negligible relative to the massive outburst flood discharge.  Similarly, we assumed that the inlet discharge upstream of the reservoir was negligible. The sea level remained constant and uniform with an elevation of zero.  We assumed an instantaneous dam failure. 
\par
For comparison with \citet{george2011adaptive} and the results produced by the CADAM project, we used a Manning coefficient of 0.033 for our simulations.

\subsection{Simulation Refinement Flags}
GeoFlood allows users to control AMR refinement, setting limits or enforcing minimum and maximum refinement levels based on multiple criteria. These criteria include flow features (depth, velocity, depth gradients) or geographic regions. For real-world problems, this has been found to provide more reliable behavior with better computational efficiency than error estimation algorithms (\emph{e.g.,} Richardson extrapolation) \citep{leveque2011}. Additionally, users may have interest in particular locations or times. 
\par
\begin{figure}[H]
	\centering
	\begin{subfigure}[b]{4.0cm}
		\centering
		\includegraphics[width=\textwidth]{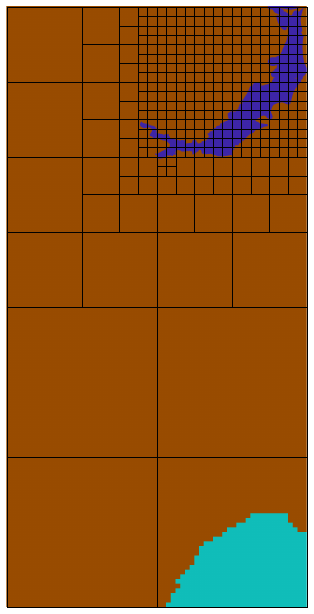}
		\caption{At $t=0~s$}
		\label{fig:a}
	\end{subfigure}
	\hfil
	\begin{subfigure}[b]{4.0cm}
		\centering
		\includegraphics[width=\textwidth]{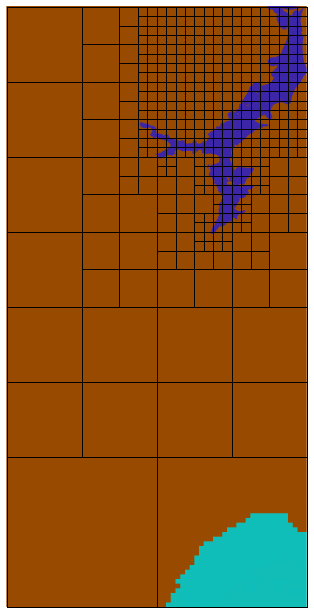}
		\caption{At $t=200~s$}
		\label{fig:b}
	\end{subfigure}
	\hfil
	\begin{subfigure}[b]{4.0cm}
		\centering
		\includegraphics[width=\textwidth]{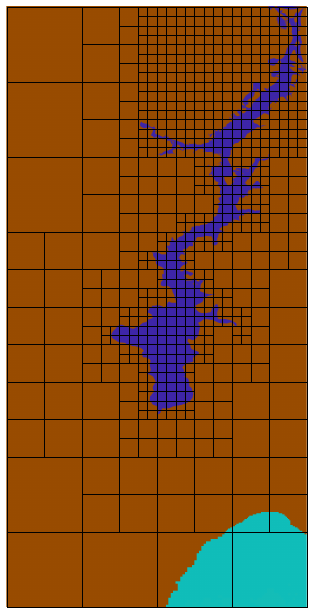}
		\caption{At $t=1000~s$}
		\label{fig:c}
	\end{subfigure}
	   \caption{Adaptive mesh refinement (AMR) applied in the simulation of the Malpasset dam failure is depicted at times of $t=0, 200,$ and $1000$~\unit{seconds}. Lines indicate the edges of blocks, each containing $32$ x $32$ individual cells. The reservoir was refined to level $l=5$ throughout the computation. As time progressed, the expanding flood was refined to $l=5$, while the unaffected areas remained at the coarsest refinement level $l=1$.  The AMR procedures are designed to allow optimizing the balance of feature resolution with computational efficiency.}
    \label{fig:amr}
  \end{figure}

For our simulation of the Malpasset flood, we enforced the highest level of refinement (level 5) for rapidly flowing water. We also enforced refinement of the reservoir based on its geographic location, by specifying its bounding box as input parameters. (Note that enforcing refinement to the highest level everywhere that water is present would result in unnecessary refinement of the Mediterranean Sea (\Fig{amr}) where large depths would lead to severe time-step restrictions over the duration of the simulation.) We chose to enforce the highest refinement level for the reservoir in order to represent the total outflow discharge at the dam as accurately as possible (the flow dynamics in the lake would not otherwise need a high spatial resolution). More elaborate schemes, such as refining flow at a variety of levels based on flow criteria are possible.
\subsection{Comparison and validation of model results}
\begin{figure}[ht]
    \centering
    \begin{subfigure}[b]{7.5cm}
        \centering
        \includegraphics[width=\textwidth,clip=true,trim=0cm 0cm 0cm 2cm]{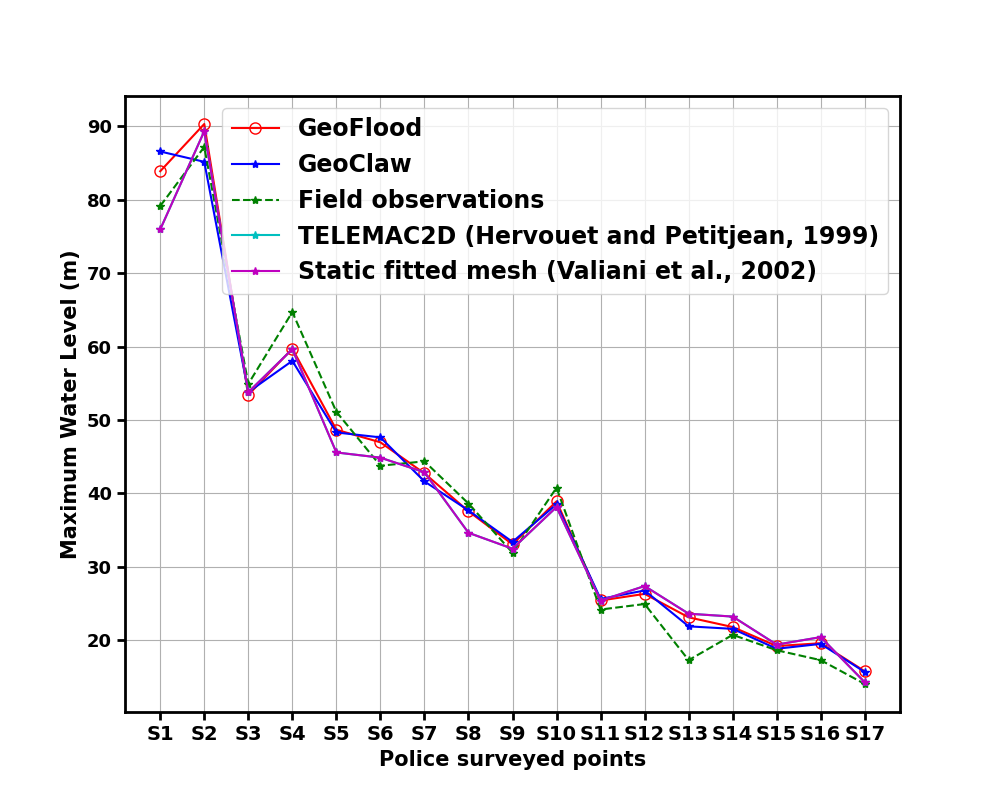}
        \caption{Police-surveyed points}
        \label{fig:police}
    \end{subfigure}
    \begin{subfigure}[b]{7.5cm}
        \centering
        \includegraphics[width=\textwidth,clip=true,trim=0cm 0cm 0cm 2cm]{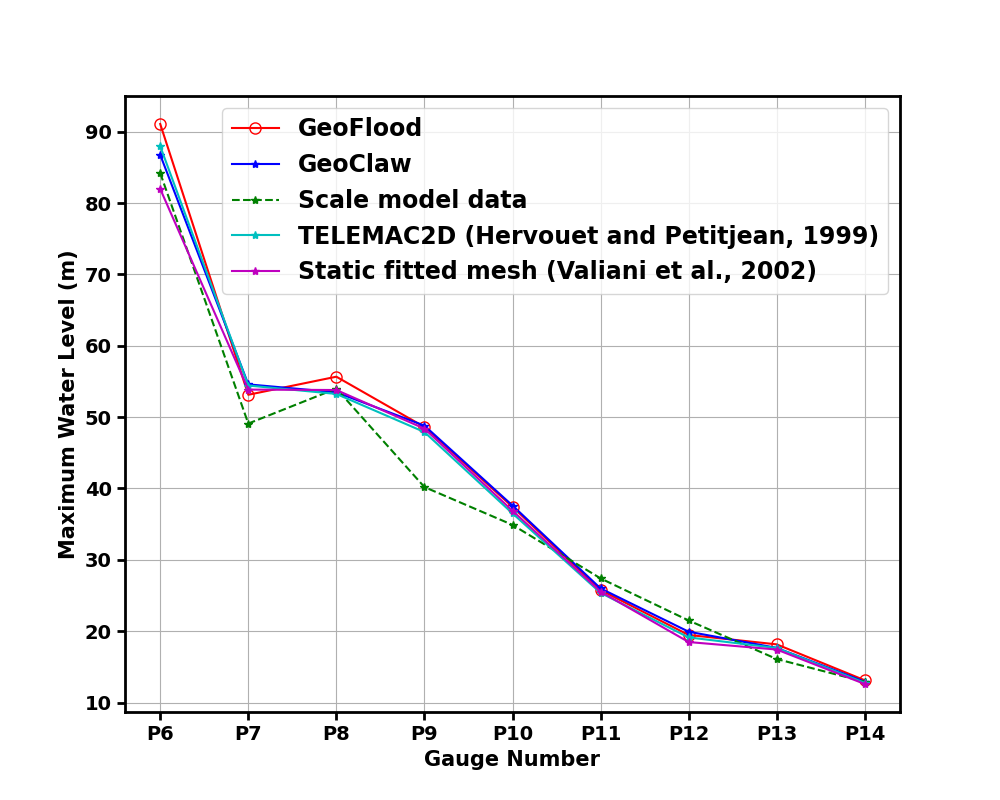}
        \caption{Gauge points}
        \label{fig:gauge}
    \end{subfigure}
    \caption{Maximum water levels simulated by GeoFlood, GeoClaw, and other model simulations compared to observations from (a) $17$ field-surveyed points (referred to as ``police-surveyed points" in related literature) and (b) $9$ gauge locations from a laboratory scale model \citep{frazaodam}. \label{fig:policegauge}}
\end{figure}

The Malpasset dam break and flood serves as an excellent real-world test case for model verification and validation, thanks to the availability of field survey data on high-water marks and laboratory-scale model data. These datasets enable verification of the shallow-water equations and facilitate model comparison and validation. We compare GeoFlood results against empirical data and previous results from other models, all of which are based on the shallow-water equations. Despite the steep terrain, where shallow-water assumptions are most uncertain, GeoFlood performs well, as do the other models. \Fig{policegauge} presents a comparison of GeoFlood, GeoClaw \citep{george2011adaptive}, and two additional models from the 1999 CADAM workshop \citep{frazaodam} against observations from 17 field-surveyed locations and 9 model gauge stations. 
\subsection{Computational efficiency benchmarks}
GeoFlood simulations performed on a laptop ($2.3$\unit{GHz} Quad-Core Intel Core $i7$ processor with $16$ \unit{GB} of RAM), required approximately $1500$ \unit{seconds} of wall clock time to complete the entire scenario, a total simulated time of  $4000$~\unit{seconds}. Our simulation used lowest level computational grids measuring $64 \times 128$~\unit{m}, with the levels of grid refinement ranging from $1$ to $5$, yielding an effective resolution of $1024 \times 2048$~ \unit{m} and a cell resolution of $6$~ \unit{m} per grid cell. The model operated with an initial time step of $1$~\unit{second}, followed by adaptive time-stepping adhering to a maximum CFL number set at $0.75$.
\par
\begin{figure}[ht]
    \centering
     \begin{subfigure}[b]{7.8cm}
          \centering
        \includegraphics[width=\textwidth,clip=true,trim=0cm 0.25cm 0cm 0cm]{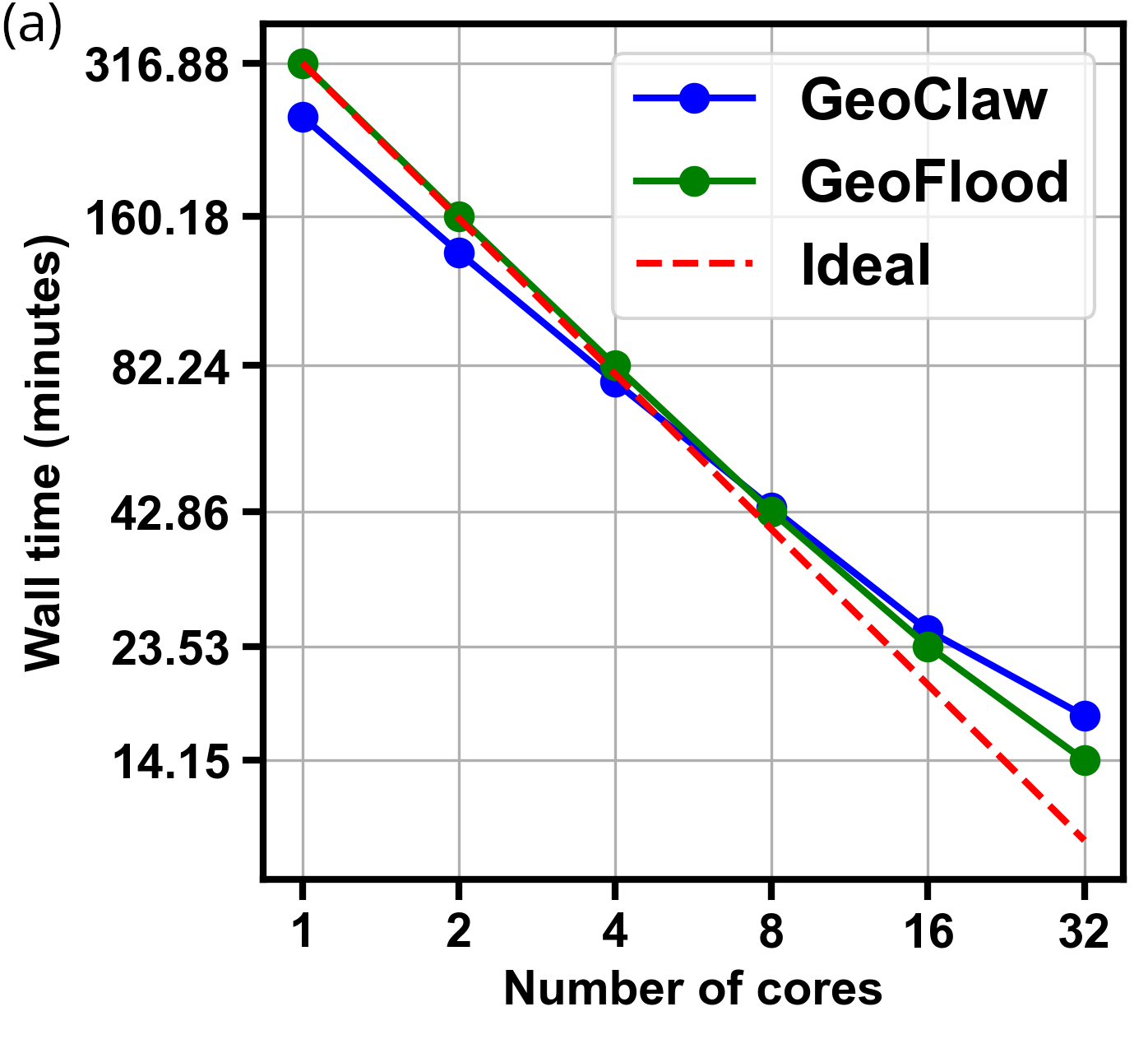}
        % \caption{Wall time}
        \label{fig:wall}
      \end{subfigure}
   \hfil
      \begin{subfigure}[b]{7.5cm} %<-- -0.3cm since the image seems bigger than the other
        \centering
        \includegraphics[width=\textwidth,clip=true,trim=0cm 0.25cm 0cm 0cm]{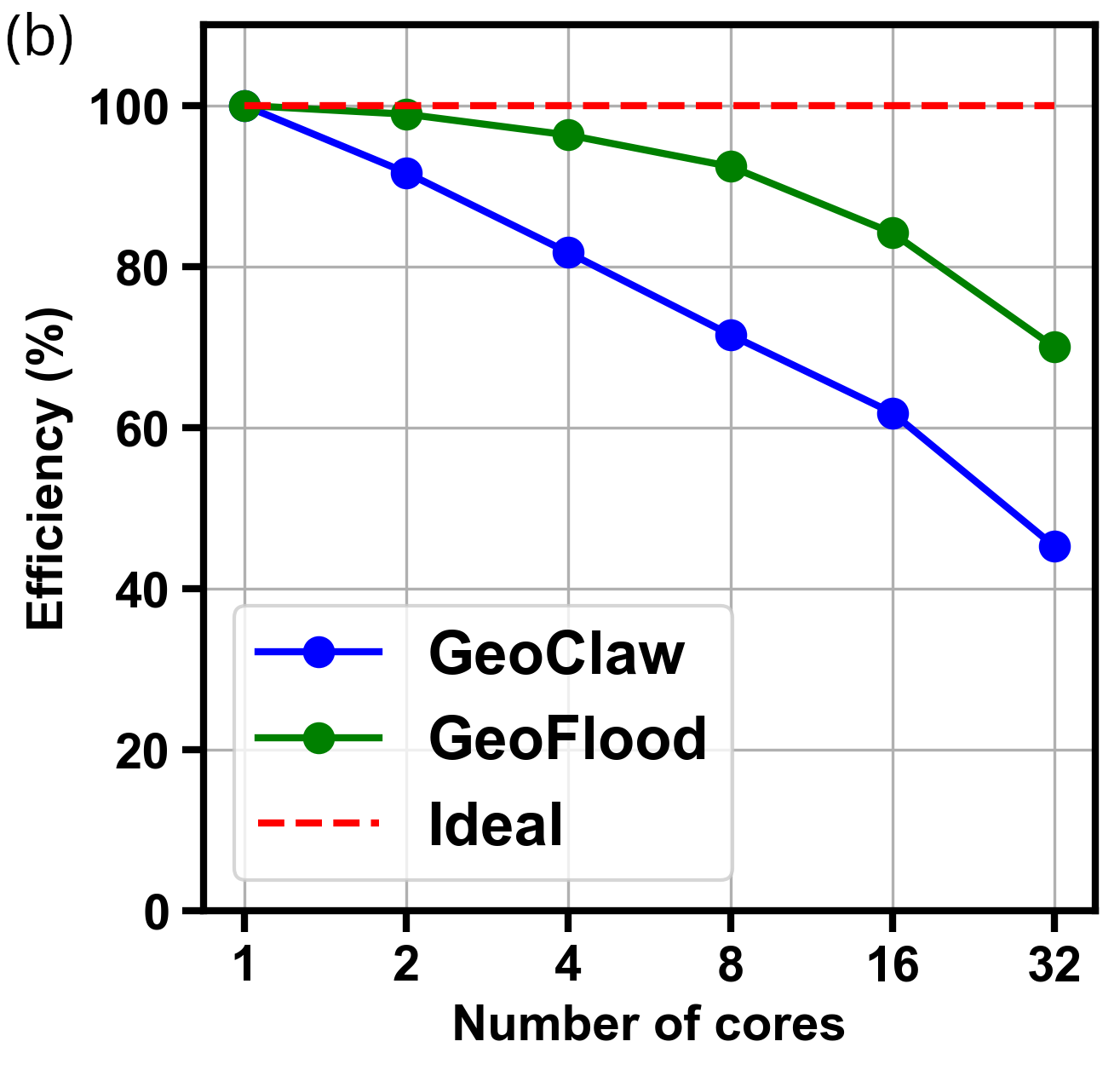}
        % \caption{Parallel Efficiency}
        \label{fig:eff}
      \end{subfigure}
      \caption{Comparisons of the wall time (a) and parallel efficiency (b) of GeoClaw and GeoFlood. For few processors (1--4), GeoClaw has a reduced wall time compared to GeoFlood. However, even beyond 8 cores, GeoFlood surpasses GeoClaw in parallel performance (a). GeoFlood outperformed GeoClaw in efficiency across all tested runs (b).}
      \label{fig:strong_scaling}
  \end{figure}

We ran multiple GeoFlood and GeoClaw simulations in order to compare computational efficiency, particularly with respect to parallel efficiency and scalability.  Experiments involving various numbers of nodes and cores per node were conducted to assess the scalability and efficiency of GeoFlood. GeoClaw runs in shared memory on  single node and uses OpenMP for parallelization (one motivation for the development of GeoFlood). Our model efficiency comparisons were therefore derived from tests conducted on a single node of the Linux-based supercomputer (Borah, Boise State University), equipped with dual Intel Xeon Gold 6252 processors, each running at 2.1 GHz with 24 cores, providing a total of 48 cores on the node. Compared to GeoClaw, GeoFlood's parallel performance and computational speed show improved scaling with even moderately higher processor counts (8, 16, and 32), and GeoFlood outperformed GeoClaw in efficiency across all tested runs (\Fig{strong_scaling}). Even on a single node, GeoFlood's dynamic regridding process is fully parallelized via the mesh management libraries ForestClaw and p4est, whereas GeoClaw performs regridding in serial. This distinction likely contributes to the observed results.

%========================================================
\begin{figure}[p]
	\centering % <-- added
	\begin{subfigure}{0.7\textwidth}
	  \includegraphics[width=\textwidth]{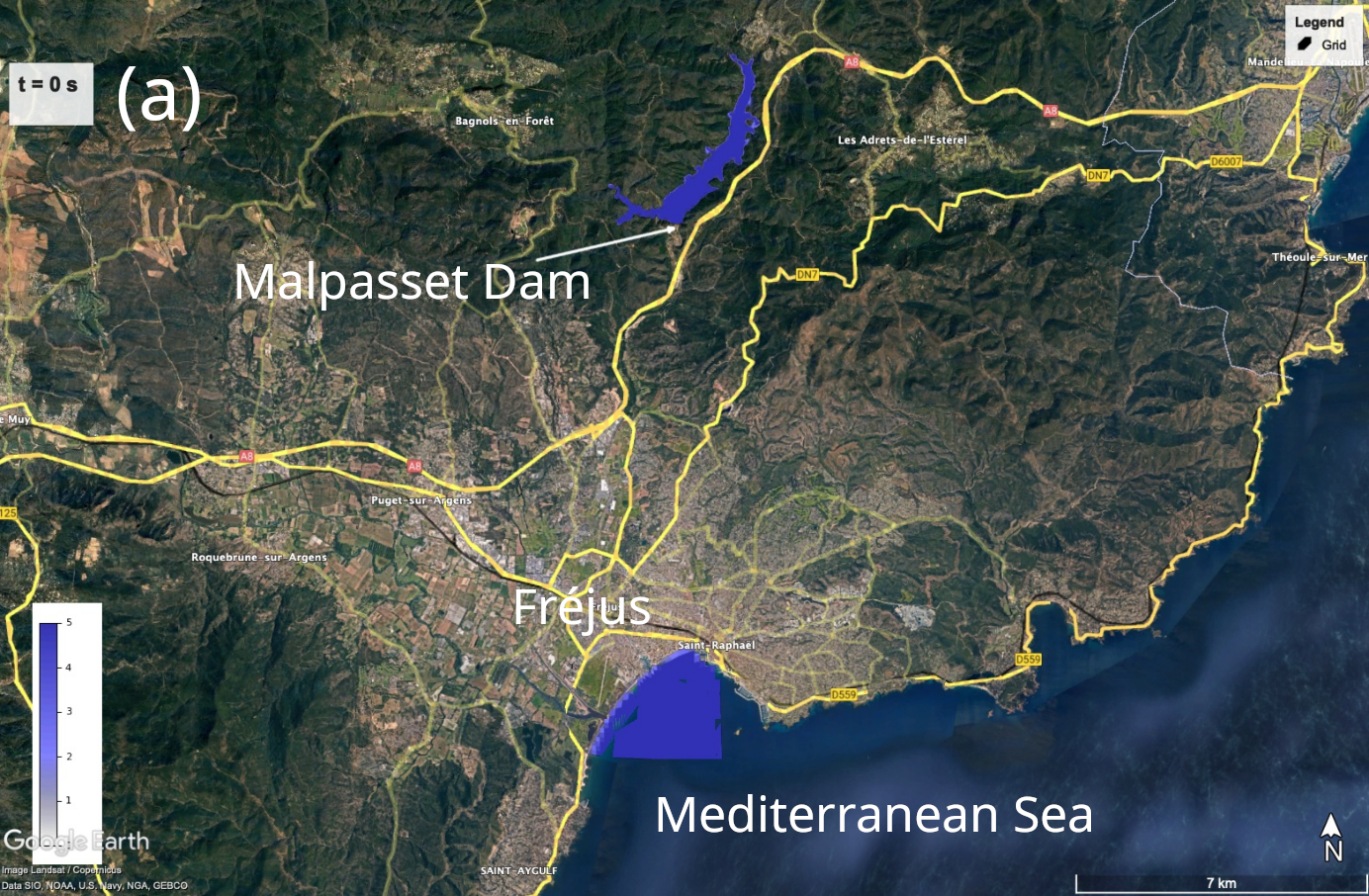}
	  % \caption{At $t=0~$seconds}
	  \label{fig:allshock0}
	\end{subfigure}\hfil % <-- added
	\begin{subfigure}{0.7\textwidth}
	  \includegraphics[width=\textwidth]{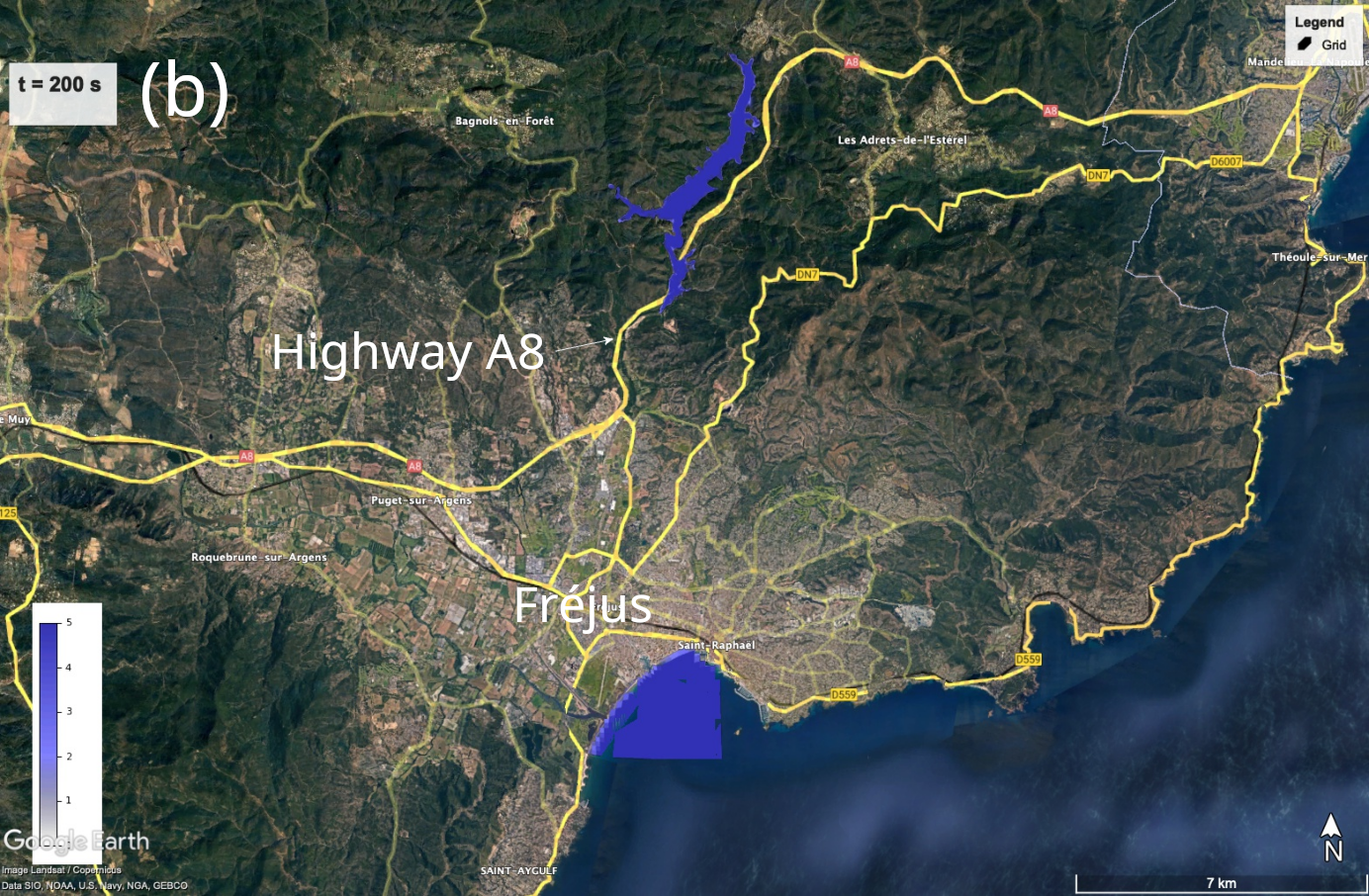}
	  % \caption{At $t=200~$seconds}
	  \label{fig:allshock3}
	\end{subfigure}\hfil % <-- added
    \caption{The sequence of the flood resulting from the Malpasset Dam failure is illustrated using Google Earth imagery at various times. At $t = 0$ \unit{seconds} (panel a), the full reservoir with the dam intact can be seen in blue. By $t = 200$ \unit{seconds} after dam failure (panel b), the floodwaters overtop the A8 highway, where the initial casualties occurred.(\textcopyright Google Earth imagery).}
\end{figure}
%-------------------------------------------------
\begin{figure}[p]\ContinuedFloat
    \begin{subfigure}[b]{0.7\textwidth}
        \includegraphics[width=\textwidth]{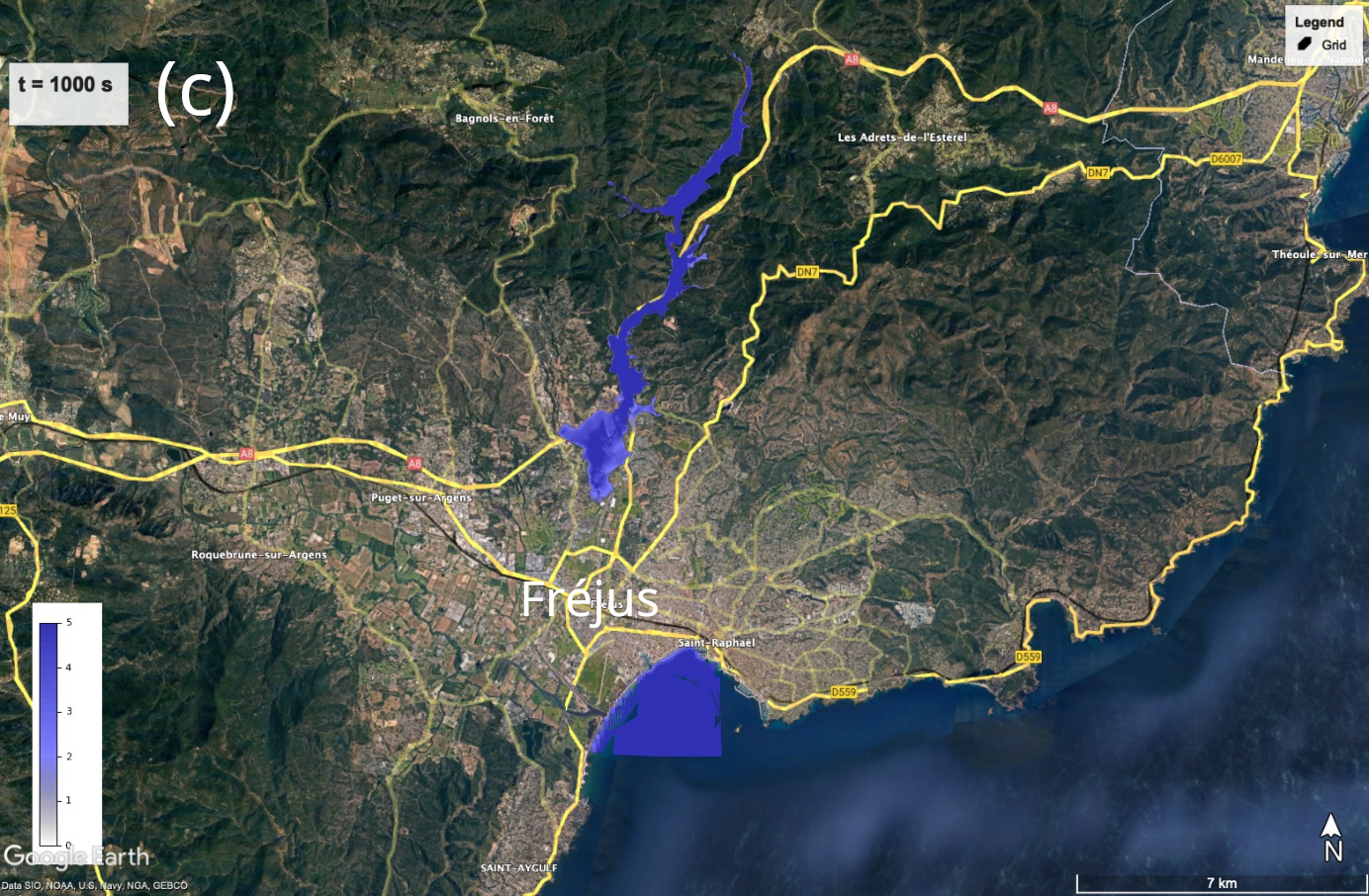}
        % \caption{At $t=1000~$seconds}
        \label{fig:allshock1}
      \end{subfigure}
	\begin{subfigure}[b]{0.7\textwidth}
	  \includegraphics[width=\textwidth]{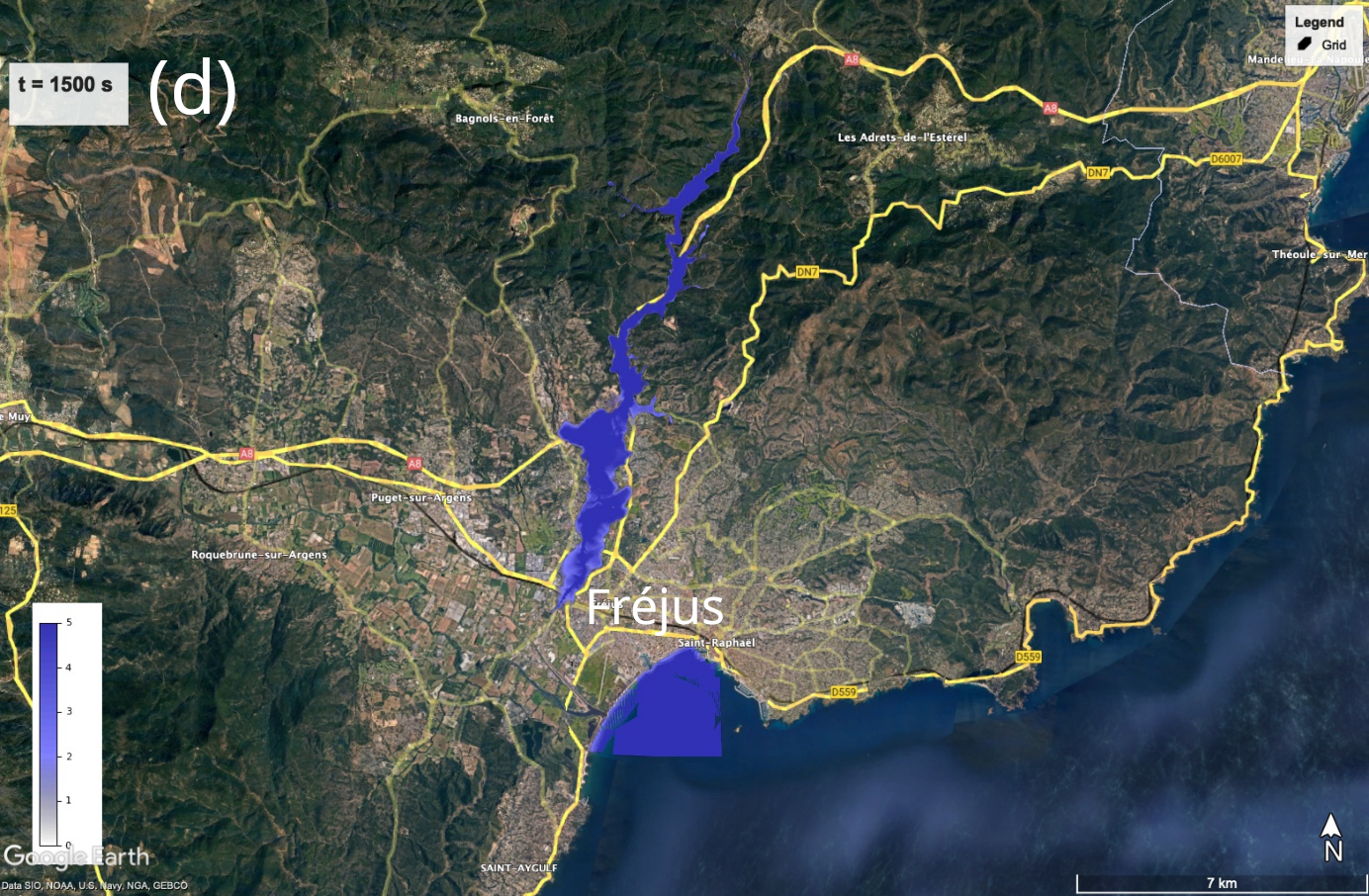}
	  % \caption{At $t=1500~$seconds}
	  \label{fig:allshockv0}
	\end{subfigure}\hfil % <-- added
	\caption[]{(cont.) The sequence of the flood resulting from the Malpasset Dam failure is illustrated using Google Earth imagery at various times. Between $t = 1000$ (panel c) and $t = 1500$ \unit{seconds} (panel d), the flood wave progressed through the valley, reaching Fr\'{e}jus approximately $21$ \unit{minutes} after the dam's collapse.(\textcopyright Google Earth imagery)}
	\label{fig:gge}
  \end{figure}
%------------------------------------------------------
  \begin{figure}[p]\ContinuedFloat
	\begin{subfigure}[b]{0.7\textwidth}
	  \includegraphics[width=\textwidth]{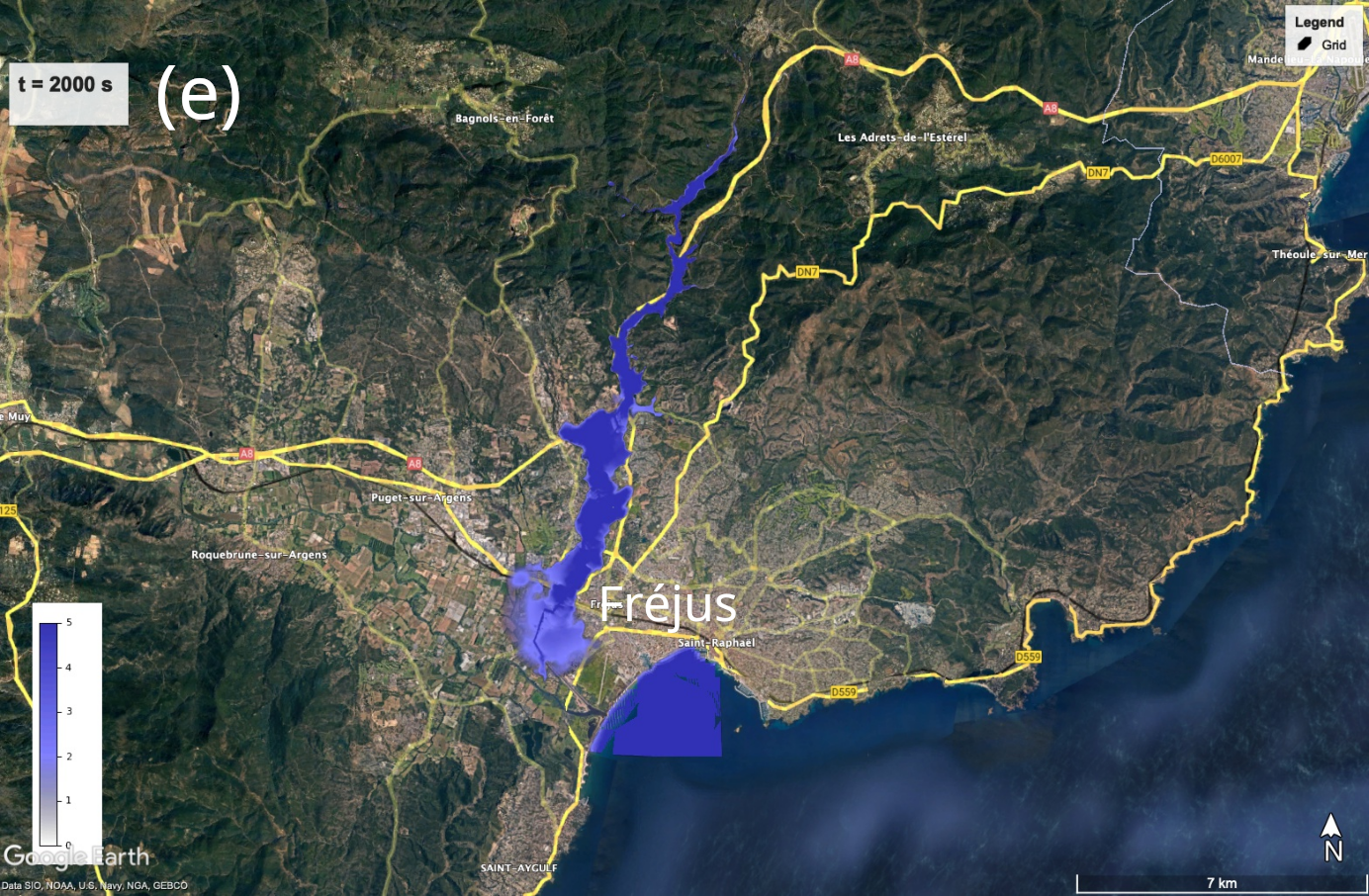}
	  % \caption{At $t=2000~$seconds}
	  \label{fig:allshockv3}
	\end{subfigure}\hfil % <-- added
	\begin{subfigure}[b]{0.7\textwidth}
	  \includegraphics[width=\textwidth]{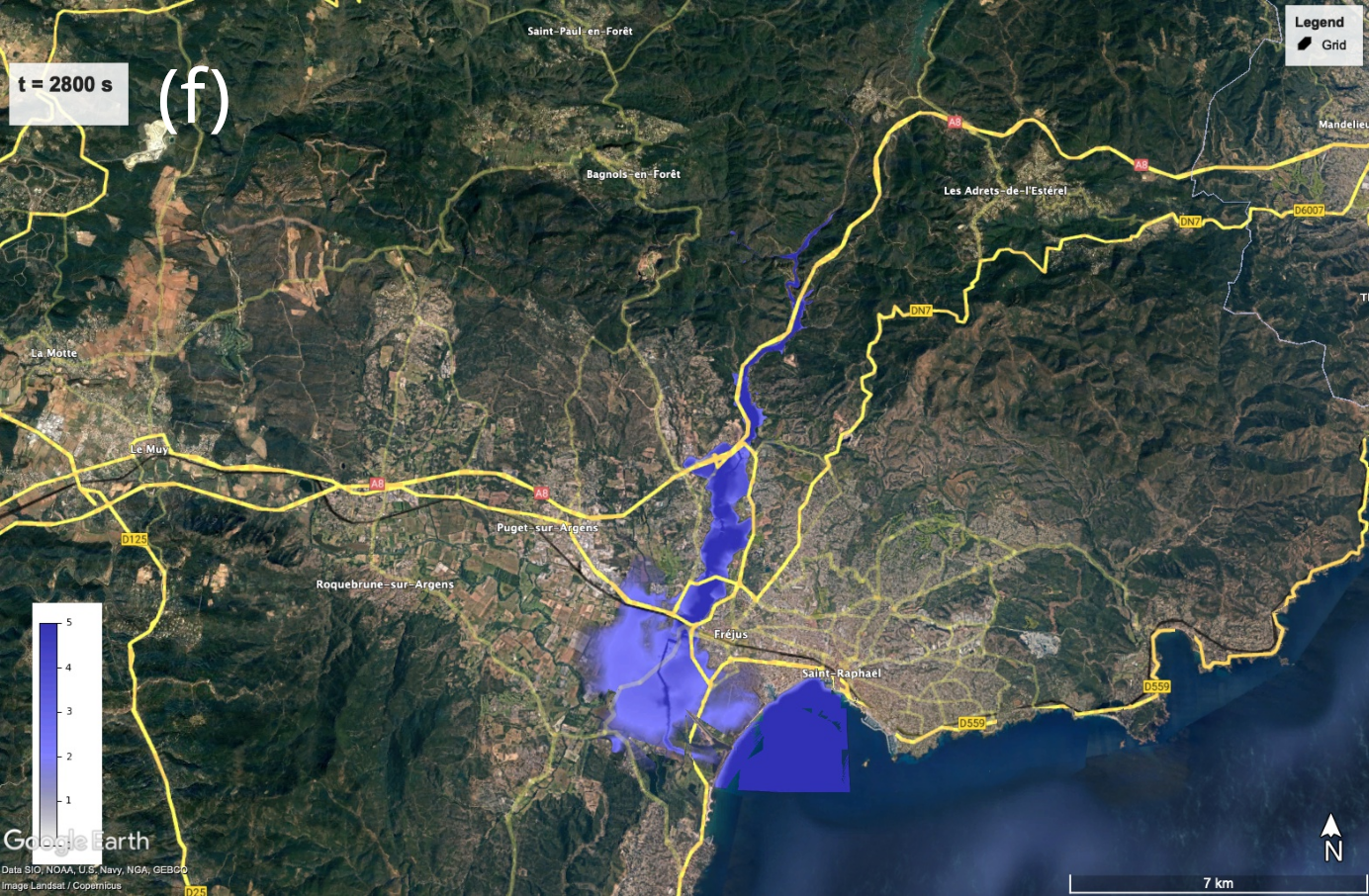}
	  % \caption{At $t=2800~$seconds}
	  \label{fig:allshockv1}
	\end{subfigure}
	\caption[]{ (cont.) The sequence of the flood resulting from the Malpasset Dam failure is illustrated using Google Earth imagery at various times. The imagery at $t = 2000$ \unit{seconds} (panel e) shows the flooded area surrounding Fr\'{e}jus. At $t = 2800$ \unit{seconds} (panel f), the flood waves enter the Mediterranean Sea. (\textcopyright Google Earth imagery).}
	\label{fig:gge}
  \end{figure}
%============================================================================

% \vspace{-0.5cm}
\subsection{GeoFlood's Google Earth graphical toolbox}
Geoflood contains a graphical toolbox that allows rendering simulation frames from an arbitrarily projected Cartesian AMR mesh into Google Earth, allowing the utilization of geo-referenced layers. This GeoFlood toolbox is designed to facilitate a user's ability to dynamically and rapidly assess a flood scenario across detailed geo-referenced topography as a simulation proceeds. \Fig{gge} shows some example Google Earth screen-capture snapshots of the Malpasset flood at various times.

\section{Conclusions}
\label{sec:conclusions}
A key motivation behind GeoFlood's development was to integrate the strengths of multiple specialized codes into a single package that is massively parallel and scalable, while still being robust and accurate for highly dynamic flood modeling applications in intricate or irregular topography. This was achieved through multi-block-based adaptive mesh refinement and PDE solvers from Clawpack and GeoClaw via ForestClaw, the creation of novel routines for data and input handling, and innovative functional linkages between code libraries. GeoFlood also includes features that facilitate model set-up and analysis (topography and other data input, graphical toolbox, etc.). Because the model is novel, the goal behind this paper is to validate GeoFlood's ability to solve the shallow-water equations for flooding applications, as well as provide a preliminary demonstration of its scalability compared to another flood modeling package, GeoClaw. For very large-scale problems spanning vast regions, we anticipate that GeoFlood’s ability to scale across many more cores (millions) will significantly reduce computational time compared to traditional flood modeling software.
\par
GeoFlood's performance was rigorously tested and confirmed through three distinct benchmark challenges. Benchmark simulation results were compared with those from the HEC-RAS and GeoClaw models. GeoFlood results on the Malpasset dam break incident were corroborated using field data, laboratory scale-model data, and numerical results from prior studies. These evaluations revealed GeoFlood's capability to effectively forecast the trajectory of flood waves validated against actual recorded events. Its validation and scalability indicate that GeoFlood may be a valuable tool for leveraging computational resources for large-scale flood risk management strategies.

\section{Acknowledgments}
The authors extend their gratitude to all contributors and developers of ForestClaw, GeoClaw, and p4est (Carsten Burstedde) for making their codes publicly available. Yu-hsuan (Melody) Shih for the initial development of the GeoClaw solver in ForestClaw, and Boise State University for providing computational resources for this study.  The authors acknowledge the financial support of the Department of Defense (DARPA AtmosSense program), NASA (ROSES Earth Surface and Interior Program), and  the National Science Foundation (NSF-DMS award \#1819257). Any use of trade, firm, or product names is for descriptive purposes only and does not imply endorsement by the U.S. Government.
%% The following commands are for the statements about the availability of data sets and/or software code corresponding to the manuscript.
%% It is strongly recommended to make use of these sections in case data sets and/or software code have been part of your research the article is based on.

% \codeavailability{The code is available on Github: \url{https://github.com/KYANJO/GeoFlood} } %% use this section when having only software code available
\vspace{-1.0cm}% \vspace{-0.5cm}

% \dataavailability{TEXT} %% use this section when having only data sets available

\codedataavailability{{The current version of GeoFlood is available  on Github: \url{https://github.com/KYANJO/GeoFlood} under the BSD 2-Clause Licence. The exact version of the model used to generate results used in this paper is archived on Zenodo \citep{geoflood-zenodo}, as are input data and scripts to run the model and produce the plots for all the simulations presented in this paper \citep{dataset-zenodo}.}}

% \sampleavailability{TEXT} %% use this section when having geoscientific samples available

% \videosupplement{TEXT} %% use this section when having video supplements available

% \appendix
% \section{}    %% Appendix A

% \subsection{}     %% Appendix A1, A2, etc.

% \noappendix       %% use this to mark the end of the appendix section. Otherwise the figures might be numbered incorrectly (e.g. 10 instead of 1).

%% Regarding figures and tables in appendices, the following two options are possible depending on your general handling of figures and tables in the manuscript environment:

%% Option 1: If you sorted all figures and tables into the sections of the text, please also sort the appendix figures and appendix tables into the respective appendix sections.
%% They will be correctly named automatically.

%% Option 2: If you put all figures after the reference list, please insert appendix tables and figures after the normal tables and figures.
%% To rename them correctly to A1, A2, etc., please add the following commands in front of them:

% \appendixfigures  %% needs to be added in front of appendix figures

% \appendixtables   %% needs to be added in front of appendix tables

%% Please add \clearpage between each table and/or figure. Further guidelines on figures and tables can be found below.

\vspace{-1.1cm}% \vspace{-0.5cm}

\authorcontribution{Brian Kyanjo drafted the manuscript, set up and carried out the simulations, and designed the code under the supervision of Donna Calhoun and David L George. The authors read and approved the final manuscript.} %% this section is mandatory

\vspace{-1cm}% \vspace{-0.5cm}

\competinginterests{The authors declare no competing interests present} %% this section is mandatory even if you declare that no competing interests are present

% \disclaimer{TEXT} %% optional section

% \begin{acknowledgements}
% TEXT
% \end{acknowledgements}

%% REFERENCES

%% The reference list is compiled as follows:

% \begin{thebibliography}{}

% \bibitem[AUTHOR(YEAR)]{LABEL1}
% REFERENCE 1

% \bibitem[AUTHOR(YEAR)]{LABEL2}
% REFERENCE 2

% \end{thebibliography}

%% Since the Copernicus LaTeX package includes the BibTeX style file copernicus.bst,
%% authors experienced with BibTeX only have to include the following two lines:
%%
\bibliographystyle{copernicus}
\bibliography{geoclaw.bib}

\end{document}